\documentclass[twocolumn]{aa} 

\usepackage{graphicx,multirow}
\usepackage{float}
\usepackage{placeins}
\usepackage{threeparttable}
\usepackage[colorlinks=true,     linkcolor=blue, citecolor=blue, filecolor=blue, urlcolor=blue]{hyperref}
\usepackage{amssymb,amsfonts,amsmath,amstext,amsgen,amsopn,amsxtra,indentfirst,times,longtable}
\usepackage{caption}

\let\ACMmaketitle=\maketitle
\renewcommand{\maketitle}{\begingroup\let\footnote=\thanks \ACMmaketitle\endgroup}
\providecommand{\orcidlink}[1]{}
\providecommand{\orcidlinkvtwo}[1]{}

\begin{document} 

   \title {Refractive indices of photochemical haze analogs for Solar System and exoplanet applications : a cross-laboratory comparative study between the PAMPRE and COSmIC experimental set-ups}

   \author{T. Drant
          \inst{1,2,3}
          \and
          E. Sciamma-O'Brien
          \inst{4}
          \and 
          L. Jovanovic 
          \inst{4}
          \and 
          Z. Perrin
          \inst{1,5}
          \and
          L. Maratrat 
          \inst{1}
          \and
          L. Vettier 
          \inst{1}
          \and 
          E. Garcia-Caurel  
          \inst{6}
          \and 
          J.-B. Brubach
          \inst{7}
          \and
          D.H. Wooden 
          \inst{4}
          \and
          T.L. Roush 
          \inst{4}
           \and 
          C.L. Ricketts  
          \inst{4}
          \and 
          P. Rannou
         \inst{8}          
       }
   \institute{ University of Paris Saclay, CNRS, LATMOS, OVSQ, 11 Boulevard d'Alembert, 78280 Guyancourt, France. 
         \and
             Ludwig Maximilian University, Faculty of Physics, Observatory of Munich, Scheinerstrasse 1, Munich D-81679, Germany.
           \and
            ETH University, Center for Origin and Prevalence of Life, Department of Earth and Planetary Sciences, 8092 Zurich, Switzerland. 
             \email{tdrant@ethz.ch}
           \and
           NASA Ames Research Center, Space Science and Astrobiology Division, Code ST, Moffett Field, CA 94035, USA.
          \and 
            LISA, Université Paris Est Creteil and Université Paris Cité, CNRS, F-94010 Créteil, France.
          \and
          Ecole Polytechnique, LPICM, Route de Saclay, 91120 Palaiseau, France.
           \and
            Synchrotron SOLEIL, L'Orme des Merisiers, 91190 Saint-Aubin, France.
            \and
            GSMA, Université de Reims Champagne-Ardenne, CNRS, 51687 Reims, France.}

   \date{ }

  \abstract
   {Previous observations of Titan, Pluto and Solar System gas giants along with recent observations of exoplanet atmospheres with the James Webb Space Telescope taught us that photochemical hazes are ubiquitous and form in a variety of temperature, gas composition and irradiation environments. Despite being crucial to understand their impact on observations and on the planetary radiative budget, the refractive indices of these haze particles are unknown and strongly influenced by changes in the gas phase chemistry. In this study, we perform a cross-laboratory investigation to assess the effect of the experimental set-up and gas composition on the refractive indices of Titan, Pluto and exoplanet haze analogs. We report new data in a broad spectral range from UV to far-IR (up to 200 $\mu$m) for future use in climate models and retrieval frameworks. 
   
   We compare the refractive indices of laboratory haze analogs produced from six different gas compositions where we varied the relative abundances N$_2$/CH$_4$ and CH$_4$/CO in the initial gas mixture using the PAMPRE (LATMOS, France) and COSmIC (NASA Ames Research Center, USA) experimental set-ups. We observed strong variations of the k values in the spectral range from UV to near-IR between the different analogs which is caused by both the experimental set-up and by changes in the gas N$_2$/CH$_4$ ratio. We found that the gas N$_2$/CH$_4$ ratio has a stronger influence on the haze refractive indices in the entire spectral range compared to the gas CH$_4$/CO ratio. The experimental set-up is the primary factor affecting the refractive indices confirming that the gas residence time, irradiation, pressure and gas temperature are important parameters influencing the composition of the solid analog. The higher n and k values in the UV-Visible and the stronger amine/alkene/aromatic/hetero-aromatic signatures in the mid-IR for the COSmIC analogs are consistent with a stronger incorporation of nitrogen into the COSmIC solid analogs compared to the PAMPRE analogs, even for similar nitrogen abundances in the gas phase. Haze analogs produced in gas mixtures without nitrogen, similar to the stratosphere of Solar System gas giants and the H$_2$-dominated atmospheres of sub-Neptunes, are generally more transparent with lower n values in the entire spectral range from UV to mid-IR and should therefore be considered carefully in climate and observational applications. The variations of IR absorption features between hazes produced with and without nitrogen could help constrain the presence of N$_2$ in exoplanet atmospheres.
}

   \titlerunning{Refractive indices of photochemical haze analogs for solar system and exoplanet applications}
   \maketitle

\section{Introduction}

Photochemical hazes are ubiquitous in the atmospheres of the different objects in the outer Solar System \citep{Kim91,Ortiz96,Wong03,Gladstone16,Horst17,Mills21,Sanchez-Lavega23}, as well as in exoplanet atmospheres \citep{Gao21}. Their formation is triggered by a complex radical organic chemistry mainly driven by vacuum ultraviolet (VUV) photons in the upper layers of planetary atmospheres, but also possible via energetic cosmic rays (e.g. suggested for Pluto's dark surface material, \cite{Krasnopolsky20}) and electrons accelerated in planetary magnetospheres (e.g. Jupiter's polar hazes and Titan hazes, \cite{Wilson04}, \cite{Wong03}). The chemical pathways leading to haze formation and controlling the composition of the resulting solid material remain largely unknown, posing the main limitation to numerically reproduce this essential piece of the atmospheric physico-chemical structure. Insights on the gas phase haze precursors, solid composition and optical properties primarily rely on laboratory experiments which provide crucial data to support modeling efforts and help interpret observations.   \\

Laboratory experiments revealed that hazes form under a broad variety of gas compositions including CH$_4$- (e.g., \cite{Khare84} and \cite{Imanaka04}), CO$_2$- \citep{Trainer04b,Gavilan18}, CO- \citep{Horst14} and H$_2$O-rich \citep{He18} mixtures. The formation pathways and gas phase precursors are known to vary with the gas composition, for instance transitioning between aliphatic and polycyclic aromatic hydrocarbon pathways following the methane abundance \citep{Trainer04}. In N$_2$-dominated gas mixtures representative of Titan and Pluto's atmosphere, the photochemical build-up of nitrile chains contributes significantly to the formation of these aerosols \citep{Khare94,Gautier11,Sciamma14,Horst18,Perrin21,Perrin25}. Ion chemistry also leads to the formation of charged hydrocarbon and nitrile compounds acting as precursors to these haze particles during their formation in the ionosphere \citep{Vuitton09,Dubois20}. The composition of the solid analog is strongly affected by the gas phase chemistry, it is thus controlled by the initial gas composition but also by the experimental conditions, i.e., the pressure, the gas temperature and the irradiation efficiency \citep{Sciamma14,Jovanovic20,Carrasco16,Nuevo22}. A general framework has yet to be established  to link the gas phase chemistry to the composition and production rate of haze particles under a variety of atmospheric environments. This solid composition must however be considered carefully as it significantly affects the optical properties of the haze particles and thus strongly impacts observations of planetary atmospheres (see reviews of \cite{Brasse15} and \cite{Sciamma24}). \\

In radiative transfer models, the haze composition is accounted for using the material's complex refractive index, with the real part n describing dispersion of light and the imaginary part k quantifying absorption. These refractive indices are intrinsic properties controlled by the chemical composition and thus independent of the particle's size and shape. \cite{Khare84} provided the first refractive indices of laboratory-generated haze analogs for a Titan-like gas mixture with 90\% N$_2$ and 10\% CH$_4$. This pioneer dataset is widely used in the community for its broad spectral range, never reproduced to this day. There currently exists several groups able to produce laboratory haze analogs and measure their refractive indices. Among them, we identify the NASA Ames group with the COSmIC set-up \citep{Sciamma23}, the John Hopkins group with the PHAZER set-up \citep{He22b,He23}, the LATMOS group with the PAMPRE set-up \citep{Sciamma12,Mahjoub12,Gavilan17,Jovanovic21,Drant24}, and the Colorado group \citep{Hasenkopf10,Ugelow17,Ugelow18,Reed23}. This massive work developed throughout the past decades revealed that the refractive indices of haze analogs are strongly influenced by the initial gas composition but also by the properties of the set-up itself \citep{Brasse15,Sciamma24}. The lack of cross-laboratory comparative studies limits our understanding of the effect of the set-up as each group focuses on different gas compositions. In addition, data currently available are often limited to the UV-Visible and near-infrared (NIR) spectral range \citep{Ramirez02,Tran03,Mahjoub12, Sciamma12, Gavilan17, He22b, Sciamma23}. Climate calculations and data analysis of spectroscopic observations both rely on refractive indices at longer wavelengths, especially in the era of the James Webb Space Telescope (JWST) operating exclusively in the IR. \\

Observations of Titan, Pluto, Solar System gas giants and exoplanets revealed a significant influence of photochemical hazes \citep{Baines05,Kim11,Stern15,Gao21}. The imaginary part of the refractive index of the haze could be retrieved from the various Cassini-Huygens observations of Titan \citep{Rannou10,Vinatier12,Rannou22}. Comparison to laboratory analogs helped constrain the composition of the solid material. Although observations of Titan hazes at short wavelengths (UV-NIR) suggest a high incorporation of nitrogen in the solid \citep{Sciamma23}, their mid-IR (MIR) features don't show evidence of nitrogen and rather point to a solid rich in carbon and hydrogen \citep{Bellucci09,Kim11,Vinatier12,Kim13,Kim18}. The observations of Pluto's surface which revealed a strong heterogeneity in the geological terrains point to the presence of photochemical hazes mixed within ice particles \citep{Stern15,Olkin17,Grundy18,Gladstone19,Scipioni21,Lauer21,Fayolle21}. The analysis of these observations is also strongly dependent on the refractive index data used for the haze particles \citep{Protopapa20} which could be different from Titan hazes following the presence of CO, as suggested by the data in \cite{Jovanovic21}. Hazes are also observed in the Solar System gas giants, produced from a gas composition poor in nitrogen and thus rich in carbon and hydrogen compared to Titan and Pluto. Refractive index data are critically lacking for these conditions even though they are expected to differ significantly from N-rich hazes \citep{Khare87}. Early observations of exoplanet atmospheres with the Hubble space telescope in the Visible-NIR revealed evidence of scattering induced by high-altitude aerosols \citep{Ohno20,Gao21}. Expected as the main form of aerosols in exoplanetary atmospheres with temperatures <1000K \citep{Gao20}, photochemical hazes strongly influence the thermal profile by heating the upper atmosphere and cooling the lower atmosphere \citep{Lora18,Lavvas21} to an extent partially controlled by the imaginary refractive index k \citep{Adams19,Lavvas21,Steinrueck23}. Modeling and data analysis are thus sensitive to these laboratory data largely unknown for the various conditions expected in exoplanetary environments.  \\

\begin{table*}
\centering
\caption{Summary of the different haze analogs produced with the PAMPRE and COSmIC set-ups using different gas compositions.   } 
\begin{tabular}{lcccccccccccc}
 \hline
 \hline
  \rule{0pt}{2.5ex}Analog   & Gas composition & set-up  & Production$^a$ (hours) & Measurements$^b$ & Spectral range \\
 \hline 
\rule{0pt}{2.5ex}Titan 1  &  90\% N$_2$ - 10\% CH$_4$ & PAMPRE & 3 & TS, RE, FTIRS & 0.3 - 200 $\mu$m \\
\rule{0pt}{1.5ex}         &                        & COSmIC & 8.5 & RS, RE, FTIRS & 0.4 - 200 $\mu$m \\
\rule{0pt}{2.5ex}Titan 2  & 95\% N$_2$ - 5\% CH$_4$ & PAMPRE & 3 & TS, RE, FTIRS(*) & 0.3 - 200 $\mu$m \\ 
\rule{0pt}{1.5ex}         &                    & COSmIC & 10 &  RS, RE, FTIRS(*) & 0.4 - 200 $\mu$m \\ 
\rule{0pt}{2.5ex}Pluto & 95\% N$_2$ - 4.95\% CH$_4$ - 0.05\% CO & PAMPRE & 3 & TS, FTIRS & 0.27 - 200 $\mu$m \\ 
\rule{0pt}{1.5ex}      &                                     & COSmIC & 11 & RS, RE, FTIRS(*) & 0.4 - 200 $\mu$m \\ 
\rule{0pt}{2.5ex}Exoplanet 1 & 95\% N$_2$ - 4\% CH$_4$ - 1\% CO & PAMPRE & 5 & TS, FTIRS & 0.285 - 200 $\mu$m \\
\rule{0pt}{1.5ex}        &                         & COSmIC & 18 & RS, FTIRS & 0.35 - 30 $\mu$m \\
\rule{0pt}{2.5ex}Exoplanet 2 & 95\% Ar - 5\% CH$_4$ & PAMPRE & 5 & TS, RE, FTIRS(*) & 0.3 - 200 $\mu$m \\
\rule{0pt}{1.5ex}        &                         & COSmIC & 18 & RS, RE, FTIRS & 0.4 - 15 $\mu$m \\
\rule{0pt}{2.5ex}Exoplanet 3 & 95\% Ar - 4\% CH$_4$ - 1\% CO & PAMPRE & 7 & TS, FTIRS & 0.25 - 200 $\mu$m \\
\rule{0pt}{1.5ex}        &                         & COSmIC & 18 & RS, FTIRS & 0.3 - 12 $\mu$m \\
\hline
\end{tabular}
\vspace{0.1cm}
\begin{tablenotes}
      \small
      \item \textbf{Notes.} 
      \item $^a$ The production refers to the duration of the experiment.
      \item $^b$ The following measurements were carried out to determine refractive indices from UV to NIR : transmission spectroscopy (TS), reflection spectroscopy (RS), and reflection ellipsometry (RE). Fourier-Transform IR spectroscopy (FTIRS) was used to obtain refractive indices from NIR to FIR. FTIRS measurements were also used to characterize the MIR absorption properties of haze analogs at low temperatures (40 - 288 K, see Section 3.4). These low-temperature measurements were performed on a few analogs here marked with (*).
\end{tablenotes}
\label{tab:samples}
\end{table*}

In the present work, we aim to assess the influence of the gas composition and experimental set-up on the refractive indices of haze analogs in a broad spectral range from UV to far-IR (FIR). We cover a wide range of gas compositions to characterize the effect of N$_2$ and CO, as well as, to provide state-of-the-art data for Titan, Pluto, Jupiter and exoplanet applications.  
The different gas compositions and experimental set-ups used for the production of our haze analogs are described in Section 2. Sections 3 and 4 present the measurements and models used to derive the refractive indices from UV to FIR. The results are summarized in Section 5 along with implications for observation and modeling studies.

\section{Production of laboratory haze analogs}

\subsection{Comparative study of different gas compositions}

We produced haze analog samples using various gas compositions to mimic the atmospheres of Titan, Pluto, Solar System gas giants (e.g., Jupiter) and exoplanets. Table \ref{tab:samples} summarizes the six different gas mixtures used, where we varied the relative abundances N$_2$/CH$_4$ and CH$_4$/CO. 

The Titan analogs are produced from initial gas mixtures made of N$_2$ and CH$_4$. Previous work has shown that the nitrogen abundance in the gas phase can strongly influence the refractive indices of the solid analogs from UV to NIR \citep{Mahjoub12,Sciamma23}. We thus considered two different cases where we only varied the gas relative abundance N$_2$/CH$_4$ to assess the influence on the solid analog refractive indices from UV to FIR. The Titan 1 experiment reproduces the gas composition chosen by \cite{Khare84} in their pioneering work whereas the Titan 2 conditions, with a higher N$_2$/CH$_4$, match the abundances observed in Titan's atmosphere \citep{Niemann05,Niemann10,Gautier24}. 

For the Pluto analog, we added 500 ppm of CO in the N$_2$-CH$_4$ gas mixture following the known abundances in Pluto's atmosphere previously constrained by observations with the Very Large Telescope \citep{Lellouch11} and the Atacama Large Millimeter/Submillimiter Array \citep{Lellouch16}. Comparison between the Titan 2 and Pluto analog refractive indices is crucial to understand the effect of CO and predict variations between the optical properties of Titan and Pluto aerosols. 

For the first exoplanet analog (Exoplanet 1, Table \ref{tab:samples}), we further decreased the relative CH$_4$/CO gas abundance to reproduce the lower C/O environments expected in exoplanet atmospheres (e.g. \cite{Liggins20} or \cite{Liggins23}). Previous investigations focused on the effect of the CH$_4$/CO$_2$ gas relative abundance \citep{Gavilan17,Gavilan18,Drant24} since this molecular pair likely prevailed in the atmosphere of early Earth \citep{Arney16} and could encode a biotic origin of methane in exoplanet atmospheres \citep{Mikal22}. On the other hand, the CO-CH$_4$ pair is expected in rocky exoplanet atmospheres produced from a reduced planetary interior via outgassing \citep{Liggins23,Tian24,Drant25} as well as in the H$_2$-dominated atmospheres of sub-Neptunes \citep{Wogan24,Shorttle24}. A CH$_4$/CO gas relative abundance of 4 was chosen for the Exoplanet 1 analog, equivalent to the previous CH$_4$/CO$_2$ ratio used in \cite{Drant24}, to directly compare the effect of CO vs. CO$_2$ on the refractive indices of haze analogs produced in N$_2$-dominated gas mixtures. Variations of chemical pathways and gas phase haze precursors are expected between CO- and CO$_2$-rich mixtures in response to the strong differences in their photo-dissociation efficiency. Previous investigations by \cite{Horst14} suggested that CO plays a key role on the production and growth rate of haze particles, the influence on the refractive indices, however, is yet to be explored. 

Finally, two additional exoplanet analogs were produced to mimic the N-poor environments found in the upper atmosphere of Jupiter and expected in most exoplanet sub-Neptunes \citep{Madhusudhan23,Wogan24}. For these analogs (Exoplanet 2 and 3 in Table \ref{tab:samples}), the gas compositions are similar to the Titan 2 and Exoplanet 1 cases with Ar replacing N$_2$ as the dominant gas. The inert Ar gas was chosen for its molar mass closer to N$_2$ than He or H$_2$ and the resulting effect on the properties of the plasma discharge. We aimed to avoid strong variations in the electron energy distribution between the six different experiments hence our choice to not use H$_2$ plasmas which are known for their highly energetic properties. The purpose of the Exoplanet 2 and Exoplanet 3 samples is to obtain the refractive indices of pure C-H-O hazes formed from hydrocarbon gas precursors without the contribution of nitriles. Comparison between the Titan 2 and Exoplanet 2 samples will further help understand the effect of N$_2$ on the refractive indices of photochemical haze analogs. Exoplanet atmospheres, such as sub-Neptunes, and the Solar System gas giants are H$_2$-dominated, a molecule that likely contributes to the chemistry and may for instance modify the C/H ratio of the hazes generated. \cite{Khare87} showed that the imaginary part of the complex refractive index is affected by the abundance of H$_2$ relative to CH$_4$. Future work will provide additional data for hazes produced in H$_2$-dominated gas mixtures that will be compared to the reference data provided in this study in the absence of H$_2$.  

\begin{figure}
\centering
\includegraphics[width=1.\columnwidth]{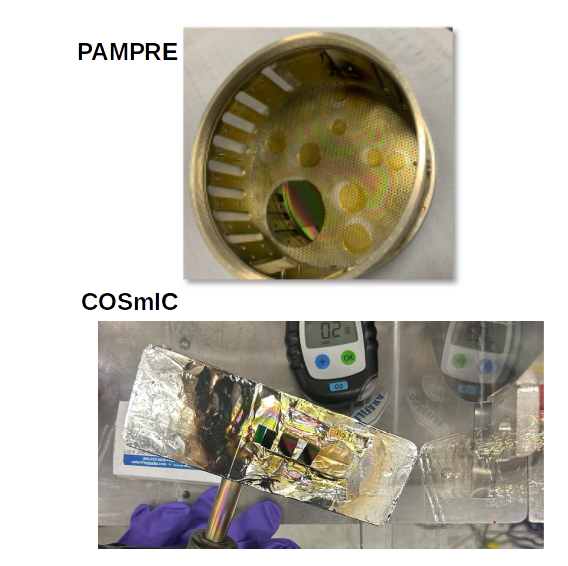}
\caption{Images of the Titan haze analogs produced with the PAMPRE (top) and COSmIC (bottom) set-ups from a mixture of 95\% N$_2$ and 5\% CH$_4$. The analog organic material is deposited onto MgF$_2$ windows and silicon wafers.    }
\label{fig:Figure1}
\end{figure}

\subsection{Comparative study of different experimental set-ups}

The PAMPRE and COSmIC facilities based at LATMOS (France) and NASA Ames (USA), respectively, were used to produce analogs from similar gas compositions (see Table \ref{tab:samples}).  \\ 

The PAMPRE (french acronym for production of aerosols in micro-gravity by a reactive plasma) set-up described in detail in \cite{Szopa06} was designed to produce Titan aerosol analogs. A 13.56 MHz radio-frequency (RF) plasma discharge sustained with a power of 30 W enables the dissociation and ionization of the main gas molecules (N$_2$ and CH$_4$ for Titan) via electron impact. This initiates a complex radical chemistry similar to the one driven by VUV photons in the upper layers of planetary atmospheres. The properties of the plasma source were characterized in detail through experimental measurements and modeling which revealed that the electron energy distribution is relatively similar to the Solar spectrum with a more significant contribution at high energies boosting dissociation and ionization mechanisms \citep{Alves12}. The gas mixture (e.g. 95\% N$_2$ - 5\% CH$_4$ for the Titan 2 conditions, see Table \ref{tab:samples}) is injected continuously in the PAMPRE chamber at a controlled flow rate of 55 sccm (standard cubic centimeter per minute). A pumping system is connected to the chamber to stabilize the pressure around 1 hPa. In these conditions, the initial species and the chemical products remain in the reactive medium for $\approx$ 30 s before being pumped out of the chamber. We will refer to this parameter as the gas residence time. The low-pressure environment prevents strong temperature variations within the reactive medium \citep{Alcouffe10}. Chemical reactions therefore occur at room temperature. A steel-aluminum cage is placed around the upper electrode of the PAMPRE chamber to confine the plasma in a smaller volume. We place optical windows and silicon wafers at the bottom of the cage which is, here, acting as the grounded electrode. During the experiment, the proximity to the reactive medium will allow the growth of an organic film directly onto the window/wafer acting as the sample substrate. The samples are exposed to the atmosphere for a short time during collection at the end of the experiment. Fig. \ref{fig:Figure1} (top) shows the inside of the cage at the end of the experiment with the different substrates now covered by an organic film. After collection, the samples are kept in a glove box filled with pure N$_2$ (O$_2$ : 0.1 ppm, H$_2$O : 1 ppm). Finally, the PAMPRE chamber is pumped to a secondary vacuum ($\approx$ 10$^{-6}$ hPa) and the walls are heated to remove water vapor and prevent contamination for the next experiment.     \\ 

The COsmic SImulation Chamber (COSmIC) facility described in \cite{Sciamma14} and \cite{Salama18} was designed to study the solid and gas phase chemistry for applications to planetary and astrophysical environments. The chosen gas mixture is injected in the chamber at a rate of 2000 sccm under atmospheric pressure and at room temperature. The gas flows through a thin slit (127 $\mu$m x 10 cm) using a pulse of 1.28 ms creating a supersonic jet expansion cooling the temperature adiabatically to $\approx$ 150 K and reducing the pressure to $\approx$ 30 hPa. In the stream of the expansion, a pulsed direct-current (DC) plasma discharge is created along the slit to trigger dissociation of the main gas species and initiate chemical reactions leading to the formation of haze analogs. Spectroscopic measurements in emission confirmed that the temperature remains at 200 K in the plasma discharge \citep{Sciamma17}. The supersonic speed of the gas limits the residence time to $\approx$ 3 $\mu$s in the reactive medium which is still sufficient to produce small organic particles with sizes ranging from 20 to 500 nm depending on the initial gas composition \citep{Sciamma17, Gavilan20,Sciamma20}. These small particles are transported by the jet expansion and deposited onto the optical windows and silicon wafers placed vertically on the walls, 5 cm downstream of the electrodes. The continuous stacking of particles at jet speed onto the windows and wafers creates a film-like layer similar to the analogs produced in the PAMPRE set-up. Fig. \ref{fig:Figure1} (bottom) shows the COSmIC samples collected at the end of the Titan 2 experiment. \\

The main differences between the experimental conditions of PAMPRE and COSmIC are summarized in Table \ref{tab:set-ups}. The variations of temperature, pressure, plasma discharge (irradiation efficiency) and gas residence time affect chemistry and more specifically the formation of gas phase haze precursors. As a result, differences in composition and thus optical properties are expected between the PAMPRE and COSmIC analogs. Our aim is to quantify these variations with the refractive indices (n and k) used as input parameters in radiative transfer calculations.

\begin{table}
\centering
\caption{Main differences between the PAMPRE and COSmIC experimental conditions.} 
\begin{tabular}{lcccccccccccc}
 \hline
 \hline
  \rule{0pt}{2.5ex} Properties   & PAMPRE & COSmIC \\
 \hline 
\rule{0pt}{2.5ex}Plasma discharge &  Radio frequency & Direct-current \\ 
\rule{0pt}{1.5ex}Pressure & 1 hPa & 30 hPa \\
\rule{0pt}{1.5ex}Gas flow rate & 55 sccm & 2000 sccm \\ 
\rule{0pt}{1.5ex}Gas residence time & $\approx$ 30 s & $\approx$ 3 $\mu$s \\ 
\rule{0pt}{1.5ex}Gas temperature & $\approx$ 300 K & $\approx$ 200 K \\
\hline
\end{tabular}
\vspace{0.1cm}
\label{tab:set-ups}
\end{table}

\section{Measurements on the haze analogs from UV to far-IR}

Various spectroscopic and ellipsometric measurements were carried out in the range from UV to FIR to obtain highly reliable data, and to optimize the data analysis and the accuracy of the retrieved refractive indices. The haze analogs were characterized using four types of measurements : UV-Visible-NIR transmission spectroscopy, UV-Visible-NIR reflection spectroscopy, UV-Visible reflection ellipsometry, and IR transmission spectroscopy. Table \ref{tab:samples} lists the measurements performed on each analog.  \\ 

\subsection{Transmission spectroscopy in the UV-Visible-NIR}

\begin{figure}
\centering
\includegraphics[width=1\columnwidth]{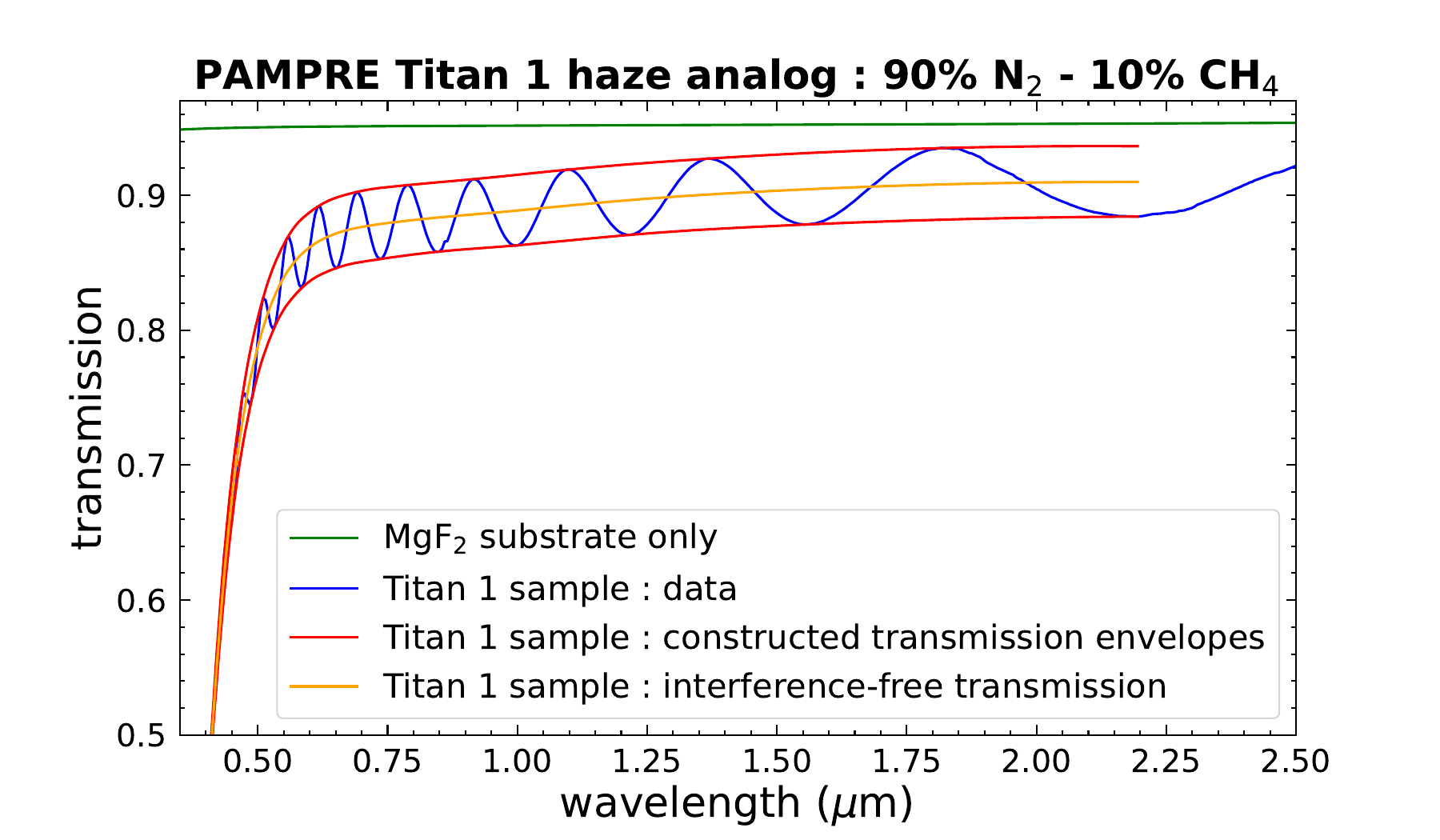}
\includegraphics[width=1\columnwidth]{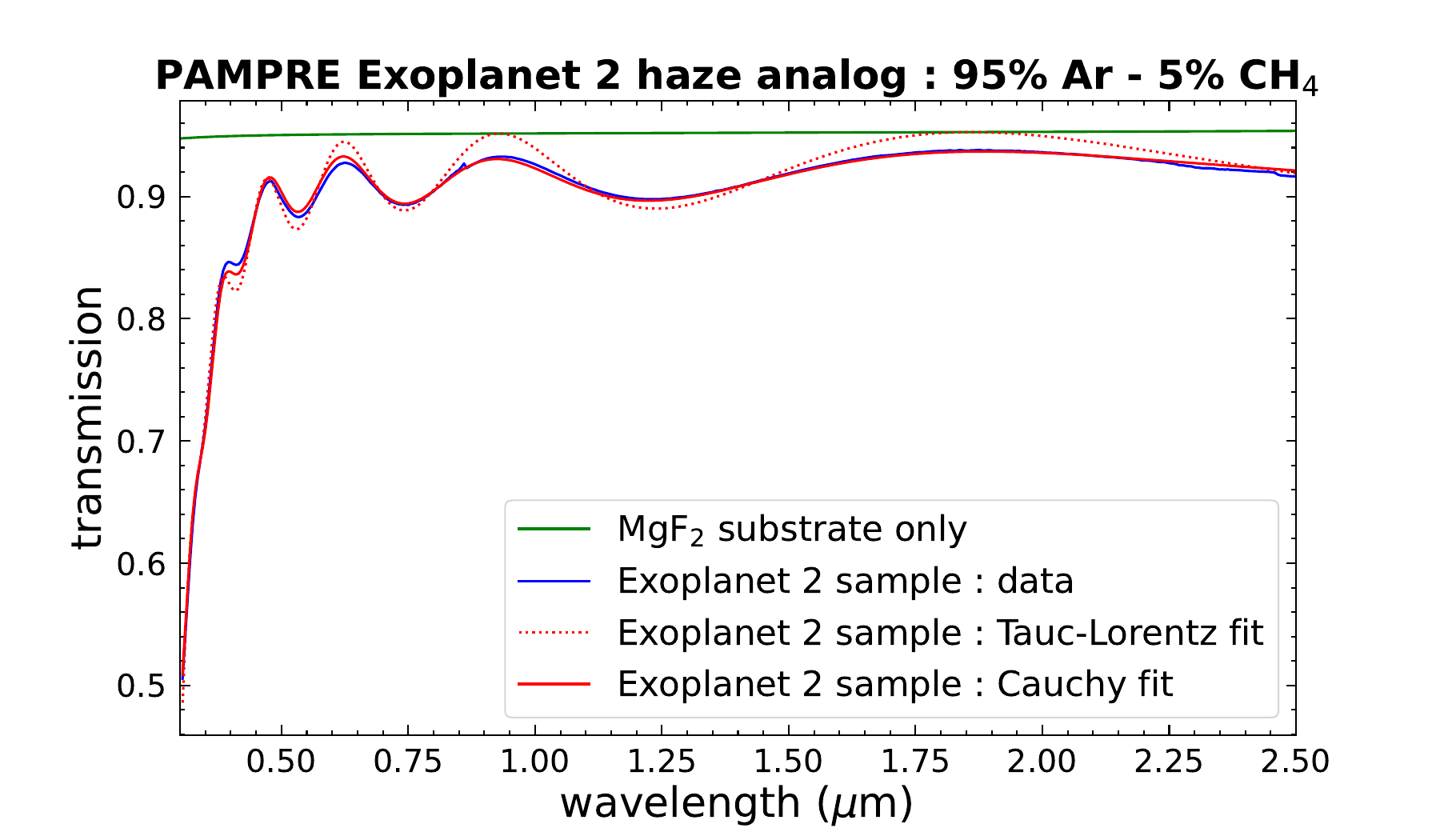}
\caption{Transmission spectra obtained on the Titan 1 and Exoplanet 2 PAMPRE haze analogs (see Table \ref{tab:samples}) from 0.3 to 2.5 $\mu$m. (top) Data and analysis using the Swanepoel method for the Titan 1 sample produced from a gas mixture of 90\% N$_2$ and 10\% CH$_4$. (bottom) Data and best fit using Cauchy and Tauc-Lorentz dispersion laws for the Exoplanet 2 analog produced from 95\% Ar and 5\% CH$_4$ in the initial gas mixture. }
\label{fig:Figure3}
\end{figure}

The transmission measurements were performed using the PerkinElmer Lambda 1050+ High Resolution Spectrophotometer based at LATMOS in Guyancourt (France) and operating from UV to NIR. We used the Three Detector Module which is designed for transmission measurements and combines both photomultiplier and InGaAs detectors to cover a broad spectral range from 0.2 to 2.5 $\mu$m. The measurement configuration used is similar to \cite{Drant24}, except that we included an optical diaphragm placed directly prior to the sample, in the sample compartment, to reduce the size of the beam spot ($\approx$ 3 - 5 mm). With this configuration, we probe a smaller area of the sample's surface and thus prevent biases caused by small thickness variability while still maintaining a high signal-to-noise ratio. We performed 2 - 3 measurements at different positions on each sample to derive an uncertainty on the refractive indices attributed to small variations of the sample thickness (see Appendix A for more details on the thickness variability of our samples). Since the PAMPRE and COSmIC samples are produced via different mechanisms (chemical growth vs. particle stacking respectively, see section 2.2), their respective film thickness varies significantly (see Appendix A and B). An important thickness variability is observed across the surface of the COSmIC samples following the particle stacking mechanism which is sensitive to spatial inhomogeneities of the plasma along the slit, during the experiments \citep{Sciamma23}. Given the millimeter-sized beam spot, it is difficult to probe a surface region with homogeneous thickness on the COSmIC samples. For that reason, another technique described in Sect 3.2 was used for the COSmIC samples and only the PAMPRE analog samples were characterized with UV-Visible-NIR transmission spectroscopy as they do not exhibit thickness variability at this scale. Fig. \ref{fig:Figure3} shows the transmission spectra (blue curves) obtained on the Titan 1 and Exoplanet 2 analogs, both produced with the PAMPRE set-up.    \\ 

\begin{figure}
\centering
\includegraphics[width=1\columnwidth]{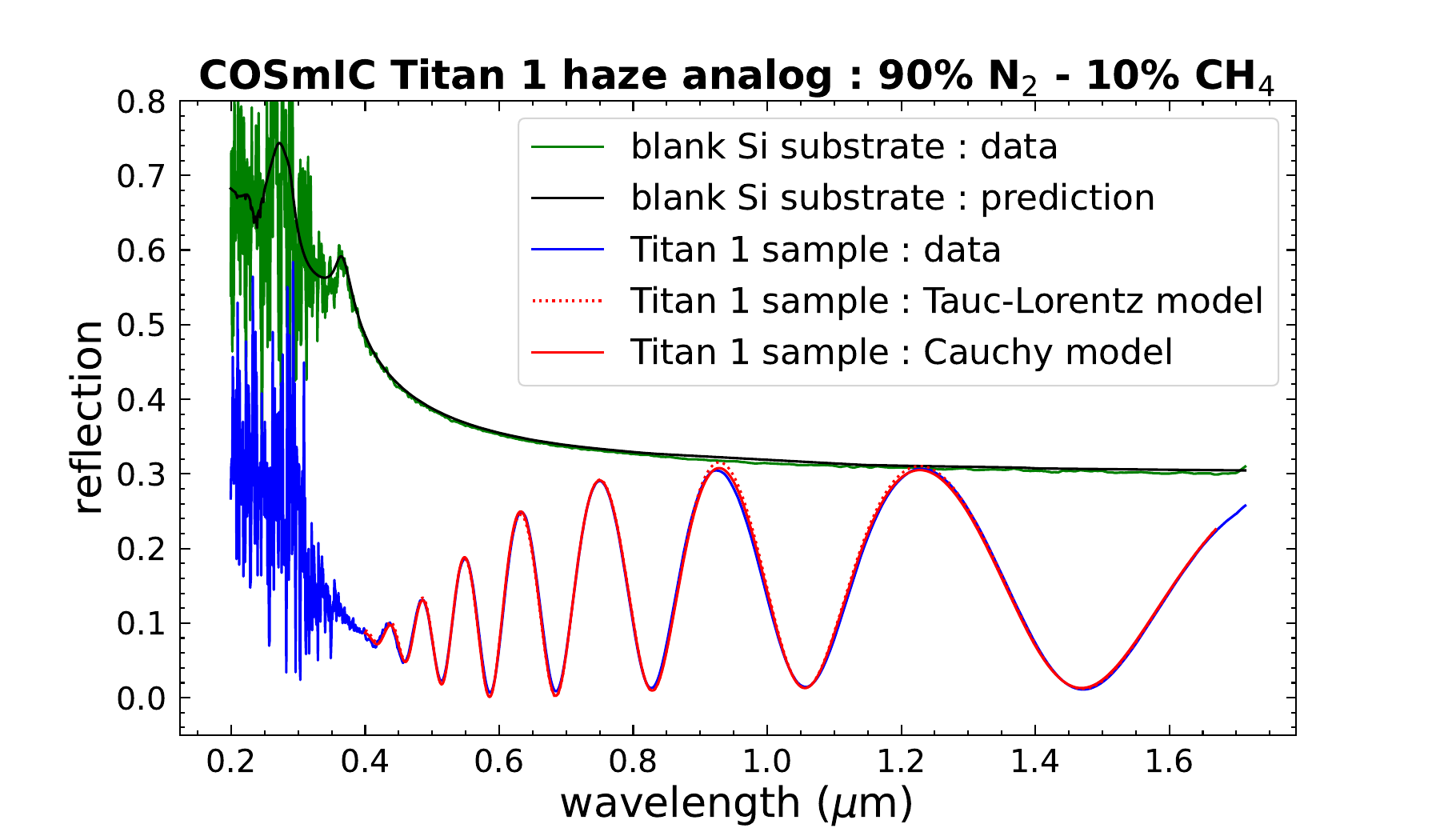}
\includegraphics[width=1\columnwidth]{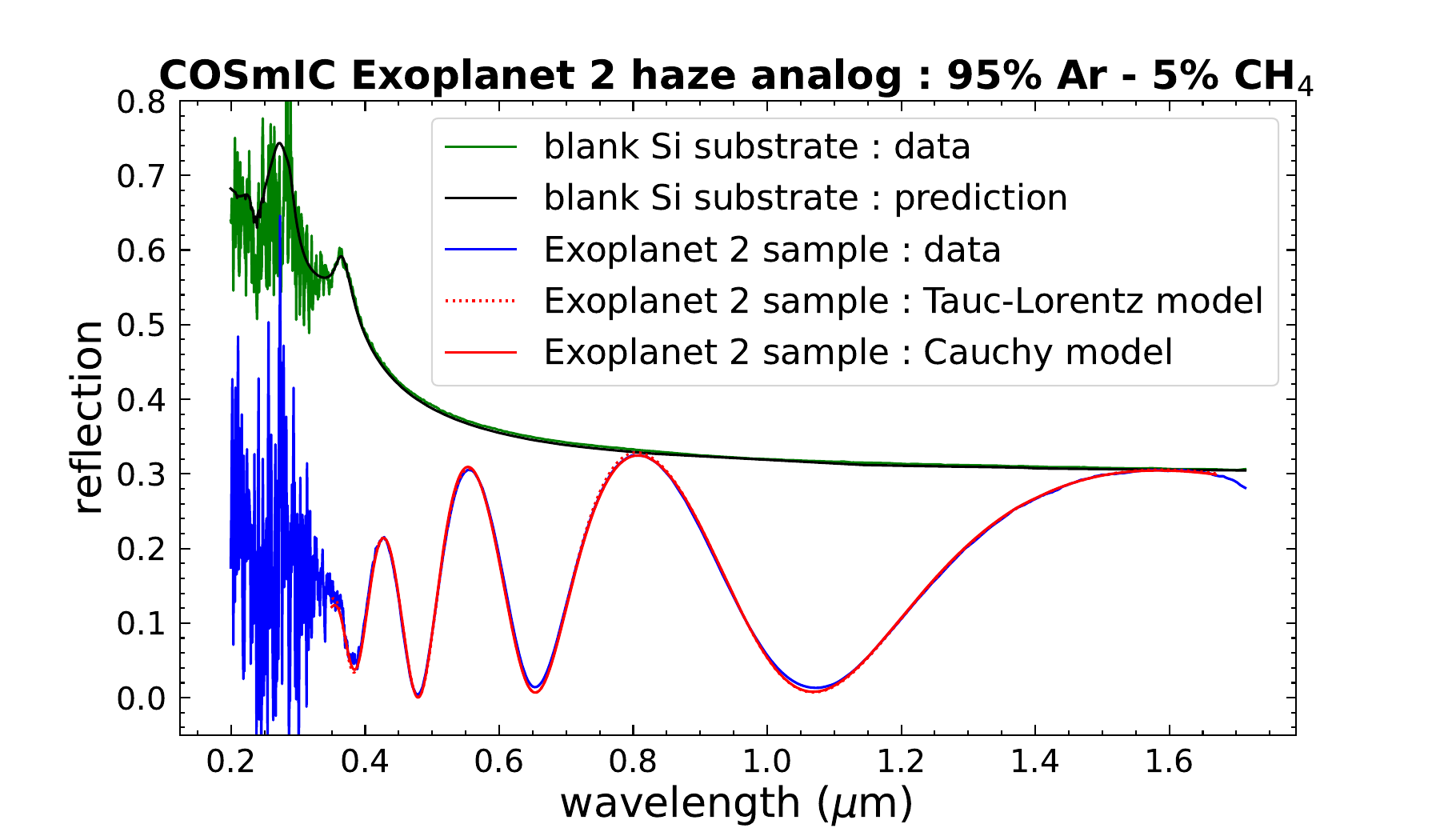}
\caption{Reflection spectra obtained on the Titan 1 and Exoplanet 2 COSmIC haze analogs (see Table \ref{tab:samples}) from $\approx$ 0.4 to 1.67 $\mu$m. The data is fitted with both Tauc-Lorentz and Cauchy functions to determine the refractive indices.   }
\label{fig:Figure4}
\end{figure}

\subsection{Reflection spectroscopy in the UV-Visible-NIR}

Spectroscopic measurements, in a reflection configuration, were carried out from 0.2 to 1.67 $\mu$m using the Filmetrics F40-UVX microscope spectrometer based at the NASA Ames Research Center. The measurements were performed at normal incidence using a 10X reflective microscope objective. A very high spatial resolution, with a spot size around 25 - 50 $\mu$m, can be reached with this instrument. It allows us to probe a region with homogeneous thickness on the COSmIC samples. These reflection measurements were also performed on a few PAMPRE analogs, we however used the refractive indices derived from the transmission data (Sect 3.1) since they cover a broader spectral range. For each COSmIC analog, 3 to 5 measurements were taken at different locations on the sample to derive an uncertainty on the retrieved refractive indices (see Appendix B for more details). In a previous study, \cite{Sciamma23} successfully used this approach to determine refractive indices of haze analogs. However accurate and positive k values could not be retrieved for all analogs in the NIR. In the present work, we improved the beam focalization of the measurements to constrain accurate k values in the NIR for each analog. Fig. \ref{fig:Figure4} presents the reflection spectra (blue curves) obtained on the Titan 1 and Exoplanet 2 COSmIC haze analogs.

\subsection{Reflection ellipsometry in the UV-Visible}

\cite{Drant24} recently reported a reasonable agreement between the refractive indices of haze analogs derived using reflection ellipsometry and transmission spectroscopy in the UV-Visible. Only a small discrepancy around 3-4\% on the real part of the refractive index n was observed. To ensure the reliability of our refractive indices in the present work, reflection ellipsometric measurements were performed on several analogs (see Table \ref{tab:samples}) in a wavelength range limited to the UV-Visible. Other transmission and reflection spectroscopic measurements described in the previous sections (Sect 3.1 and 3.2) provided data at longer wavelengths in the NIR. The main purpose of these additional ellipsometric measurements is to ensure reasonable agreement between the refractive indices determined and thus assess the robustness of our results. \\

\begin{figure}
\centering
\includegraphics[width=1\columnwidth]{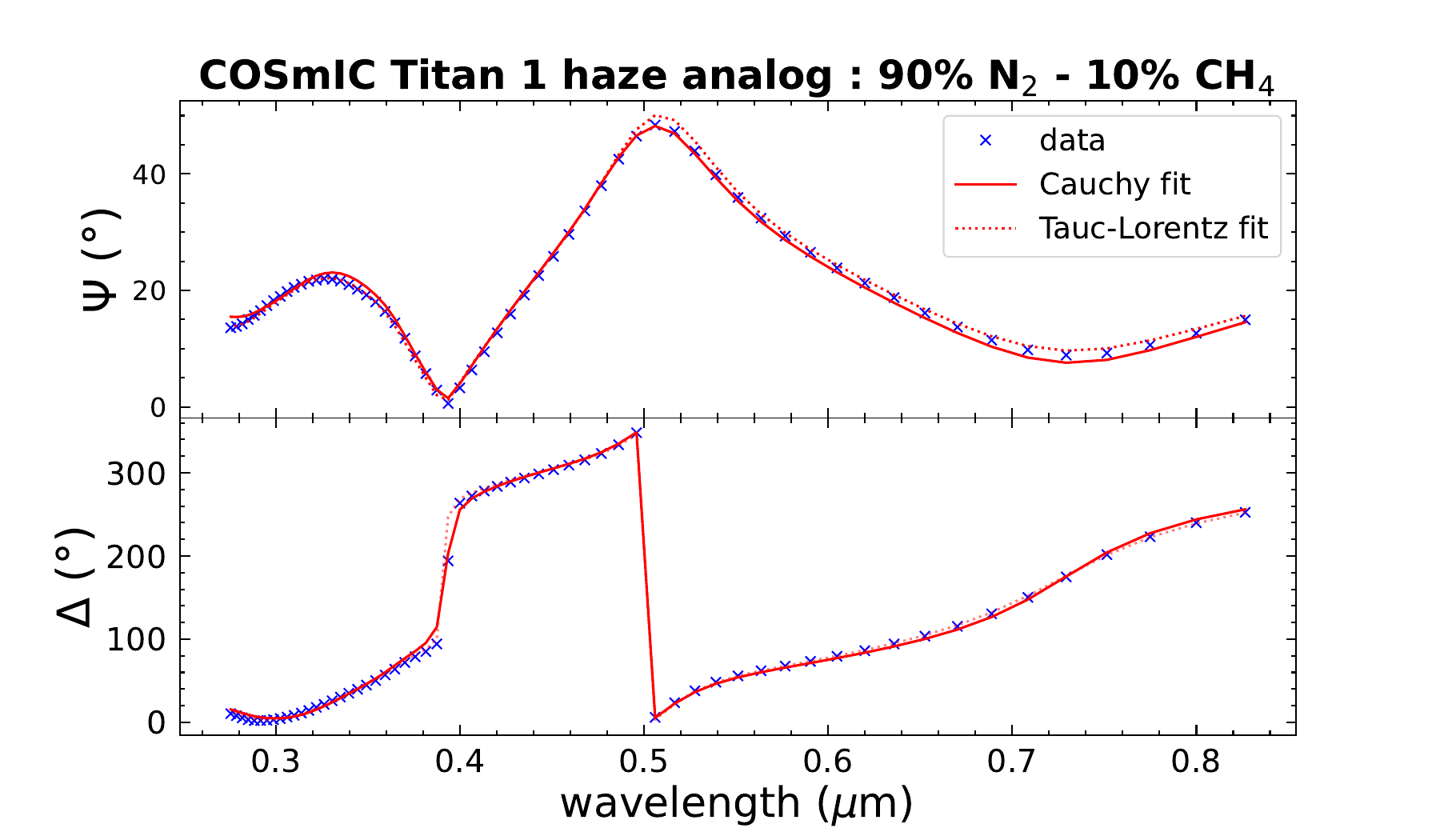}
\includegraphics[width=1\columnwidth]{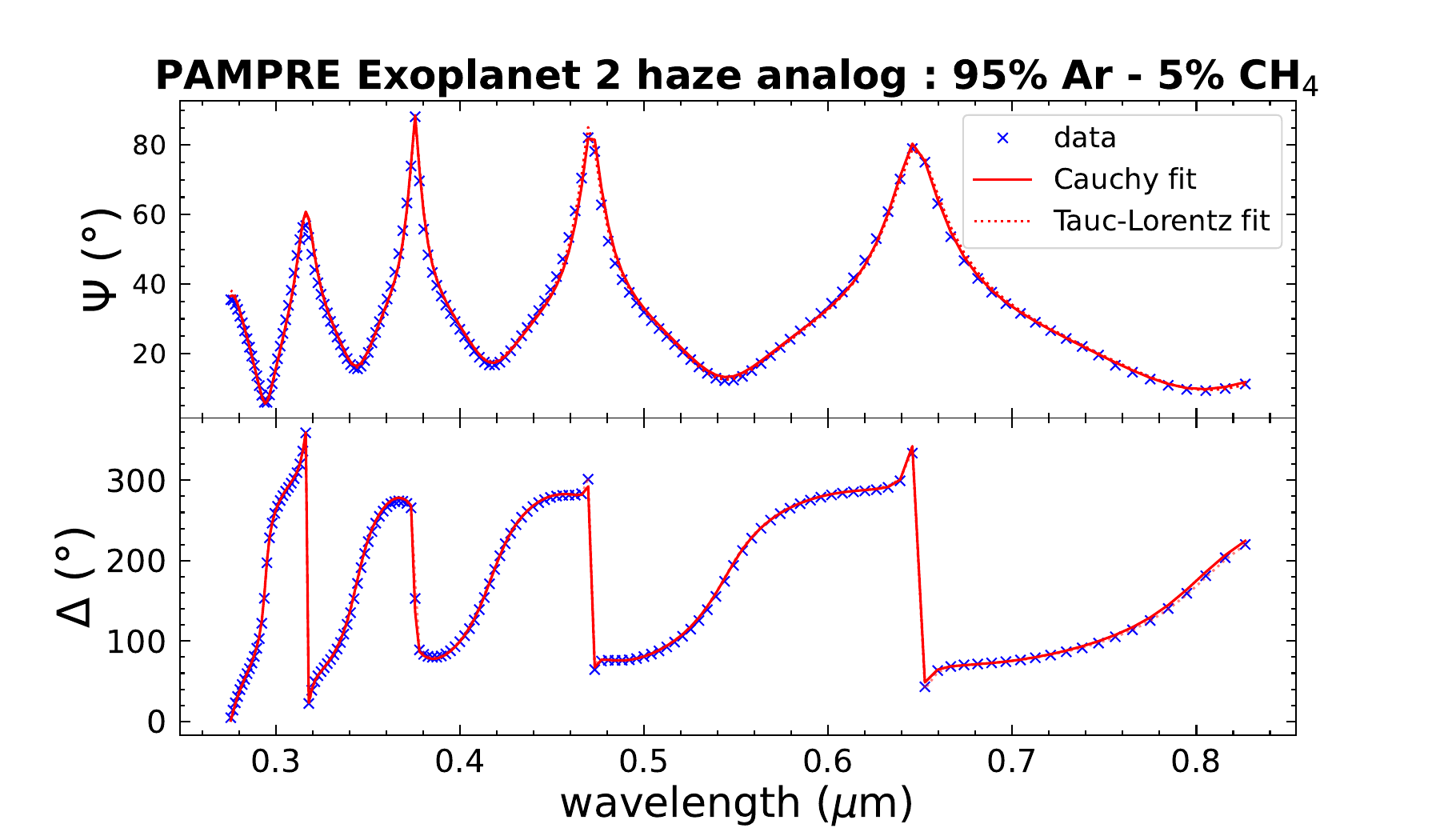}
\caption{Ellipsometric data obtained in reflection configuration for the Titan 1 COSmIC haze analog (top panel) and Exoplanet 2 PAMPRE haze analog (bottom panel). The data are fitted to the theoretical model considering both Cauchy and Tauc-Lorentz parametrized functions for the dispersion of the refractive indices.   }
\label{fig:Figure2}
\end{figure}

We used the Jobin-Yvon UVISEL ellipsometer at the LPICM (Laboratoire de Physique des Interfaces et des Couches Minces) laboratory, in Palaiseau (France), to measure the reflecting properties of our samples, from 0.27 to 0.83 $\mu$m, with two polarization states. The instrument and the operating configurations are described in \cite{Drant24}. The ellipsometric data are shown in Fig. \ref{fig:Figure2} (blue points) for the Titan 1 COSmIC analog and Exoplanet 2 PAMPRE analog (see Table \ref{tab:samples}). For each analog, we performed measurements at three different positions on the sample to assess the thickness variability and derive an uncertainty on the determined refractive indices.   \\

\subsection{Transmission spectroscopy from near-IR to far-IR}

\begin{figure}
\centering
\includegraphics[width=1\columnwidth]{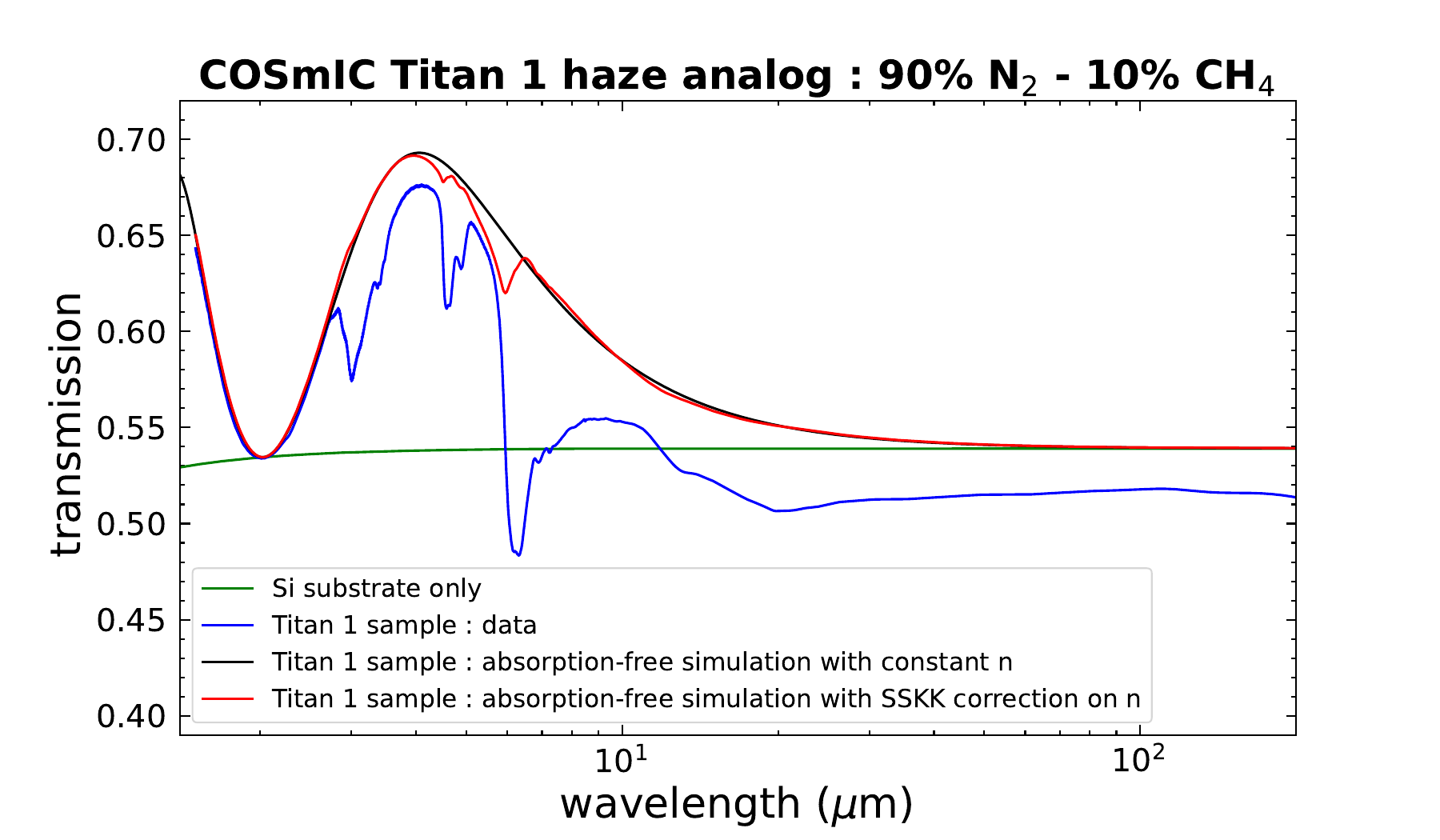}
\includegraphics[width=1\columnwidth]{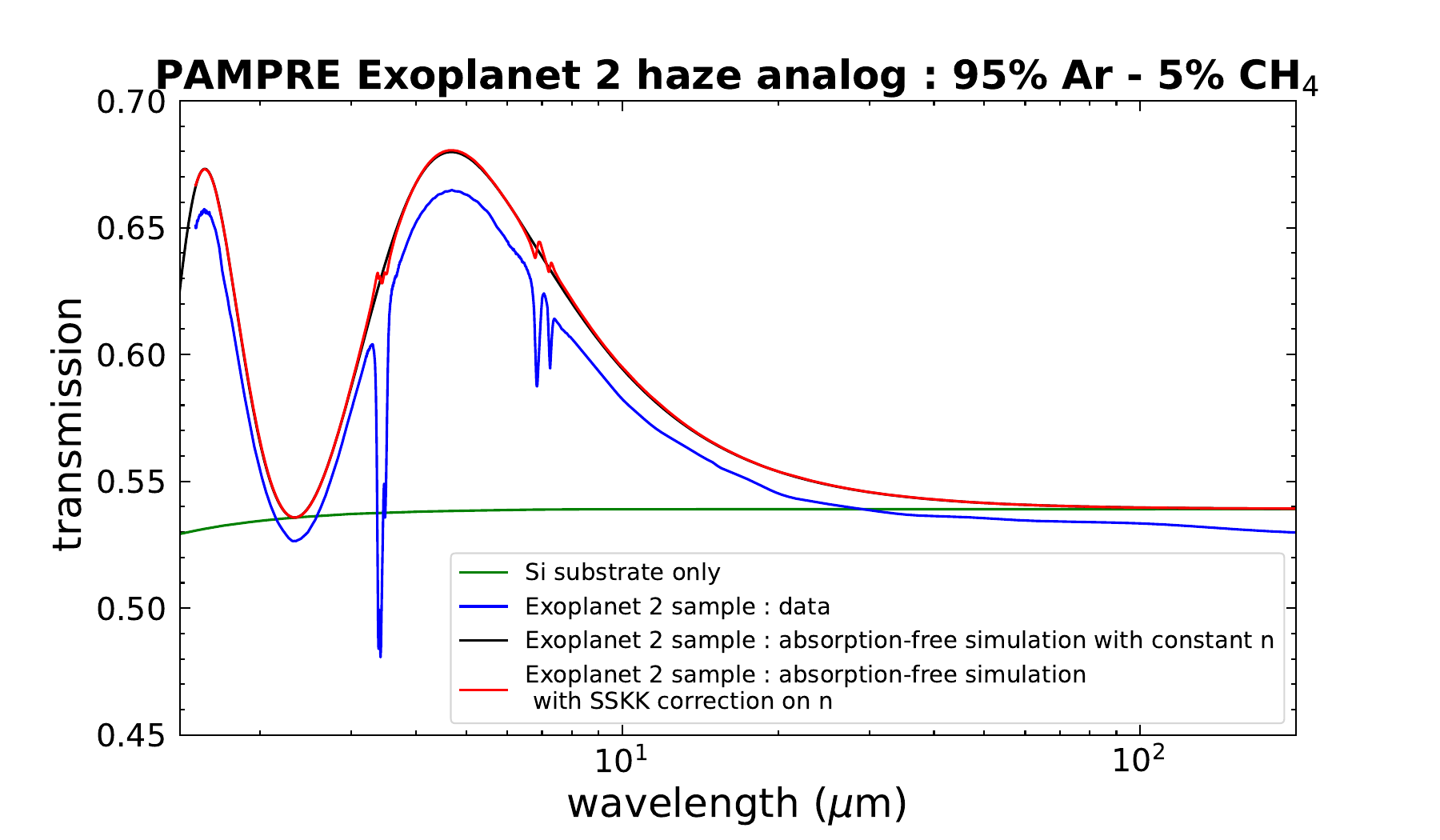}
\caption{Spectroscopic transmission spectra from 1.5 to 200 $\mu$m for the Titan 1 COSmIC haze analog (top) and the Exoplanet 2 PAMPRE haze analog (bottom). The simulated absorption-free transmission assuming a constant real refractive index n is shown (black curve). The film thickness is fitted to match the observed interference fringes in the NIR. The simulated absorption-free transmission spectrum now considering spectral variations of n in the IR (with SSKK model) is also shown (red curve). }
\label{fig:FigureS}
\end{figure}

Transmission measurements were performed from NIR to FIR (1.5 to 200 $\mu$m) using the Bruker IFS125HR Fourier-Transform IR (FTIR) spectrometer at the Ailes beamline of synchrotron SOLEIL. Two independent measurements were needed to cover the entire IR since the beamsplitter and the detector must be changed to ensure the accuracy of our data. For the NIR and MIR (1.5 - 15 $\mu$m), we used a KBr beamsplitter and a MCT detector cooled with liquid nitrogen. In the FIR (15 - 200 $\mu$m), a 6-micron multilayer mylar beamsplitter and a 4.2K bolometer detector (Irlabs) cooled with liquid helium were used. The aperture in the optical compartment, prior to the sample compartment, was set to 1 mm to ensure a small beam spot size on the haze analog sample and thus prevent biases caused by thickness variability. The transmission measurements were performed at normal incidence. During the measurement, the entire instrument is under vacuum, pumped down to $\approx$ 5.10$^{-3}$ hPa, to optimize the stability of the source and avoid the presence of atmospheric gaseous signatures in the spectra. We acquired spectra at two different positions on each analog sample in the entire spectral range. For the MIR and FIR measurements, we used similar positions on the sample to avoid thickness variations and ensure reliable concatenation of the data sets at 15 $\mu$m. Fig. \ref{fig:FigureS} shows the transmission spectra (more details on the spectra treatment in Section 4) obtained from 1.5 to 200 $\mu$m for the Titan 1 COSmIC analog and Exoplanet 2 PAMPRE analog. \\

\begin{figure}
\centering
\includegraphics[width=1\columnwidth]{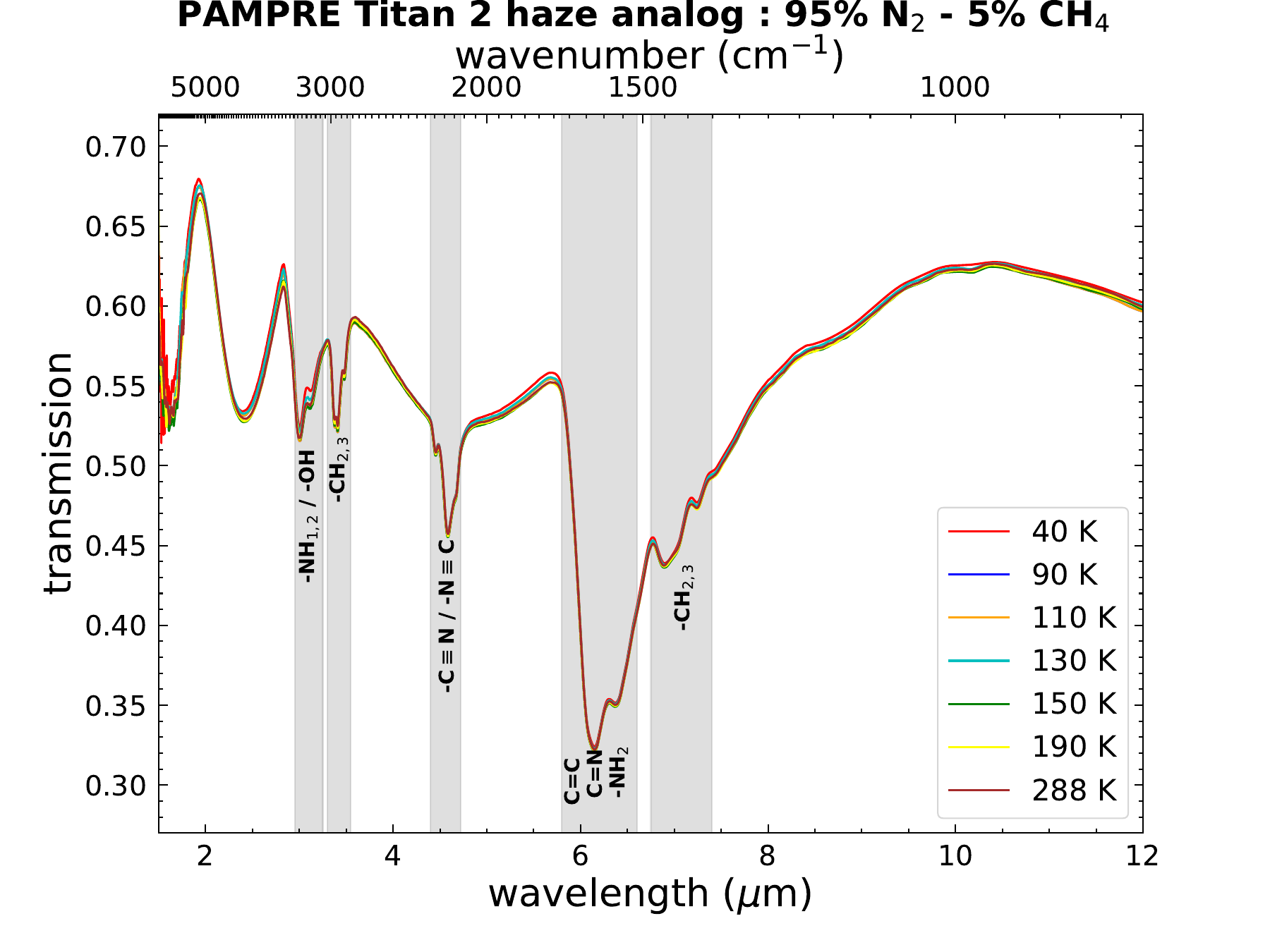}
\caption{NIR-MIR transmission spectra of the Titan 2 PAMPRE haze analog produced from a gas mixture of 95\% N$_2$ and 5\% CH$_4$ (Table \ref{tab:samples}). The different spectra were obtained at different temperatures using the cryostat set-up (see Sect 3.4).   }
\label{fig:Figure6}
\end{figure}

In addition to these data acquired at room temperature, we performed NIR and MIR transmission measurements (1.5 - 12 $\mu$m) at low temperature using a similar FTIR instrument coupled to 4K Two-Stage Pulse Tube Cryocoolers (Cryomech). The aim of these measurements is to assess the influence of the sample temperature on its refractive indices and more specifically on the relative strengths of IR vibrational modes. In contact with the sample holder, the cryostat system was used to cool the temperature of the sample down to 40 K and thus covered temperature conditions relevant for the surface and atmosphere of Titan and Pluto. With this set-up, the instrument is pumped down to a higher vacuum, $\approx$ 6.10$^{-6}$ hPa, to minimize the condensation of water on the sample that tends to add unwanted features in the spectra. These experiments are expensive in time as the temperature must be stabilized accurately at each temperature step. For that reason, these measurements were performed only on a few analogs (see Table \ref{tab:samples}) and we limited the number of temperature steps to 7. A transmission spectrum was taken at 40, 90, 110, 130, 150, 190 and 288 K. Fig. \ref{fig:Figure6} presents the transmission spectra acquired from 1.5 to 12 $\mu$m at low temperature on the PAMPRE Titan 2 haze analog. The contribution of water condensation could not be prevented completely, despite the high vacuum, which explains the variations of the transmission between 2.7 and 3.2 $\mu$m caused by the wide O-H feature overlapping with the amine N-H bands of our haze analogs.   \\ 

\section{Determination of the haze analogs refractive indices from UV to far-IR}

\subsection{General description of the model and theory}

The different spectroscopic and ellipsometric measurements used (Sect 3) monitor the reflection and/or transmission properties of our haze analog samples. The theoretical expressions of transmission and reflection for our samples, consisting of a thin film overlaying a substrate, follow \citep{Harbecke86,Stenzel91,Imanaka12} : 

\begin{equation}
\begin{split}
& T_{fs} = \frac{|t_{afs}|^2 |t_{sa}|^2 e^{- 2 \beta_{s,im}}}{1 - |r_{sfa}|^2 |r_{sa}|^2 e^{- 4 \beta_{s,im} }} & \\
\end{split}
\label{eq:T_theory}
\end{equation}
\begin{equation}
\begin{split}
& R_{fs} = |r_{afs}|^2 +  \frac{|t_{afs}|^2 |r_{sa}|^2 |t_{sfa}|^2 e^{- 4 \beta_{s,im}}}{1 - |r_{sfa}|^2 |r_{sa}|^2 e^{- 4 \beta_{s,im}}} & \\
\end{split}
\label{eq:R_theory}
\end{equation}
\begin{equation}
\begin{split}
& \textrm{With : } \ \beta = \frac{2 \pi d}{\lambda} \sqrt{n^2 - \sin({\theta_i})^2} & \\
\end{split}
\label{eq:phase_theory}
\end{equation}
\noindent where $a$, $f$ and $s$ refer to the different media, i.e., air, film and substrate respectively; T and R refer to the measured transmission and reflection respectively, whereas t and r are their respective Fresnel coefficients; $\beta$ is the optical phase, with $\beta_{s,im}$ referring to its imaginary part for the substrate; d is the thickness of the medium (film or substrate); $\theta_i$ is the angle of incidence on the sample.  \\

These expressions in Eq. \ref{eq:T_theory} - \ref{eq:phase_theory} make up the basis of our theoretical model used to fit the experimental data. The model parameters include the incident angle on the sample during the measurement ($\theta_i$), the refractive indices of the substrate (n$_s$ - k$_s$), the thickness of the substrate (d$_s$), the refractive indices of the haze film analog (n$_f$ - k$_f$), and the thickness of the haze film (d$_f$). Some of these parameters must a priori be known to simplify the data analysis and retrieve the refractive indices of the haze film. Only the thickness and/or refractive indices of the film are unknown and fitted during the data analysis. The materials used as substrates were chosen specifically for their known n and k values in the different spectral ranges. More details on the data analysis associated to each measurement are provided in the following sections. We emphasize that the theoretical description used in the model (Eq. \ref{eq:T_theory} - \ref{eq:phase_theory}) is fully accurate only for an ideal sample structure and measurement where the film thickness is homogeneous and the light beam is collimated. In practice, the film thickness varies across the surface of the sample and the light beam of the instrument is focalized. These properties of the samples and instrument must be carefully considered to avoid biases on the refractive indices determined. This motivated the use of different techniques / measurements described in Section 3 to identify the most reliable, i.e., the one providing conditions as close as possible to this ideal theoretical description.  \\ 

The typical approach to retrieve the refractive indices is to iteratively fit the data to the theoretical model (i.e., Eq. \ref{eq:T_theory} - \ref{eq:phase_theory}) within a given spectral range. We developed an in-house python model based on non-linear iterative least-square fitting (scipy packages) to retrieve the film thickness and refractive indices using both spectroscopic and ellipsometric data. The iterative numerical core is common for the different types of data. A first library contains the different theoretical descriptions to consider ellipsometry and spectroscopy. It includes the general equations of reflection and transmission in Eq. \ref{eq:T_theory} - \ref{eq:phase_theory} as well as simplifications discussed in the following sections. A second library compiles different parametrized functions, each with their own assumptions and limitations, used to describe the spectral dispersion of the refractive indices. The different coefficients in these functions are the parameters fitted during the data analysis, in addition to the film thickness.

Among the large number of existing physical descriptions providing parameterized functions for the refractive indices, we used the Tauc-Lorentz equations to describe the spectral dispersion of n and k in the transition between electronic and atomic polarization at UV-Visible wavelengths. The Tauc-Lorentz formalism expresses the imaginary part of the dielectric constant $\varepsilon_2$ as the product of Tauc's equation with the classical Lorentz oscillator \citep{Tauc66,Campi88} : 

\begin{equation}
\varepsilon_2(E) = \left\{
    \begin{array}{ll}
        \hspace{1.4cm} 0 \hspace{2.65cm} \mbox{for} \ E \leqslant E_g  \\
        \frac{A \ Eo \ C \ (E-E_g)^2}{E \ ((E^2-Eo^2)^2 \ + \ C^2 \ E^2)} \hspace{1cm} \mbox{for} \ E > E_g 
    \end{array}
\right.
\label{eq:tauc_lorentz}
\end{equation}
\noindent where E is the energy (eV), E$_g$ is Tauc's bandgap energy (eV). A, Eo and C are respectively the strength (eV), position (eV) and width (eV) of the Lorentz oscillator.  \\

The real part of the dielectric constant $\varepsilon_1$ can be expressed analytically from $\varepsilon_2$ \citep{Jellison96} following Kramers-Kronig causality \citep{Kronig26,Kramers27}. An additional $\varepsilon_\infty$ coefficient scales $\varepsilon_1$ at infinite energy (or wavenumber). n and k are derived directly from the dielectric constant ($\varepsilon$ = $\varepsilon_1$ + i $\varepsilon_2$) following \citep{Fujiwara07}: 

\begin{equation}
\begin{split}
& n = [ \frac{1}{2} [\varepsilon_1 + (\varepsilon_1^2+\varepsilon_2^2)^{1/2}]  ]^{1/2} & \\
& k = [ \frac{1}{2} [-\varepsilon_1 + (\varepsilon_1^2+\varepsilon_2^2)^{1/2}]  ]^{1/2} & \\
\end{split}
\label{eq:diel_func}
\end{equation} \\ 

Since k $\rightarrow$ 0 below the bandgap energy, by definition, in the Tauc-Lorentz description (Eq. \ref{eq:tauc_lorentz}), the low k values of our weakly absorbing haze analogs in the Visible-NIR cannot be retrieved accurately \citep{Gavilan17,Jovanovic21,Drant24}. Another approach relies on the use of monotonic power law functions to quantify the spectral slope in the UV-Visible caused by electronic transitions at higher energies. This so-called Cauchy formalism expresses n and k as \citep{Jenkins81} : 

\begin{equation}
\begin{split}
& n(\lambda) = n_\infty + \frac{\lambda_{n,1} }{\lambda^2} + \frac{\lambda_{n,2}} {\lambda^4}, &\\
& k(\lambda) = k_\infty + \frac{\lambda_{k,1}}{\lambda^2} + \frac{\lambda_{k,2}}{\lambda^4}, & \\
\label{eq:Cauchy_abs}
\end{split}
\end{equation}
\noindent where $n_\infty$ and $k_\infty$ are constants used to scale n and k respectively. $\lambda_{n,1}$, $\lambda_{n,2}$, $\lambda_{k,1}$, $\lambda_{k,2}$ are additional constants to describe the dispersion of n and k at shorter wavelengths. The units of these parameters ensure dimensionless n and k values.   \\

The Cauchy and Tauc-Lorentz functions have both advantages and limitations. The Cauchy description enables the quantification of low k values essential to interpret observations in the Visible-NIR (e.g., \cite{Sciamma23}). The Tauc-Lorentz expression however ensures an accurate physical correlation between n and k since it satisfies Kramers-Kronig causality. Cauchy on the other hand expresses n and k with independent coefficients (Eq. \ref{eq:Cauchy_abs}) and thus does not necessarily ensure the validity of the Kramers-Kronig correlation. \\ 

\begin{figure*}
\centering
\includegraphics[scale=0.6]{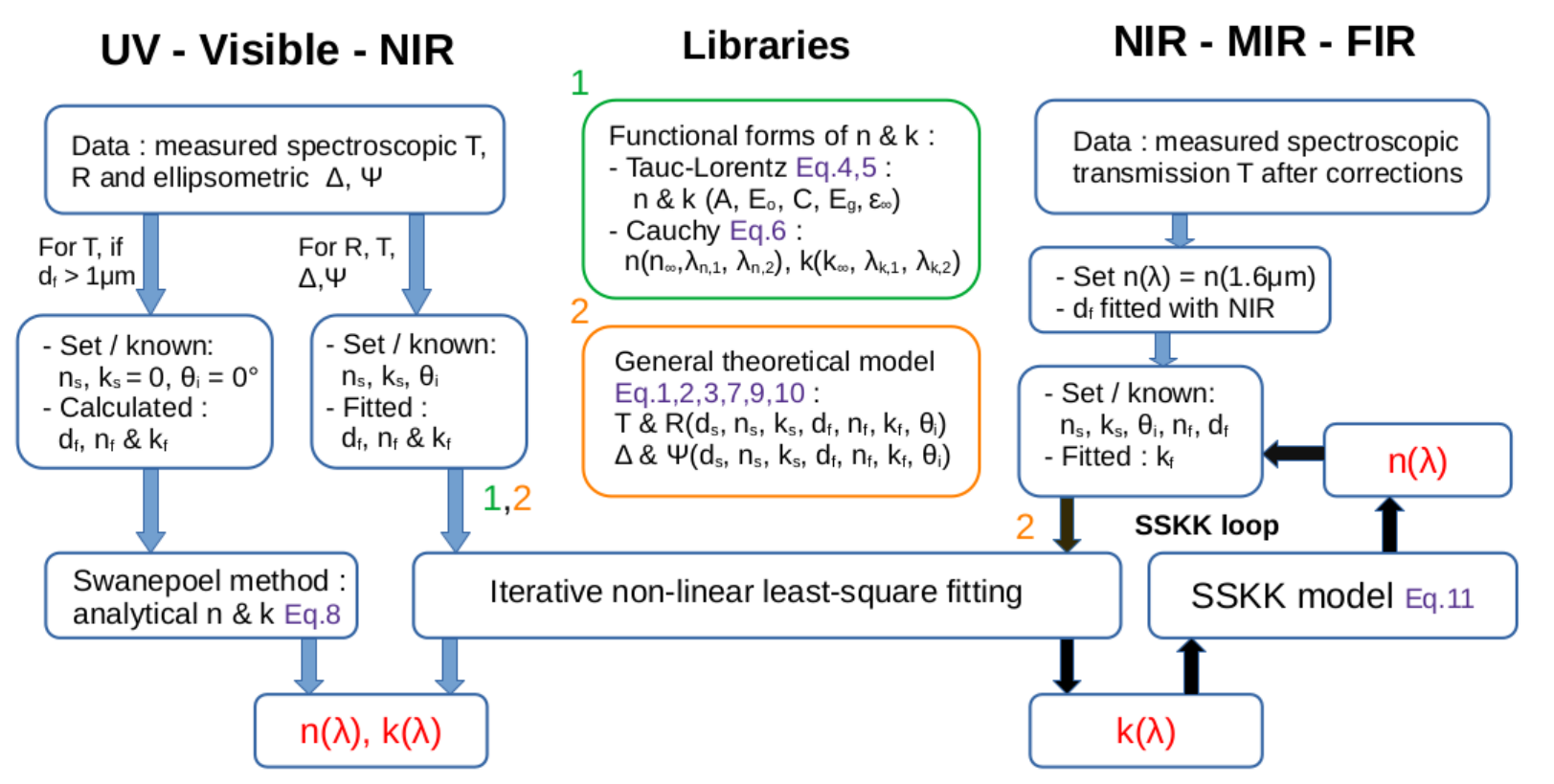}
\caption{Schematic summarizing the different procedures to determine refractive indices using spectroscopic and ellipsometric data from UV to far-IR. The data include transmission (T), reflection (R) and the ellipsometric angles ($\Delta$ and $\Psi$). $f$ and $s$ refer to film and substrate respectively. $\theta_i$ is the angle of incidence on the sample during the measurement. d refers to the thickness of the film (d$_f$) and substrate (d$_s$). SSKK refers to the singly-subtractive Kramers Kronig equation (Eq. \ref{eq:SSKK}) describing the correlation between n and k values.  }
\label{fig:schematic_model}
\end{figure*}

The schematic in Fig. \ref{fig:schematic_model} describes the different features of our model including the different libraries and the numerical procedure to determine refractive indices from laboratory data in the UV-Visible-IR spectral range. 

\subsection{Analysis of UV-Visible-NIR transmission data}

Previous work by \cite{Drant24} revealed that the Swanepoel analytical method \citep{Swanepoel83} or iterative least-square fitting of transmission data can provide accurate refractive indices on laboratory haze analogs.  \\  

To ensure the reliability of the data analysis and apply the Swanepoel method, we used a MgF$_2$ optical window to act as the substrate of the haze analogs. Since MgF$_2$ is transparent from UV to NIR and its refractive index n$_s$ ($\approx$ 1.38, \cite{Dodge84}) is lower than the haze refractive index (expected $\approx$ 1.5-1.8), the theoretical expression of transmission in Eq. \ref{eq:T_theory} simplifies to \citep{Swanepoel83} : 

\begin{equation}
\begin{split}
& T_{fs} = \frac{A \ x}{B - C \ x \ \cos{(2 \beta)} + D \ x^2} & \\
& \textrm{With : } \ A = 16 \ n_f^2 \ n_s & \\
& \hspace{1cm} B = (n_f+1)^3 \ (n_f+n_s^2) & \\
& \hspace{1cm} C = 2 \ (n_f^2-1) \ (n_f^2-n_s^2) & \\
& \hspace{1cm} D = (n_f-1)^3 \ (n_f-n_s^2) & \\
& \hspace{1cm} x = e^{- \alpha_f d_f} & \\
& \hspace{1cm} \alpha_f = \frac{4 \ \pi \ k_f}{\lambda} & \\
\end{split}
\label{eq:T_swanepoel}
\end{equation}
\noindent where $\alpha$ is the absorption coefficient. \\

\cite{Swanepoel83} showed that n and k can be calculated analytically from the data in this configuration following : 

\begin{equation}
\begin{split}
& n_f = \sqrt{C + \sqrt{(C^2-n_s^2)}} & \\
& k_f = \frac{-\lambda}{4 \ \pi \ d_f} \ln(\frac{E_M-\sqrt{E_M^2 - (n_f^2-1)^3 (n_f^2-n_s^4)}}{(n_f-1)^3 \ (n_f-n_s^2)}) & \\
& \textrm{With : } \  C = 2 \ n_s \ \frac{T_M - T_m}{T_M \ T_m} + \frac{n_s^2+1}{2} & \\
&  \hspace{1.1cm} E_M = \frac{8 \ n_f^2 \ n_s}{T_M} + (n_f^2-1) (n_f^2-n_s^2) & \\
\end{split}
\label{eq:nk_analytical_swanepoel}
\end{equation}
\noindent where T$_m$ and T$_M$ are transmission envelopes constructed using the data. \\

During the data analysis, the analog refractive indices and the film thickness are the only unknown parameters since the angle of incidence is set to 0° and the refractive indices of MgF$_2$ are known \citep{Dodge84} and compiled from the Refractive Index database\footnote{\url{https://refractiveindex.info/}}. 

Depending on the thickness of the haze film, two different procedures can be used to determine the refractive indices from the transmission data. For most analogs, we used the Swanepoel method based on analytical calculations. This approach is reliable as long as a large number of interference fringes are observed on the spectrum. The different fringe extrema are used to construct the transmission envelopes (see Fig. \ref{fig:Figure3}, top panel). The accuracy of the calculations strongly rely on the construction of these envelopes. A low film thickness will be characterized by a spectrum with fewer fringes, which does not provide enough information to accurately constrain T$_m$ and T$_M$. We generally found that a film thickness above $\approx$1 $\mu$m is sufficient to construct reliable transmission envelopes. Fig. \ref{fig:Figure3} (top panel) shows the analysis of the transmission data with the Swanepoel method for the Titan 1 PAMPRE haze analog. A film thickness below 1 $\mu$m was only seen on the Exoplanet 2 PAMPRE analog leading to larger error bars on the retrieved refractive indices with the Swanepoel method (see Appendix A for the analyses of all the PAMPRE analogs). For that reason, we used a different approach where the transmission data is fitted directly to the theoretical model, i.e., Eq. \ref{eq:T_swanepoel}, in order to retrieve the refractive indices with the iterative approach described in Sect 4.1. The transmission spectrum of the Exoplanet 2 PAMPRE analog is fitted with the Tauc-Lorentz and Cauchy dispersion functions in Fig. \ref{fig:Figure3} (bottom panel). The two procedures used to analyze the transmission spectra and retrieve n and k, i.e., Swanepoel method and iterative fitting, are summarized in Fig. \ref{fig:schematic_model}. The results are discussed in Section 5. 

\subsection{Analysis of UV-Visible-NIR reflection data}

For the reflection measurements, we used the sample where the film is deposited on a doped silicon wafer. Si was chosen for its highly absorbing properties in the UV-Visible \citep{Aspnes83} preventing incoherent multiple reflection within the substrate. In addition, the back face of the Si wafer is rough thus inducing scattering and preventing multiple reflection even above 1 $\mu$m, when Si becomes transparent. By eliminating the multiple reflection within the substrate, we reduce the parameter space of the model since the substrate thickness is no longer required, and the general expression of reflection in Eq. \ref{eq:R_theory} simplifies to :  \\

\begin{equation}
\begin{split}
& R_{fs} = \frac{|r_{af}|^2 + |r_{fs}|^2  \ e^{- 2 \beta_{f,im}}}{1 + |r_{fa}|^2 \ |r_{fs}|^2 \ e^{- 2 \beta_{f,im}}} & \\
\end{split}
\label{eq:R_theory_onesiderough}
\end{equation} \\

The data analysis was performed with the iterative approach described in Sect 4.1 using Eq. \ref{eq:R_theory_onesiderough} as the theoretical expression of reflection. The thickness and refractive indices of the film analog are the only unknowns. The refractive indices of silicon are needed to consider reflection at the film/substrate interface. These refractive indices are well known \citep{Aspnes83} and used as an input parameter during the data analysis. The data are fitted to the theoretical model with the Cauchy and Tauc-Lorentz expressions of n and k in Fig. \ref{fig:Figure4}. The iterative procedure used on reflection data to determine n and k values is summarized in Fig. \ref{fig:schematic_model}. We limited the fit to the spectral range with visible interference fringes and high signal to noise ratio. The short-end of the wavelength range depends on the film thickness and the absorbing properties of the sample, it therefore varies depending the analog. Since the COSmIC samples exhibit high absorption properties, the fit is often unreliable in the UV thus limiting the retrieval of refractive indices to the Visible-NIR. The short-end wavelength limit taken for the analysis of each sample is indicated in Table \ref{tab:samples}. The analyses of all the COSmIC samples reflection data are provided in Appendix B along with the film thicknesses determined. The results are discussed in Section 5.

\subsection{Analysis of UV-Visible ellipsometric data} 

 Standard ellipsometry relies on the measurement of reflection using two polarization states to retrieve the ellipsometric angles $\Delta$ and $\Psi$ describing the phase shift and amplitude ratio respectively. The ratio of parallel (p) and perpendicular (s) reflectance can be expressed using these ellipsometric angles : 

\begin{equation}
\begin{split}
& \rho = \frac{r_p}{r_s} = \tan(\Psi) \ e^{i \Delta}  & \\
\end{split}
\label{eq:stand_ellipso}
\end{equation}

Eq. \ref{eq:R_theory_onesiderough} and Eq. \ref{eq:stand_ellipso} were added in the library of our model for theoretical expressions (see Fig. \ref{fig:schematic_model}). \\

For the ellipsometric measurements, we also used the sample where the film is deposited on a doped silicon wafer to prevent incoherent multiple reflection within the substrate. For the data analysis, only the refractive indices and thickness of the haze film layer are unknown following Eq. \ref{eq:R_theory_onesiderough} and Eq. \ref{eq:stand_ellipso}.  A first measurement performed on the blank Si substrate constrained the angle of incidence precisely to 69.9°. The ellipsometric data is fitted to the theoretical model using both Cauchy and Tauc-Lorentz functions in Fig. \ref{fig:Figure2}. The results are discussed in Section 5.1.

\subsection{Analysis of NIR-FIR transmission data}

For the IR transmission measurements, we used the analog sample where the film is deposited onto the intrinsic Si substrate with both sides polished. Si was chosen for its transparency in the entire IR spectral range allowing to constrain the absorbing properties of the film alone without significant contribution from the substrate. As stated in \cite{Drant24}, FTIR measurements do not provide absolute data directly, the measured intensity is very sensitive to variations of the source intensity over time. Several corrections must be performed on the data to obtain an absolute transmission and retrieve reliable n and k values. First, the measured intensity spectra of the blank substrate and samples are divided by an open-beam reference intensity spectrum to obtain an 'uncorrected' transmission. A reference spectrum was acquired regularly (every three hours approximately) to account for variations of the source intensity and detection efficiency over time. At this stage, the transmission is not yet absolute and said 'uncorrected'. The different procedures to treat the transmission data are detailed in the following.

First, interference fringes caused by multiple reflection within the substrate are removed from the spectra following the procedure with transmission envelopes described in \cite{Swanepoel83}. Given the 300-$\mu$m thick substrate, the spectral frequency of the fringes caused by the substrate is significantly higher than those caused by the film which makes them easily distinguishable. The transmission data shown in Fig. \ref{fig:FigureS} and \ref{fig:Figure6} have already been treated to remove the fringes caused by the substrate, and only the fringes caused by the film remain. 

Since the theoretical transmission of a silicon wafer can be easily calculated using its known refractive indices \citep{Salzberg57,Edwards80}, a first comparison between the measured and theoretical transmission of silicon tells us if additional corrections are needed to obtain absolute data. We use the ratio between the measured and theoretical transmission of Si as a first correction factor that we apply to our data, to correct any vertical shift caused by variations of the lamp intensity and/or detection efficiency between the measurements. This correction also removes the small absorption features observed in the spectrum of Si, which are attributed to silicon oxides and caused by a small oxidation layer at the surface of the substrate. In theory, this first correction on the data should suffice if there were no variations of the source intensity between the measurement of the blank Si substrate and the haze analog sample. In practice, this assumption is not always valid and another correction is required to further improve the data. 

For this second and last correction, we first fit the haze analog film thickness using the fringes observed on the transmission data from 1.5 to 2.5 $\mu$m (see Fig. \ref{fig:FigureS}). Only the film thickness is fitted using the refractive indices previously derived with the UV-Visible-NIR measurements (see Sect 4.2 and 4.3). Once the film thickness is determined, we can simulate the absolute transmission of our haze analog sample in the NIR and compare it to the measured data. If variations of the source intensity occurred between the measurement of the blank substrate and analog sample, a vertical shift is observed between the simulated and measured transmission spectra of the haze analog sample. If not corrected, this shift will lead to strong biases in the retrieved k values. The ratio of the measured and simulated NIR transmission provides the last correction factor to obtain an actual absolute transmission unaffected by measurement biases.  \\

For the data analysis, the procedure consists of fitting the absolute IR transmission of the haze analog sample with the theoretical expression in Eq. \ref{eq:T_theory}. For the fit, one could describe spectral variations of n and k using a set of Lorentz oscillators following the procedure in \cite{Drant24} used to analyze MIR Mueller ellipsometric data. In practice, too many model parameters have to be fitted depending on the number of absorption features and strong model degeneracy can bias the results \citep{Drant24}. We therefore used a different approach where only k is retrieved from the fit and n is calculated using a singly-subtractive Kramers-Kronig (SSKK) model to ensure an accurate physical correlation between both real and imaginary part of the refractive index. 

The model used for the IR resembles the approach of \cite{Imanaka12}. At the first iteration, k is the only unknown fitted at each data point (at each wavelength) since n is assumed constant in the entire IR following the value derived at 1.6 $\mu$m with the UV-Visible-NIR measurements (see Sect 4.2 and 4.3). The film thickness was already derived using interference fringes in the NIR. A first spectrum of k is obtained using iterative least-square fitting between the absolute transmission spectrum and the model (Eq. \ref{eq:T_theory}). Fig. \ref{fig:FigureS} shows the first simulated absorption-free (k = 0) transmission spectrum for the Titan 1 COSmIC analog and the Exoplanet 2 PAMPRE analog (black curves). This simulation does not consider absorption by the film but considers its thickness and the resulting interference fringes. The difference between the data (in blue) and this first absorption-free simulation (black) constrains k. 

Using the retrieved k values and an anchor refractive index n at a specific wavelength, the n spectrum can be derived from the SSKK equation following \citep{Hawranek76}:

\begin{equation}
\begin{split}
n(\nu_{i}) = n_r + \frac{2}{\pi} \ [ P \int_{0}^{\infty} \frac{\nu k(\nu)}{\nu^2 - \nu_{i}^2} \ d\nu - P \int_{0}^{\infty} \frac{\nu k(\nu)}{\nu^2 - \nu_r^2} \ d\nu \ ] \
\end{split}
\label{eq:SSKK}
\end{equation}
\noindent where n$_r$ and $\nu_r$ are the anchor point real refractive index and wavenumber (cm$^{-1}$) respectively. $P$ is the Cauchy principal value of the integral.  \\

Our model was constructed following Maclaurin's formula which was previously identified by \cite{Ohta88} as the most reliable numerical approach to perform the integral in Eq. \ref{eq:SSKK}. For n$_r$, we used the n value derived at 1.6 $\mu$m with our UV-Visible-NIR measurements. Once n is calculated in the IR, we repeat the first procedure to derive a new k spectrum which is now considering the spectral variations of n in the IR. We repeat this procedure until the difference in n and k between two iterations is below 1\%. In practice, less than six iterations are sufficient as the main variations occur during the first iteration. This iterative procedure used to derive the refractive indices from IR transmission data is summarized in Fig. \ref{fig:schematic_model}. Fig. \ref{fig:FigureS} shows the simulated absorption-free (k = 0) transmission spectrum after the first SSKK iteration for the Titan 1 COSmIC analog and Exoplanet 2 PAMPRE analog (red curves). The difference between the first simulation (in black) assuming a constant n in the IR and the simulation with SSKK (in red) reflects the effect that spectral variations of n have on the spectra and the bias it can cause on the retrieved k values. For the Exoplanet 2 PAMPRE analog, the SSKK correction has little impact on the transmission spectrum and the data analysis since the C-H features are not strong enough to cause significant variations of n (Fig. \ref{fig:FigureS}, bottom panel). On the other hand, the SSKK correction has a significant effect on the transmission spectrum and data analysis for the Titan 1 COSmIC analog (Fig. \ref{fig:FigureS}, top panel). This stems from the strong (hetero-)aromatic C=C and C=N features causing important variations of n around 6 $\mu$m. Generally, the data analysis of the haze analogs produced from an Ar-dominated mixture (Exoplanet 2 and 3 samples in Table \ref{tab:samples}) is rather insensitive to spectral variations of n contrary to the analysis of the other samples produced from a N$_2$-dominated gas mixture (see Table \ref{tab:samples}). The use of an iterative model satisfying Kramers-Kronig correlation between n and k is thus essential to retrieve accurate refractive indices in the entire IR range. Appendix C presents the analysis of the IR data for all the COSmIC and PAMPRE haze analogs. The results are discussed in Section 5.

\section{Results and Discussions}

\subsection{Identifying the best technique to retrieve refractive indices from UV to near-IR}

\begin{figure}
\centering
\includegraphics[width=1\columnwidth]{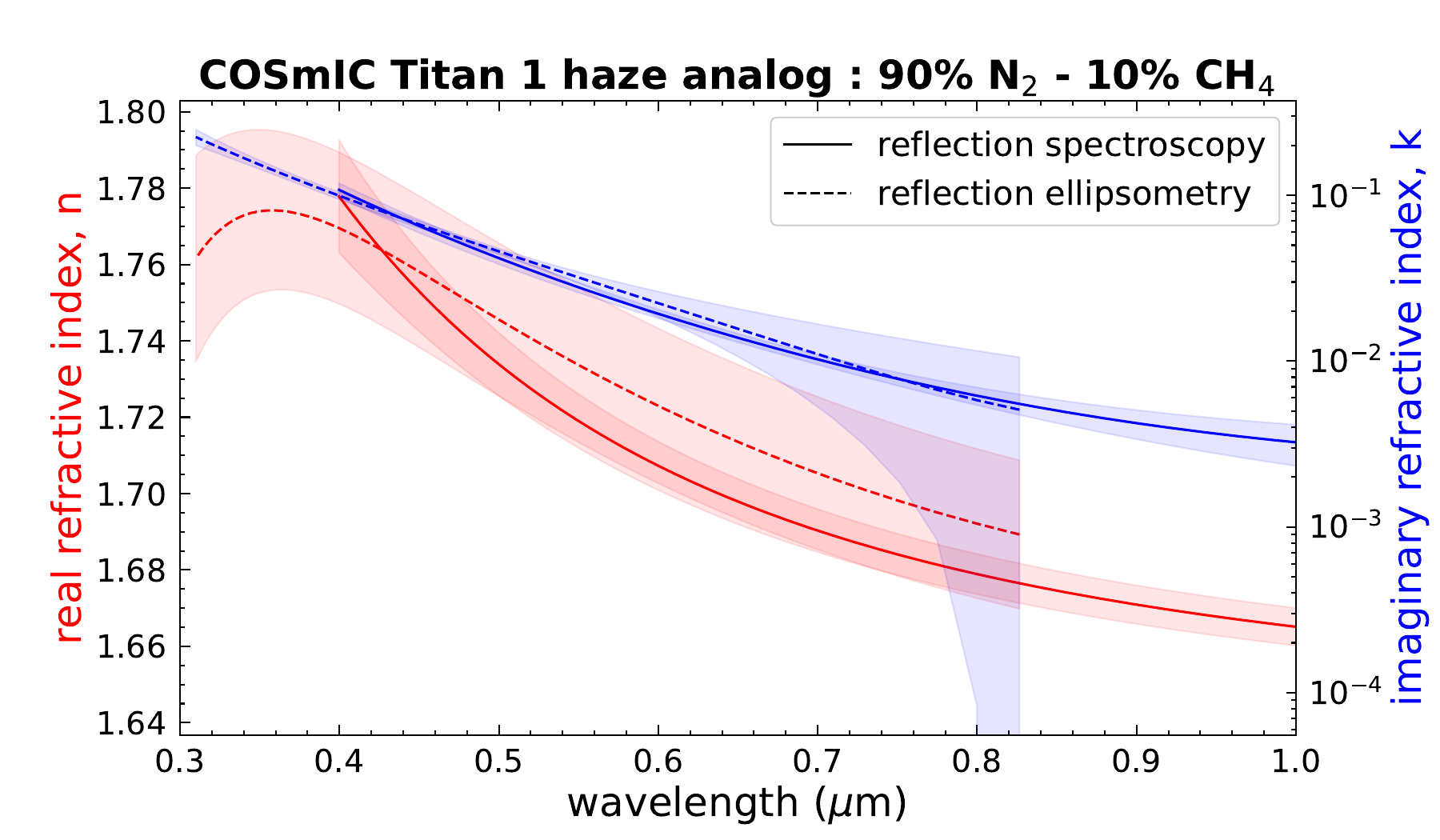}
\includegraphics[width=1\columnwidth]{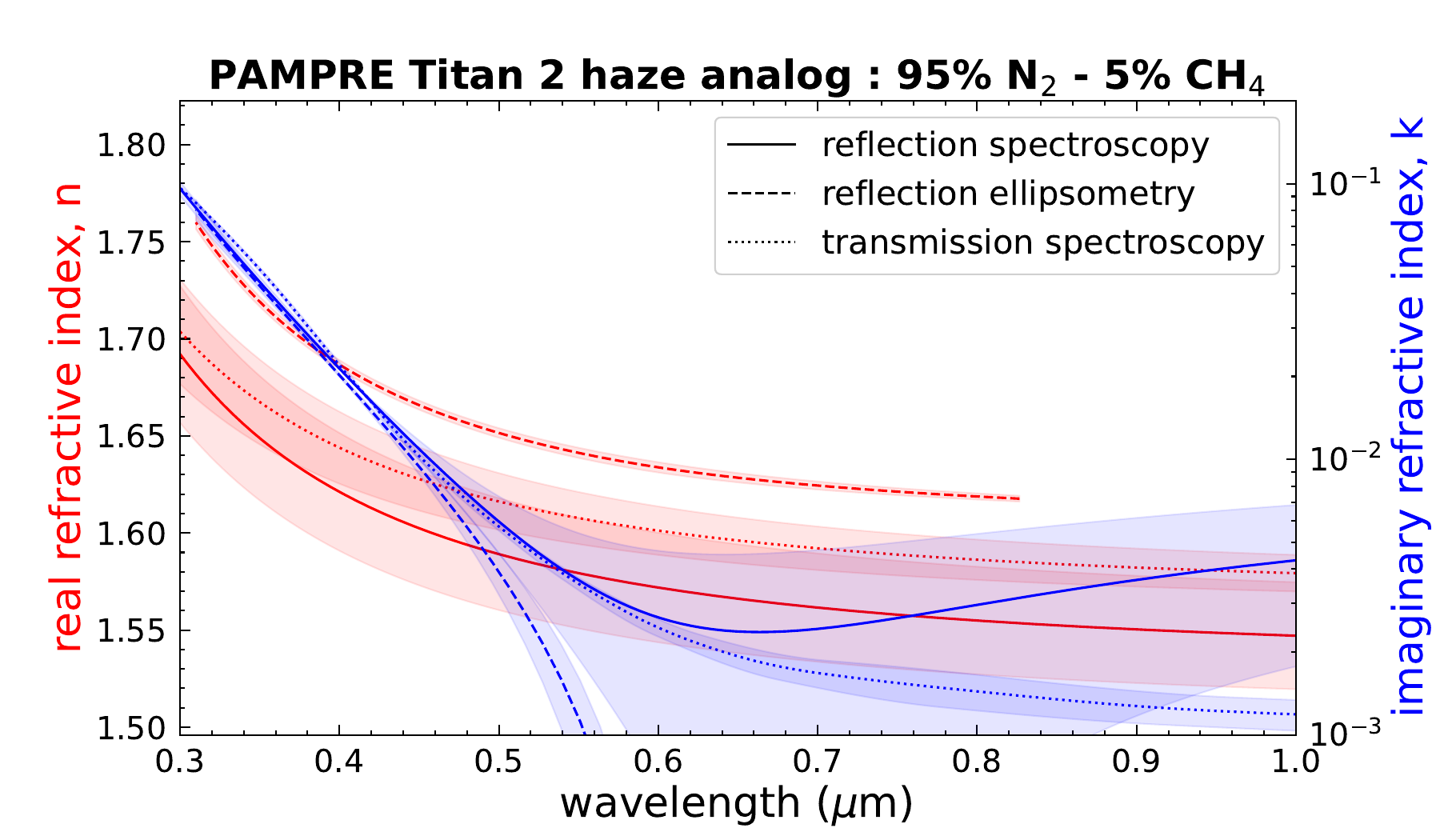}
\caption{Influence of the measurement technique used on the retrieved refractive indices n and k in the UV-Visible and near-IR. (top) Comparison between the refractive indices determined by reflection ellipsometry and reflection spectroscopy for the COSmIC Titan 1 analog. (bottom) Comparison between the refractive indices determined by reflection ellipsometry, reflection spectroscopy and transmission spectroscopy for the PAMPRE Titan 2 analog.   }
\label{fig:Figure7}
\end{figure}

The previous studies of \cite{Khare84} and \cite{Tran03} revealed discrepancies in the derived real refractive index between different optical techniques. A discrepancy around 3\% between transmission spectroscopy and reflection ellipsometry was reported more recently by \cite{Drant24}. In the present work, we performed a similar comparison to \cite{Drant24} and found that the discrepancy in n is caused by the difficulty in retrieving accurate k values from our reflection measurements in the Visible-NIR spectral range. Fig. \ref{fig:Figure7} (top panel) compares the UV-Visible refractive indices derived from reflection spectroscopy and reflection ellipsometry on the COSmIC Titan 1 haze analog produced from a gas mixture of 90\% N$_2$ and 10\% CH$_4$. We find a good agreement between both data sets within the range of uncertainty. The bottom panel of Fig. \ref{fig:Figure7} now compares the UV-Visible-NIR refractive indices derived from transmission spectroscopy, reflection spectroscopy and reflection ellipsometry on the PAMPRE Titan 2 haze analog produced with 95\% N$_2$ and 5\% CH$_4$. We generally found that transmission spectroscopy and reflection spectroscopy provide similar refractive indices within the error bars, although reflection spectroscopy is, in the case of our measurements and samples, more affected by biases and requires careful beam focalization. The absorption slope below 0.5 $\mu$m obtained from ellipsometric data is well constrained and in agreement with the transmission data, but the k values retrieved from ellipsometry are unconstrained in the region of weak absorption, above 0.5 $\mu$m, regardless of the functional forms of n and k (i.e., Cauchy or Tauc-Lorentz) used during the data analysis. Our reflection ellipsometric measurements are not sensitive enough to constrain these low k values on the PAMPRE analogs, in the Visible spectral range. This lack of sensitivity also influences the fitted n values. In Fig. \ref{fig:Figure7} (bottom), the n values obtained with reflection ellipsometry are higher than those determined with transmission spectroscopy and reflection spectroscopy. This difference is explained by the lack of constraints on k in the ellipsometric data, above 0.5 $\mu$m, which likely also caused the discrepancy reported by \cite{Drant24}. The discrepancy in n between spectroscopy and ellipsometry is not observed for the COSmIC analogs since their highly absorbing properties in the entire UV-Visible range allow us to retrieve k accurately. We generally found that k is not well constrained with reflection ellipsometry for values below 5.10$^{-3}$. We emphasize that these limitations depend on the instruments used and on the structure of the samples.   \\

Unlike reflection which is mostly sensitive to variations of the real refractive index n between the film and substrate, transmission is more sensitive to absorption. k is constrained more efficiently as it directly correlates to the difference between the transmission of the substrate and the transmission envelopes of the analog (difference between red and green curves in top panel of Fig. \ref{fig:Figure3}). The choice of description to express the spectral variations of n and k, i.e. Tauc-Lorentz or Cauchy, with the iterative model (Sect 4.1) however significantly influences the fit (see Fig. \ref{fig:Figure3}, bottom panel) and the retrieved n values. Even if we have k<10$^{-2}$ in the Visible and NIR (above $\approx$ 0.6 $\mu$m), the measured transmission is still sensitive to this weak absorption. The assumption of k = 0 below the bandgap energy in the Tauc-Lorentz description leads to an artificial increase of the interference fringes' amplitude during the fitting procedure (see Fig. \ref{fig:Figure3}, bottom panel). To reproduce the fringes observed in the data while assuming k = 0, the model increases n. In other words, the assumption of k = 0 below the bandgap energy in the Tauc-Lorentz model is not appropriate for transmission data, even in the NIR, as it leads to over-estimations of n to improve the goodness of fit. Given the strong correlation between real refractive index and film thickness in the law of interference (phase parameter, Eq. \ref{eq:phase_theory}), the over-estimation of n also leads to an under-estimation of the film thickness to maintain the product n $\times$ d constant. This sensitivity analysis suggests that the use of the Cauchy expressions for n and k is generally more reliable for our weakly absorbing samples in the Visible-NIR spectral range. \\

In summary, the lack of sensitivity to low k values in our ellipsometric data or the use of simplified models assuming k = 0, such as Tauc-Lorentz, lead to similar biases on the refractive indices determined. n and k indeed both influence the amplitude of interference fringes seen in the data. If an incorrect assumption is made on one of these two parameters, e.g. k = 0, or if one of these parameters is not constrained by the data, the other parameter will not be fitted accurately. We suggest the use of transmission and reflection data analyzed with iterative fitting (with Cauchy expressions of n and k) or Swanepoel calculations to obtain more reliable n and k values in the Visible-NIR.  

\begin{figure}
\centering
\includegraphics[width=1\columnwidth]{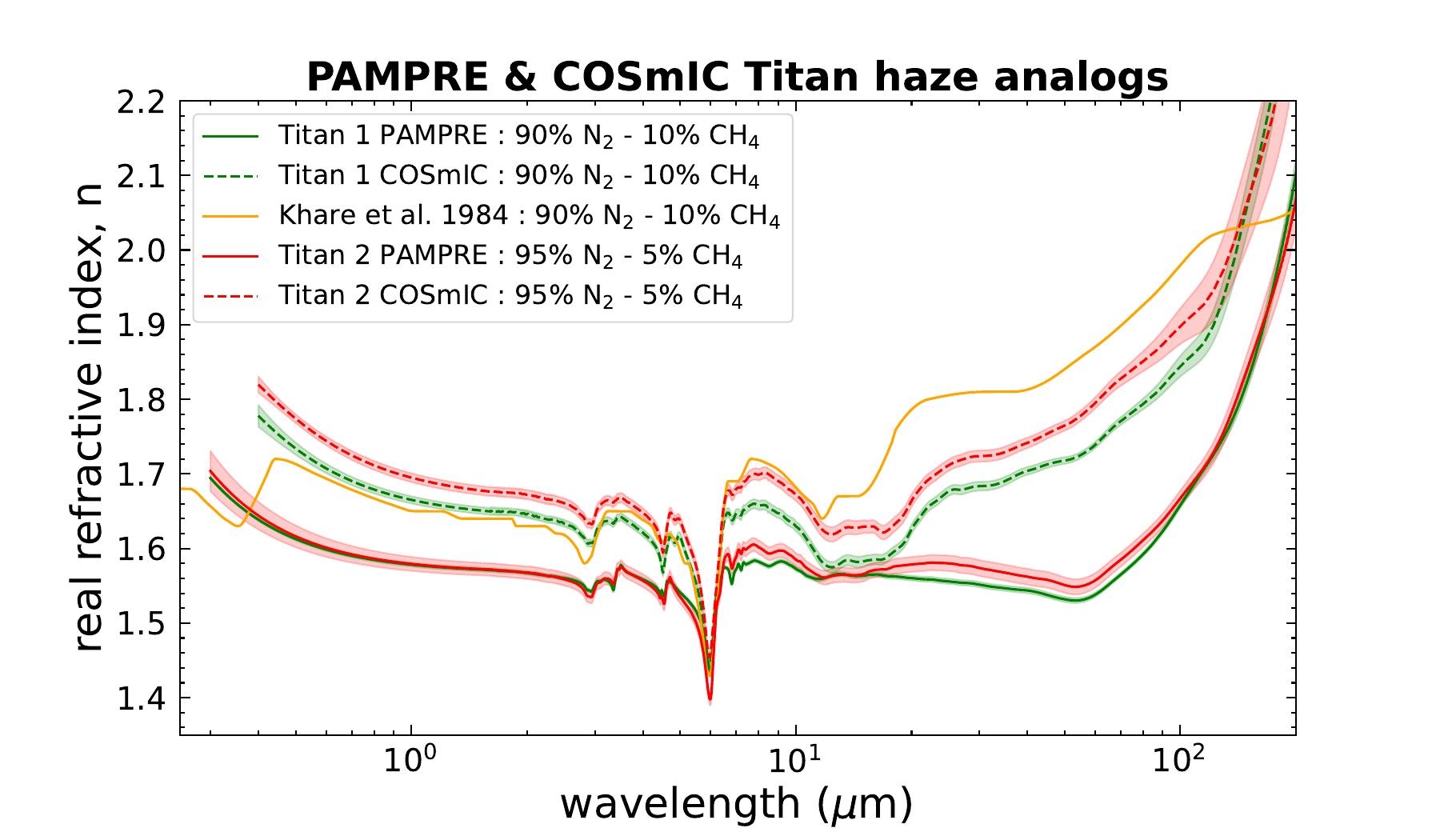}
\includegraphics[width=1\columnwidth]{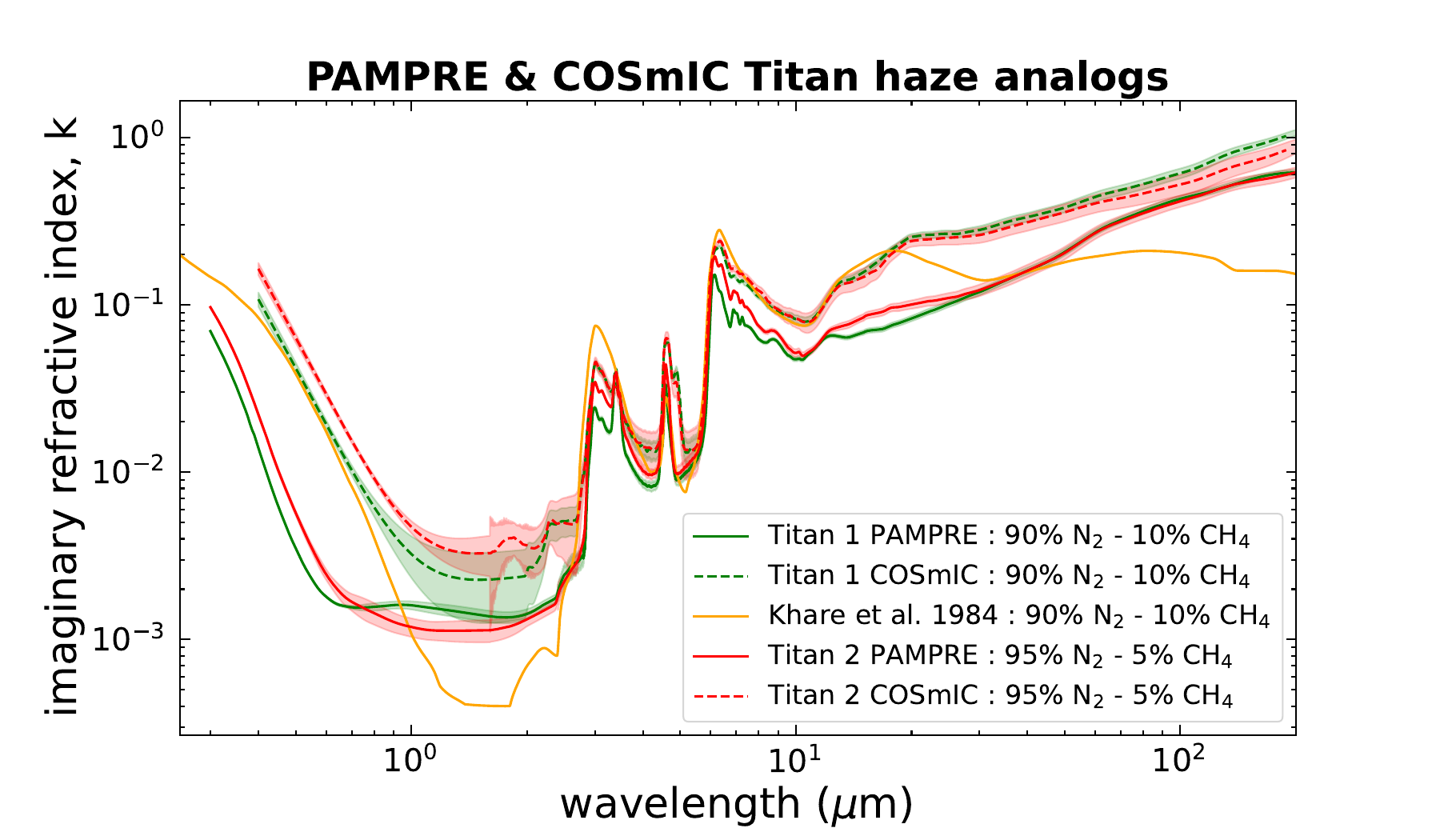}
\includegraphics[width=1\columnwidth]{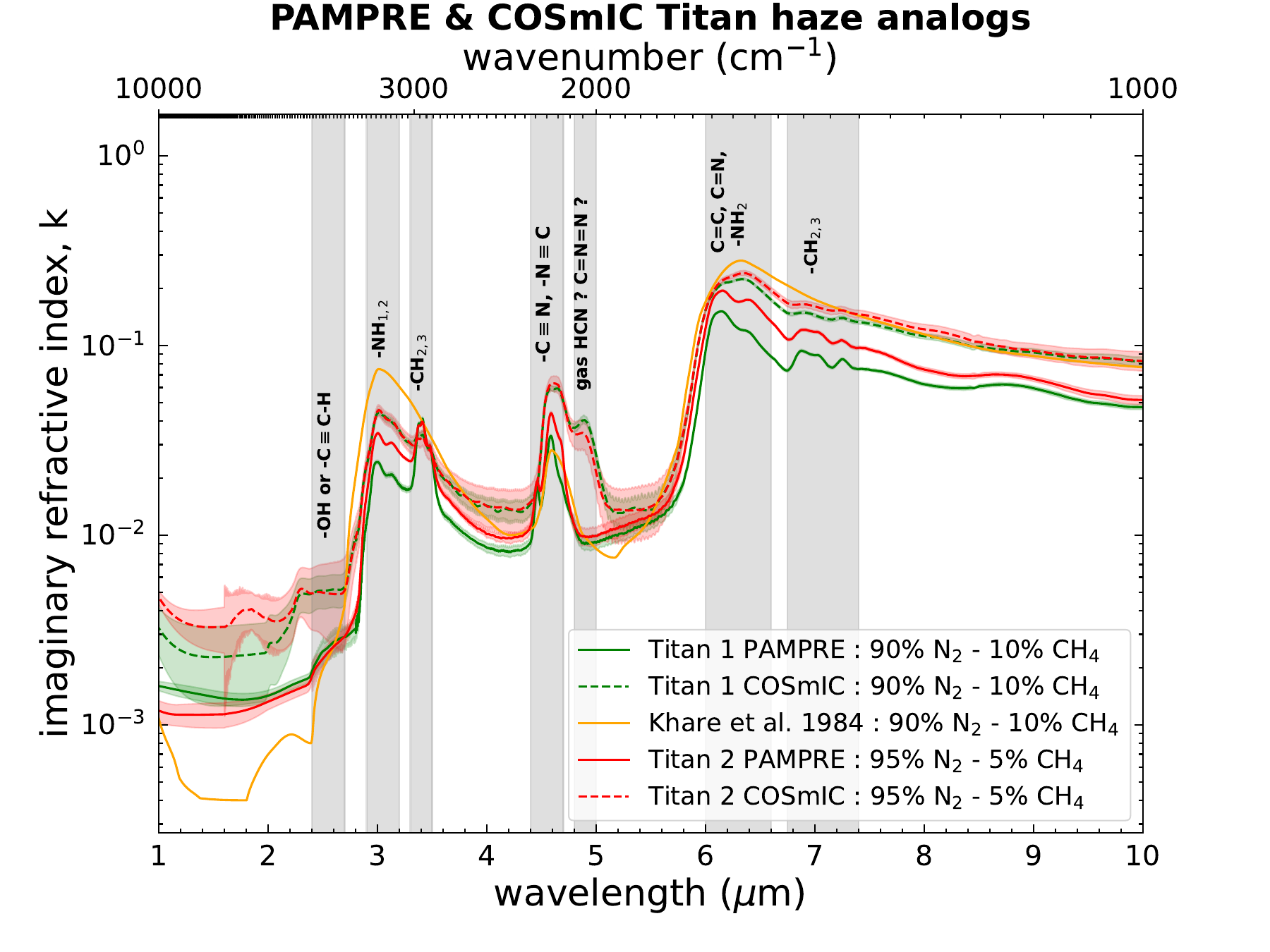}
\caption{Refractive indices n (top) and k (middle \& bottom) of PAMPRE and COSmIC Titan haze analogs from UV to far-IR. Both the Titan 1 and Titan 2 analogs produced from different gas relative abundances N$_2$/CH$_4$ (see Table \ref{tab:samples}) are shown. The refractive indices of \cite{Khare84} obtained for a Titan analog with an initial gas composition similar to our Titan 1 analog are also plotted for comparison. The attribution of IR resonances is shown on the bottom panel.    }
\label{fig:Figure9}
\end{figure}

\begin{figure}
\centering
\includegraphics[width=1.1\columnwidth]{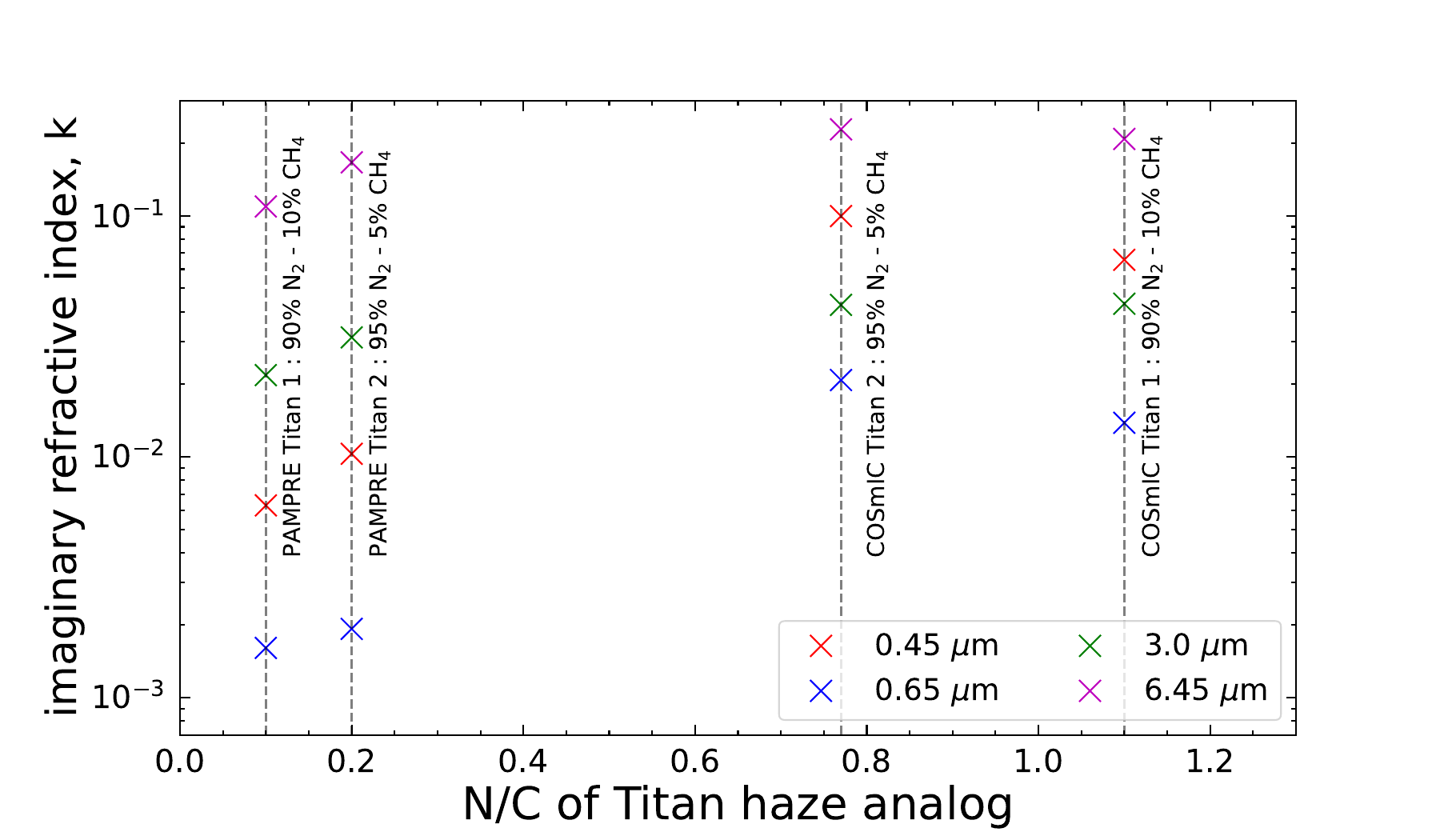}
\caption{Correlation between the N/C ratio of the solid haze analog and the imaginary refractive index k. The trend is shown at different wavelengths relevant for absorption involving nitrogen (Visible wavelengths, N-H at 3 $\mu$m and C=N at 6.45 $\mu$m). }
\label{fig:FigureNC}
\end{figure}

\subsection{Refractive indices of Titan haze analogs }

Fig. \ref{fig:Figure9} shows the refractive indices retrieved from UV to FIR for the different Titan haze analogs produced with the COSmIC and PAMPRE set-ups. We only show the refractive indices determined using the Cauchy expressions of n and k in the iterative model, in the UV-Visible-NIR spectral range (see Sect 4.2 and 4.3), since they provide more reliable estimates (see Sect 5.1). For the NIR k values, higher uncertainties are reported for the COSmIC analogs compared to the PAMPRE analogs. It stems from the reflection measurements used which were found to be very sensitive to beam focalization. In addition, an important thickness variability is observed on the COSmIC samples (see more details in Appendix B). \\ 

Fig. \ref{fig:Figure9} reveals that the refractive indices of the Titan haze analogs are strongly affected by both the gas composition (gas relative abundance N$_2$/CH$_4$) and the experimental set-up used for their production. Since previous studies have performed elemental analyses on similar COSmIC and PAMPRE Titan analogs and reported their N/C ratio \citep{Carrasco16,Nuevo22}, we can use these previous data and emphasize the correlation between the N/C ratio of the solid analogs and their refractive indices in a broad spectral range. \cite{Nuevo22} reported N/C = 1.1 $\pm$ 0.37 for their COSmIC Titan analog produced from a gas mixture of 90\% N$_2$ and 10\% CH$_4$ (Titan 1, Table \ref{tab:samples}) and N/C = 0.77 $\pm$ 0.12 for their analog produced from a gas mixture of 95\% N$_2$ and 5\% CH$_4$ (Titan 2, Table \ref{tab:samples}). It therefore seems that the bulk N/C ratio in the solid COSmIC Titan analogs decreases with increasing N$_2$/CH$_4$ in the initial gas mixture. In contrast, an opposite trend was reported for the PAMPRE Titan analogs. \cite{Carrasco16} observed an increase of the N/C ratio of their Titan 1 and Titan 2 haze analogs correlated with the increase of the N$_2$/CH$_4$ ratio chosen in the initial gas mixture. They reported N/C = 0.1 and 0.2 for their Titan 1 and Titan 2 analogs respectively. 

Fig. \ref{fig:Figure9} (middle panel) tells us that the UV-Visible imaginary refractive index of the COSmIC and PAMPRE Titan analogs changes with the relative abundance N$_2$/CH$_4$ in the initial gas mixture. A similar trend is observed for the PAMPRE and COSmIC Titan analogs where k increases with increasing N$_2$/CH$_4$ ratio in the initial gas mixture and the bandgap energy (E$_g$, Eq. \ref{eq:tauc_lorentz}) shifts towards lower energies (longer wavelengths) when the initial gas composition is richer in nitrogen. A similar behavior was observed by \cite{Mahjoub12} on their PAMPRE Titan analogs. Based on the known N/C ratio for the PAMPRE Titan analogs, the refractive indices data suggest an increase of k in the UV-Visible and a shift of the bandgap energy towards lower energies when the solid haze material is richer in nitrogen. Since the n$_N$ $\rightarrow$ 3sa and n$_N$ $\rightarrow$ 3pa electronic transitions  (5.7 - 8 eV) associated with amine functional groups peak at lower energies than the $\sigma$-$\sigma$* and n-$\sigma$* transitions (10 - 12 eV) associated with methyl groups (e.g. \cite{Gavilan18} and references therein), a material richer in amine groups, and more generally richer in nitrogen, should indeed exhibit higher k values and a shift of E$_g$ towards longer wavelengths as seen for our analogs (Fig. \ref{fig:Figure9}, middle panel). In other words, the incorporation of nitrogen and formation of amine structures in the solid material lead to stronger absorbing properties in the UV-Visible range in response to electronic transitions peaking at shorter wavelengths, not covered by our measurements. The trend between k and the gas N$_2$/CH$_4$ ratio is similar for the COSmIC Titan analogs, but this means that the correlation between UV-Visible k values and the N/C ratio in the solid is different compared to the interpretation made for the PAMPRE Titan analogs. We note, however, that the bulk elemental analyses of \cite{Nuevo22} were made on COSmIC analogs produced with a voltage of -800V for the plasma discharge whereas the analogs used for the present study were made with a voltage of -700V. Since previous work has shown that changes in the plasma voltage between -1000V and -700V in the COSmIC set-up can significantly influence the UV-Visible k values and thus the haze analog composition \citep{Sciamma23}, it is possible that the N/C ratios of our COSmIC Titan analogs are different than those reported by \cite{Nuevo22}. In addition, the N/C ratios determined by \cite{Nuevo22} on both COSmIC Titan haze analogs could in fact be similar within the range of uncertainty. Additional elemental analysis measurements on our COSmIC Titan analogs are needed to better understand the correlation between refractive indices and compositional N/C ratio.

The real part of the refractive index n is similar in the NIR, around 1.58, for both PAMPRE Titan analogs (Fig. \ref{fig:Figure9}, top panel) in agreement with previous ellipsometric measurements by \cite{Mahjoub12} and \cite{Sciamma12}. For the COSmIC Titan haze analogs, n increases in the NIR for a higher initial N$_2$/CH$_4$ ratio in the gas phase (Fig. \ref{fig:Figure9}, top panel). It would suggest that the n values decrease following a higher N/C ratio in the solid analog. This result is in agreement with the data of \cite{Mahjoub12} on PAMPRE Titan analogs although this behavior was only observed when the N$_2$/CH$_4$ ratio of the initial gas mixture is above 49, which corresponds to a gas mixture with an abunance of N$_2$ equal to or above 98 \%. 

In the MIR, Fig. \ref{fig:Figure9} (middle \& bottom panels) shows that the k values of the PAMPRE Titan analogs, reflecting bending and stretching resonances, also vary with the N$_2$/CH$_4$ ratio in the initial gas mixture. The amine -NH$_{1,2}$ features (3.1 - 3.3 $\mu$m) are stronger for the PAMPRE Titan 2 analog compared to the PAMPRE Titan 1 analog in agreement with previous IR analyses by \cite{Mahjoub12} and \cite{Gautier12}. These higher k values, attributed to amine modes, for the PAMPRE Titan 2 analog correlate to a higher N/C ratio in the solid. The strength of -CH$_{2,3}$ aliphatic/alkane stretching features (3.37 - 3.5 $\mu$m) is similar for both PAMPRE Titan analogs (Fig. \ref{fig:Figure9}, middle \& bottom panel). \cite{Carrasco16} indeed reported a similar H/C ratio, around 0.1, for the Titan analogs produced using different N$_2$/CH$_4$ ratio in the gas phase. The k values associated to alkene, aromatic, hetero-aromatic and -NH$_2$ features (6 - 6.6 $\mu$m) are also higher for the PAMPRE Titan 2 analog (Fig. \ref{fig:Figure9}, middle \& bottom panels) which, following the higher N/C ratio in the solid, points to stronger C=N and -NH$_2$ modes. For the COSmIC Titan analogs, on the other hand, the MIR k values are very similar for the two N$_2$/CH$_4$ conditions (see Fig. \ref{fig:Figure9}, middle \& bottom panels). This likely indicates that the composition of the COSmIC Titan analogs is not as sensitive to small variations of the gas composition as are the PAMPRE Titan analogs. The N/C ratio in the solid material indeed varies by a factor of 2 between the two PAMPRE Titan analogs \citep{Carrasco16} whereas it only varies by a factor of $\approx$ 0.5 between the two COSmIC Titan analogs \citep{Nuevo22}. 

At 4.88 $\mu$m, between nitrile (4.4 - 4.7 $\mu$m) and alkene/(hetero-)aromatic/-NH$_2$ (6 - 6.6 $\mu$m) features, a strong MIR resonance is observed on the COSmIC Titan analogs but not on the PAMPRE Titan analogs. The nature of this feature, previously observed in \cite{Sciamma17}, is unclear. It is likely caused by C=C=N or C=N=N \citep{Carlson16}, or by adsorbed gaseous HCN trapped in the pores of the sample. Further investigations are required to understand the nature of this resonance. We also note that the Titan analog produced with 90\% N$_2$ - 10\% CH$_4$ (Titan 1 analog, Table \ref{tab:samples}) with the COSmIC set-up presents a UV-Visible k slope and NIR n values very similar to the analog of \cite{Khare84} produced using a similar gas composition (Fig. \ref{fig:Figure9}, top and middle panels). The MIR k values attributed to alkene/aromatic/hetero-aromatic C=C and C=N and -NH$_2$ (6 - 6.6 $\mu$m) features are also very similar between the COSmIC Titan 1 analog and the analog of \cite{Khare84}. This could suggest a similar solid composition between these analogs. A variation of the k values between these analogs is however observed around 3.1 - 3.3 $\mu$m (Fig. \ref{fig:Figure9}, middle \& bottom panels). It could be explained by stronger amine -NH$_{1,2}$ features for the analogs of \cite{Khare84} or by a stronger contribution of O-H in the Khare analogs. This latter hypothesis is possible since the gas bottles used by \cite{Khare84} to produce their haze analogs were not as pure as those used in the present study. They thus contained water vapor at higher abundances which could have influenced the gas chemistry and the composition of the solid analogs.

If we now compare the PAMPRE and COSmIC Titan haze analogs, Fig. \ref{fig:Figure9} reveals significant differences in both n and k. n and k are generally higher for the COSmIC analogs in the entire spectral range. The higher k values and lower bandgap energy of the COSmIC Titan analogs are explained, following our previous interpretation, by the higher N/C ratio in the COSmIC solid material \citep{Carrasco16,Nuevo22}. There is, in fact, a change of N/C ratio by a factor of ten between the PAMPRE and COSmIC Titan 1 haze analogs which likely explains the strong variations of k values in the UV-Visible-NIR. It also explains the higher k values for the COSmIC Titan analogs at amine NH$_{1,2}$ (3.1 - 3.3 $\mu$m), nitriles/iso-nitriles -C$\equiv$N/-N$\equiv$C (4.4 - 4.7 $\mu$m) and (hetero-)aromatic/alkene (6 - 6.6 $\mu$m) wavelengths compared to the PAMPRE Titan analogs. These differences between the refractive indices of the PAMPRE and COSmIC Titan analogs could be explained by the changes in gas residence time, gas temperature and pressure between both set-ups (see Table \ref{tab:set-ups}). \cite{Imanaka04} previously reported that the gas pressure from which hazes form strongly influences the bandgap energy and the UV-Visible k values. The shorter gas residence time in the COSmIC set-up compared to the PAMPRE set-up (see Table \ref{tab:set-ups}) results in a more simple gas phase chemistry, preventing the formation of complex carbon chains \citep{Sciamma14} which likely contributes to the higher N/C ratio observed in the COSmIC analogs compared to the PAMPRE analogs. Previous work also taught us that the change in nitrogen incorporation into the solid is sensitive to the gas temperature \citep{He22} and the properties of the plasma source \citep{Sciamma23} which both change between the PAMPRE and COSmIC set-ups. Although the main experimental parameter influencing the change in refractive indices between the COSmIC and PAMPRE Titan analogs cannot be identified, a clear correlation with the N/C ratio of the analog is observed. Fig. \ref{fig:FigureNC} illustrates the correlation between the N/C ratio of the solid analog and the imaginary refractive index k at different wavelengths. The general trend suggests an increase of k in the Visible and in the IR following the enhanced nitrogen incorporation in the solid. We emphasize however that additional measurements and more comparison with other analogs are needed to better understand this correlation. \\

\begin{figure}
\centering
\includegraphics[width=1\columnwidth]{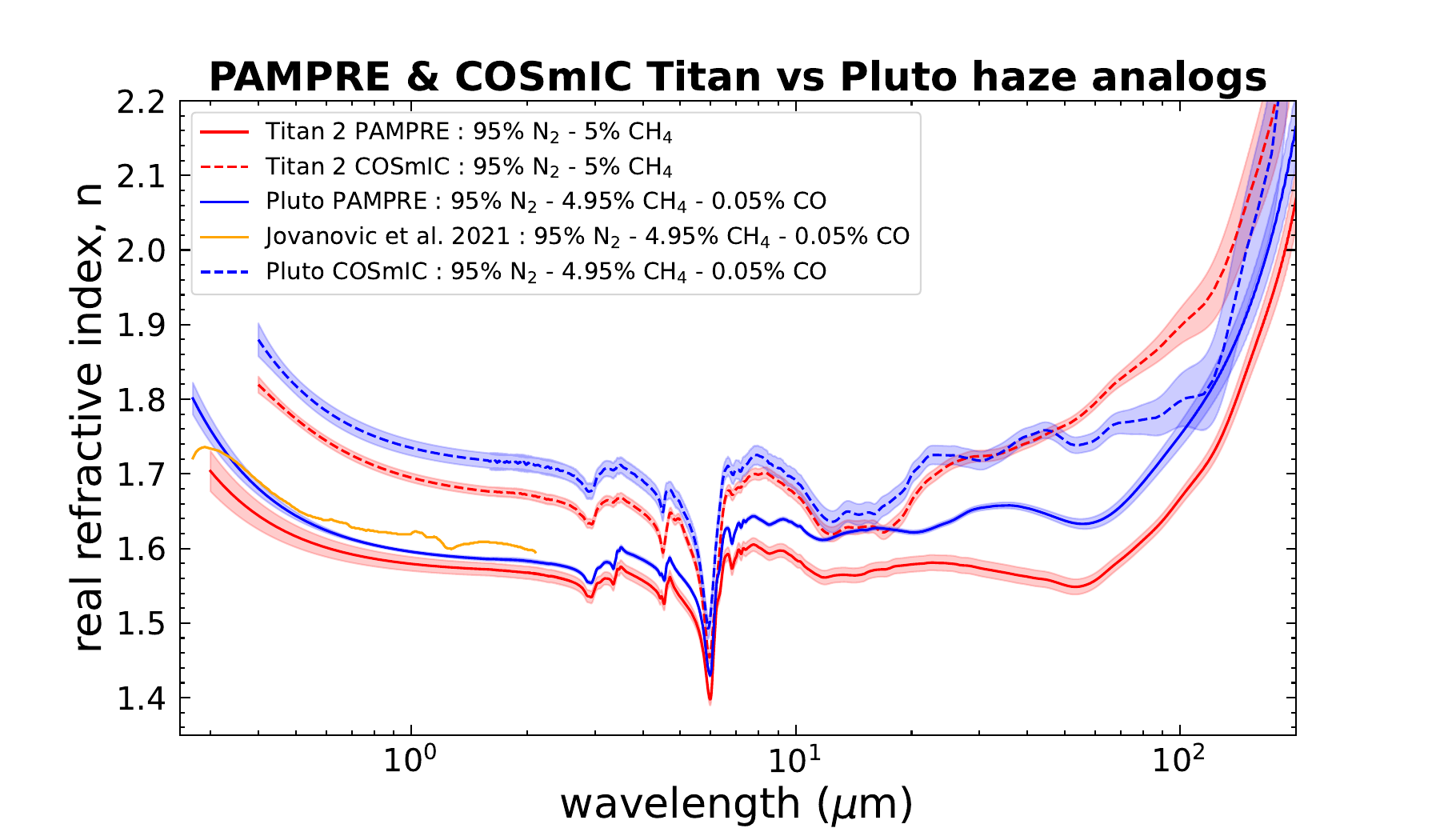}
\includegraphics[width=1\columnwidth]{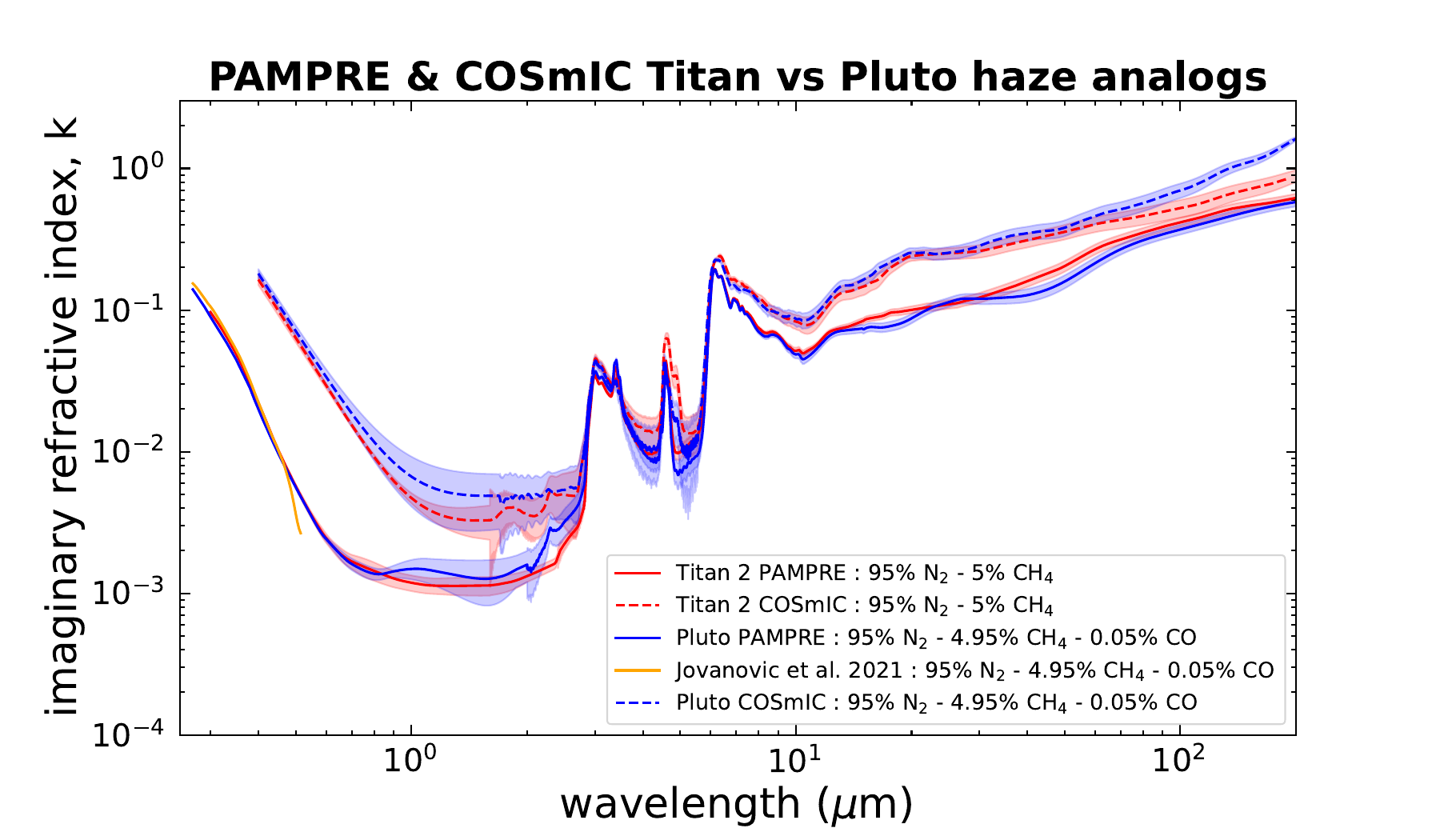}
\includegraphics[width=1\columnwidth]{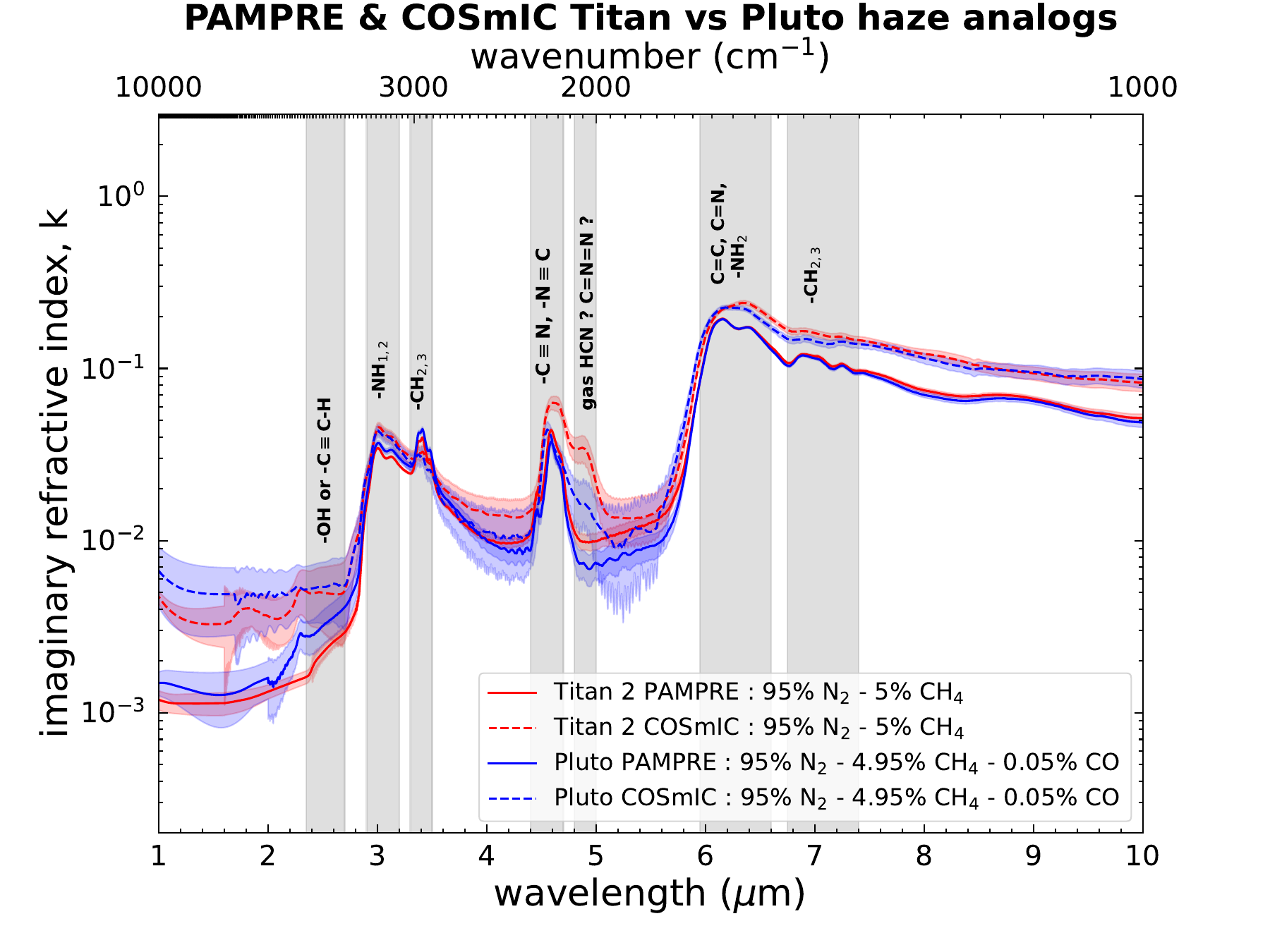}
\caption{Refractive indices n (top) and k (middle \& bottom) of PAMPRE and COSmIC Pluto haze analogs from UV to far-IR (up to 200 $\mu$m). The n and k values of the Titan 2 haze analogs produced from a similar gas relative abundance N$_2$/CH$_4$ (see Table \ref{tab:samples}) are shown for comparison.    }
\label{fig:Figure10}
\end{figure}

Previous analyses of Cassini-Huygens observations suggest that Titan hazes are highly absorbing in the UV-Visible-NIR \citep{Rannou10,Rannou22,Sciamma23} with k values higher than the data reported in the present work. It was suggested that this difference stems from a higher incorporation of nitrogen in Titan hazes compared to laboratory analogs \citep{Sciamma23}. On the other hand, MIR observations of Titan hazes do not confirm the presence of nitrogenous modes and rather points to carbon-rich materials \citep{Bellucci09,Rannou10,Kim11,Vinatier12,Kim13,Courtin15,Kim18}. More observations are needed to understand the absorbing properties of Titan hazes in a broad spectral range and how it correlates to the composition of the solid.

\subsection{Refractive indices of Pluto haze analogs}

Fig. \ref{fig:Figure10} shows the refractive indices n and k from UV to FIR of the Pluto haze analogs, produced from a gas mixture with 95\% N$_2$ - 4.95\% CH$_4$ - 0.05\% CO using the PAMPRE and COSmIC set-ups. The refractive indices of the Titan 2 haze analogs, produced from a similar gas relative abundance N$_2$/CH$_4$, are shown in comparison to assess the influence of CO on the refractive indices of the analogs. We report a very strong agreement between the UV-Visible refractive indices determined in the present study and those previously reported in \cite{Jovanovic21} for the PAMPRE Pluto analog. The comparison of the Titan 2 and Pluto k values in Fig. \ref{fig:Figure10} clearly indicate that the addition of 500 ppm of CO in the initial gas mixture does not significantly affect the refractive indices of the solid analog in the entire spectral range. The UV-Visible k slope and the bandgap energy of the Pluto analog are similar to that of the Titan 2 analog produced with a similar set-up (PAMPRE or COSmIC). Since electronic transitions associated to hydroxyl and carbonyl functional groups (between 3.5 and 5.8 eV) would likely create features at lower energies than the amine and methyl groups (\cite{Gavilan18} and references therein), they should affect the UV-Visible k values if present abundantly in the solid material. The similar k values of the Titan 2 and Pluto analogs suggest that these functional groups are not sufficiently abundant in the material, when adding 500 ppm of CO in the initial gas mixture, to influence the refractive indices. These results tell us that the differences of n and k observed by \cite{Jovanovic21} in the UV-Visible and NIR on different PAMPRE Pluto haze analogs are caused by the variations of the N$_2$/CH$_4$ ratio in the initial gas mixture and not by changes in the CH$_4$/CO ratio.

The MIR k values of the Pluto analogs are also similar to their respective Titan 2 analogs produced with the same set-up. The 6-$\mu$m C=O band previously observed in \cite{Jovanovic20} on a different type of haze analog (not film but grains) is not clearly resolved here and does not affect the k values. The relative strength of -NH$_{1,2}$ amine (3.1 - 3.3 $\mu$m) and -CH$_{2,3}$ alkane/aliphatic (3.4 - 3.7 $\mu$m) MIR modes does not vary between the Titan 2 and Pluto analogs. Although CO does not seem to influence the refractive indices of our haze analogs, it was found to influence the gas phase chemistry \citep{Moran22}, the incorporation of oxygen into the solid \citep{Jovanovic20} and the size of haze analog particles \citep{Horst14}. 

In the FIR, the reported k values of our new Pluto analogs are different than previous predictions by \cite{Khare84} for their Titan analog (see Fig. \ref{fig:Figure9}). These new data should be used to re-assess the thermal cooling properties of Pluto's atmospheric hazes previously discussed in \cite{Zhang17}, \cite{Wan21} and \cite{Wan23}. 

\begin{figure}
\centering
\includegraphics[width=1\columnwidth]{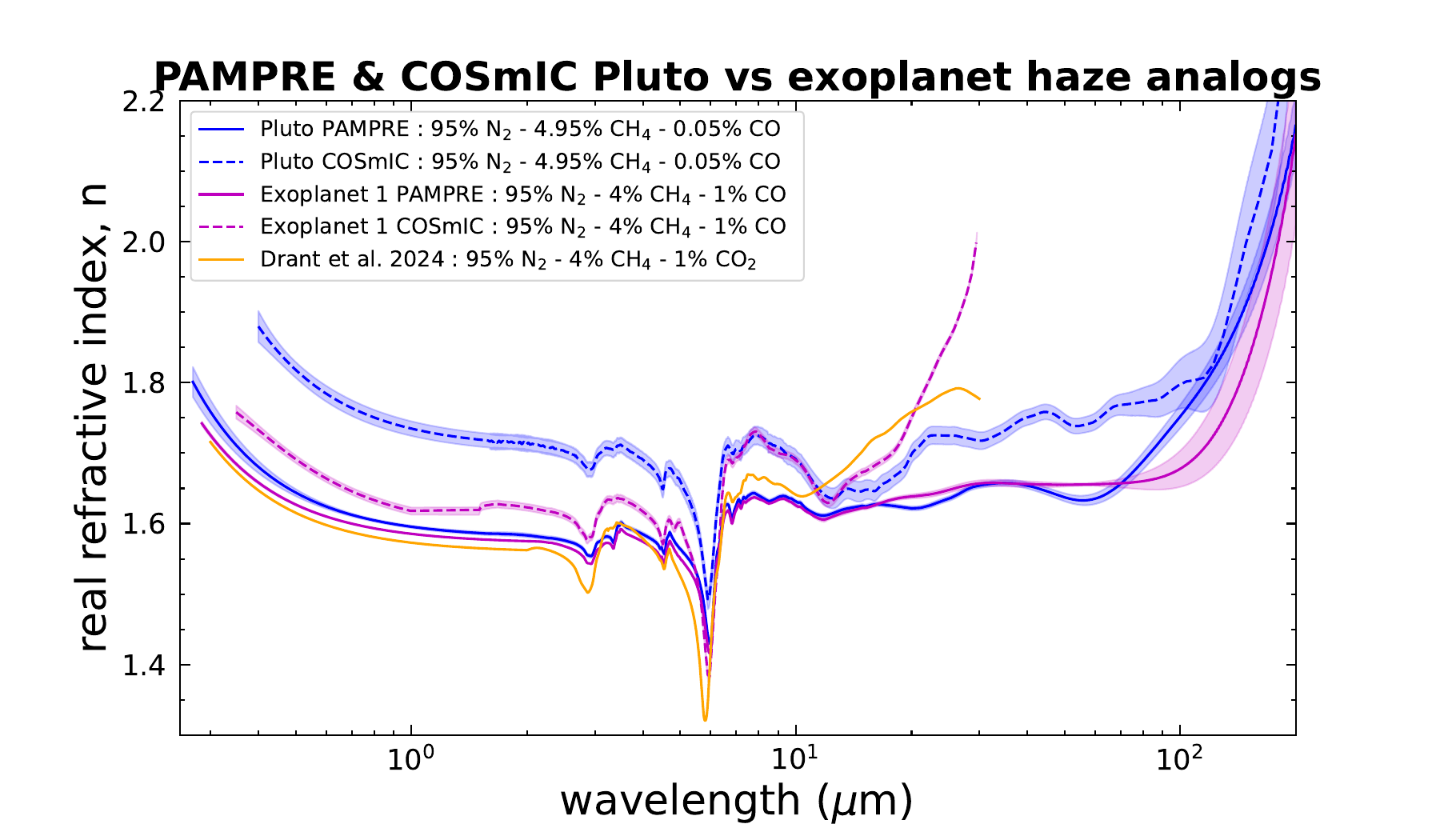}
\includegraphics[width=1\columnwidth]{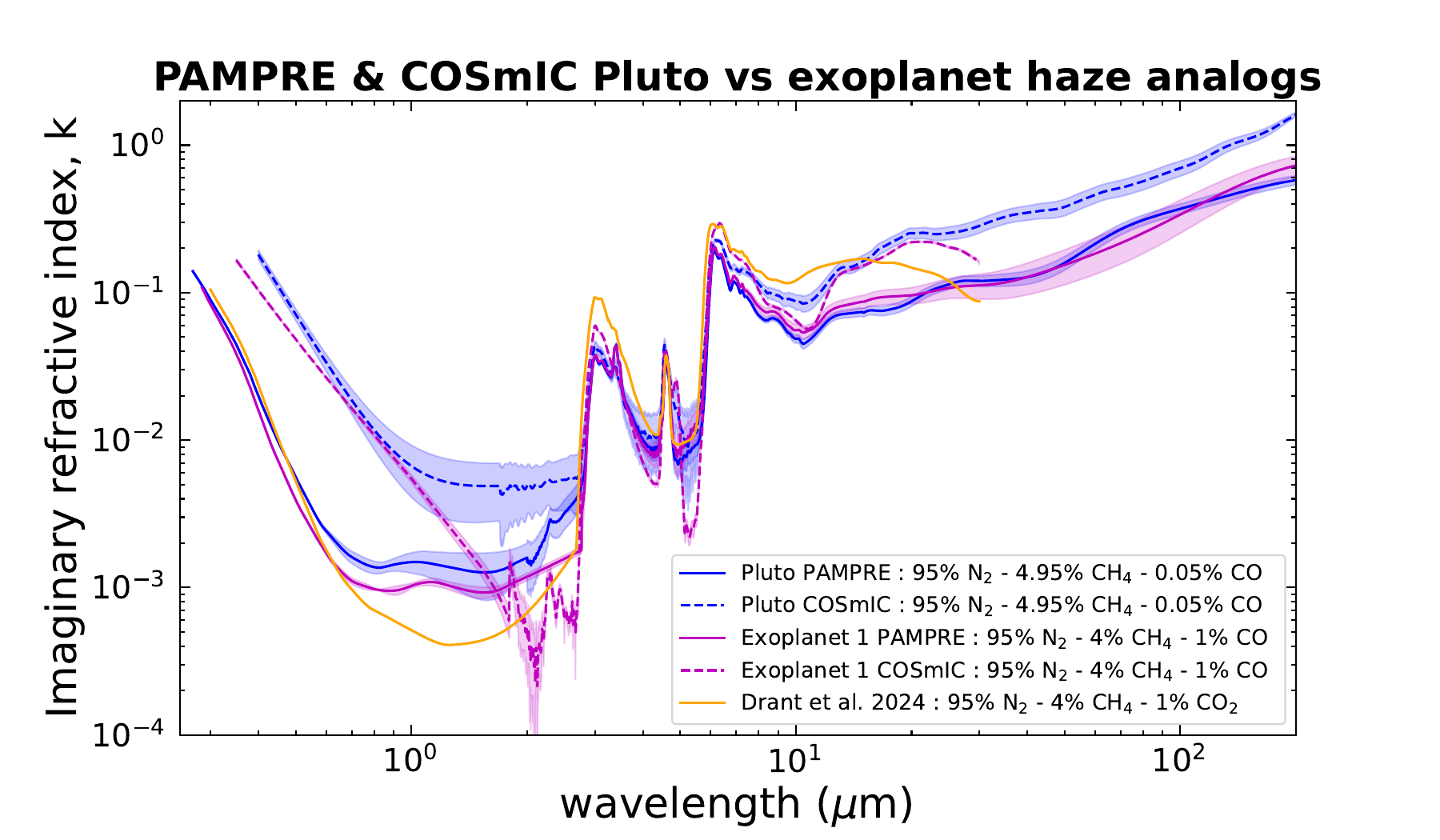}
\includegraphics[width=1\columnwidth]{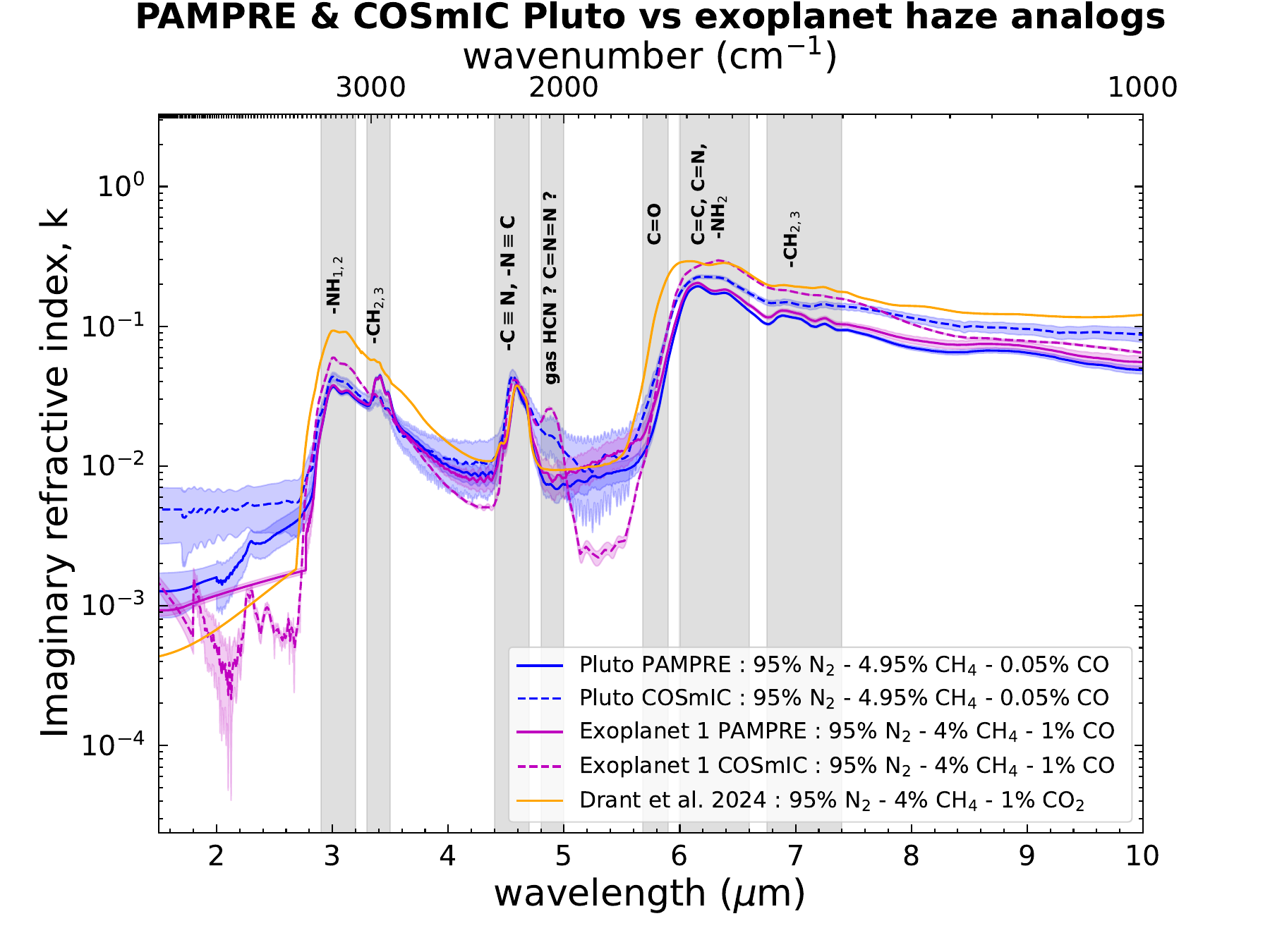}
\caption{Refractive indices n (top) and k (middle \& bottom) from UV to FIR of exoplanet haze analogs produced using the PAMPRE and COSmIC set-ups from a mixture of 95\% N$_2$ - 4\% CH$_4$ - 1\% CO. The n and k data obtained for the Pluto and Exoplanet 1 analogs produced from different CH$_4$/CO ratio in an N$_2$-dominated gas mixture (see Table \ref{tab:samples}) are shown. The previous data of \cite{Drant24} obtained for an exoplanet/early-Earth haze analog produced with 95\% N$_2$ - 4\% CH$_4$ - 1\% CO$_2$ is also plotted to compare the effect of CO vs CO$_2$ on the refractive indices of the solid haze analog material.     }
\label{fig:Figure11}
\end{figure}

The large differences in UV-Visible k values between the PAMPRE and COSmIC Pluto haze analogs, which are likely caused by variations of N/C ratio in the solid (Sect 5.2), may also explain the strong heterogeneity of Pluto's surface colors (e.g. \cite{Grundy18} and \cite{Lauer21}) if the composition of the organic material varies across Pluto's surface. 

\subsection{Refractive indices of exoplanet haze analogs produced in CO-rich gas mixtures}

Fig. \ref{fig:Figure11} presents the refractive indices from UV to FIR of the Exoplanet 1 haze analog produced from a gas mixture of 95\% N$_2$ - 4\% CH$_4$ - 1\% CO with the PAMPRE and COSmIC set-ups. The Pluto analog produced from a gas mixture with a higher CH$_4$/CO abundance ratio is plotted in comparison, to evaluate the influence of CO on the refractive indices of the solid analogs. In addition, we show the refractive indices obtained by \cite{Drant24}, with PAMPRE, for a haze analog produced from a mixture of 95\% N$_2$ - 4\% CH$_4$ - 1\% CO$_2$, to compare the effect of CO and CO$_2$ on the haze analogs refractive indices. 

Fig. \ref{fig:Figure11} (middle panel) reveals that the UV-Visible k values of the PAMPRE analogs are poorly affected by the presence of 1\% CO$_2$ or CO in the gas mixture. This result suggests, once again, that the electronic transitions associated to hydroxyl and carbonyl functional groups are not strong compared to the strength of electronic transitions attributed to amine groups in the PAMPRE analogs. Lower CH$_4$/CO gas abundance ratio may increase the oxygen content incorporated in the haze material and the number of hydroxyl and carbonyl groups which could lead to higher k values in the UV-Visible for the PAMPRE analogs. Additional experiments are required to verify this hypothesis. The comparison between the UV-Visible k values of the COSmIC Pluto and Exoplanet 1 analogs highlights a slight variation of the slope in the Visible likely caused by the change in the gas relative abundance CH$_4$/CO. For the PAMPRE analogs, the decrease of the CH$_4$/CO gas abundance ratio does not seem to influence the n values of the haze analogs (Fig. \ref{fig:Figure11}, top panel). For the COSmIC analogs, however, n seems to vary with the gas abundance ratio CH$_4$/CO. 

The MIR and FIR k values of the PAMPRE Exoplanet 1 and Pluto analogs produced from different gas relative abundances CH$_4$/CO are also similar (Fig. \ref{fig:Figure11}, middle \& bottom panels). The relative strengths of amine NH$_{1,2}$ and alkane/aliphatic CH$_{2,3}$ resonances are similar between both PAMPRE analogs, and thus similar to the PAMPRE Titan 2 analog, which suggest a poor influence of CO on the N/C ratio in the haze material. This last result brings forth a significant difference between the influence of CO and CO$_2$ on the haze refractive indices. The MIR k values obtained on the haze analog of \cite{Drant24}, produced from a gas mixture with 1\% CO$_2$ in the PAMPRE set-up, revealed a strong increase of amine N-H signatures (3.1 - 3.3 $\mu$m) relative to alkane/aliphatic C-H modes (3.4 - 3.7 $\mu$m) and higher k values for the aromatic/hetero-aromatic/alkene signatures around 6 - 6.6 $\mu$m (Fig. \ref{fig:Figure11}, middle \& bottom panels). It is also likely that the increased k values for the analog of \cite{Drant24} at 3 $\mu$m are partly due to O-H in addition to N-H. At 6 $\mu$m, the analog of \cite{Drant24} produced from a CO$_2$-rich mixture also exhibits a shift of the k slope toward shorter wavelengths which suggests the contribution of C=O at 5.9 $\mu$m in addition to the C=C and C=N features at longer wavelengths (6 - 6.6 $\mu$m). This C=O contribution is however not clearly observed for the Exoplanet 1 analog which could suggest that the haze analogs produced in CO$_2$-rich mixtures incorporate larger amounts of oxygen compared to haze analogs produced in CO-rich mixtures. The photolysis of CO and CO$_2$ is expected to produce oxygen radicals able to oxidize reduced hydrocarbons and thus inhibit the formation of complex hydrocarbon chains in the gas phase. Since CO$_2$ is photolyzed more efficiently than CO, this oxidation mechanism should be more efficient in CO$_2$-rich gas mixtures. The increased oxidizing power of CO$_2$ likely leads to an increase of N/C ratio in the solid analog by preventing the formation of complex hydrocarbons in the gas phase. \cite{Trainer04b} indeed revealed that haze analogs formed in CO$_2$-rich gas mixtures are made of more simple hydrocarbons compared to haze analogs produced without CO$_2$. Additional investigations should be performed in the future to assess the composition of the haze material (i.e., N/C ratio and O/C ratio) and the nature of the gaseous photochemical products in CO-rich and CO$_2$-rich conditions. 

These new results clearly show that the correlation between the initial gas phase composition and the haze refractive indices is not primarily sensitive to the gas C/O ratio but rather sensitive to the photo-dissociation efficiency of the oxygenated compound (e.g., CO or CO$_2$) and its reactivity with hydrocarbon species. Since similar abundances of CO and CO$_2$ can lead to different strengths in the IR haze signatures, the strategy proposed by \cite{Corrales23} to constrain the atmospheric C/O using the haze IR absorbing properties could be biased depending on which oxygenated molecule is present in the atmosphere. Previous observations of exoplanet atmospheres, as well as modeling studies, revealed that the detection of CO in transit spectra is challenging given its weak IR opacity and its main absorption feature at 4.7 $\mu$m which partly overlaps with that of CO$_2$ \citep{Wogan24,Shorttle24,Drant25}. The distinction between CO and CO$_2$ contributions was, in fact, recently proven to be challenging for rocky exoplanets given the large error bars on the observational data \citep{Hu24}. The difference in the IR absorbing properties between hazes produced in CO-rich and CO$_2$-rich gas mixtures could help constrain which molecule contributes to the absorption peak at 4.7 $\mu$m, by using the amine, aromatic, alkene and hetero-aromatic haze signatures at other wavelengths (3 - 3.3 and 6 - 6.5 $\mu$m, see Fig. \ref{fig:Figure11}, bottom panel). 

\subsection{Refractive indices of exoplanet hydrocarbon haze analogs}

\begin{figure}
\centering
\includegraphics[width=1\columnwidth]{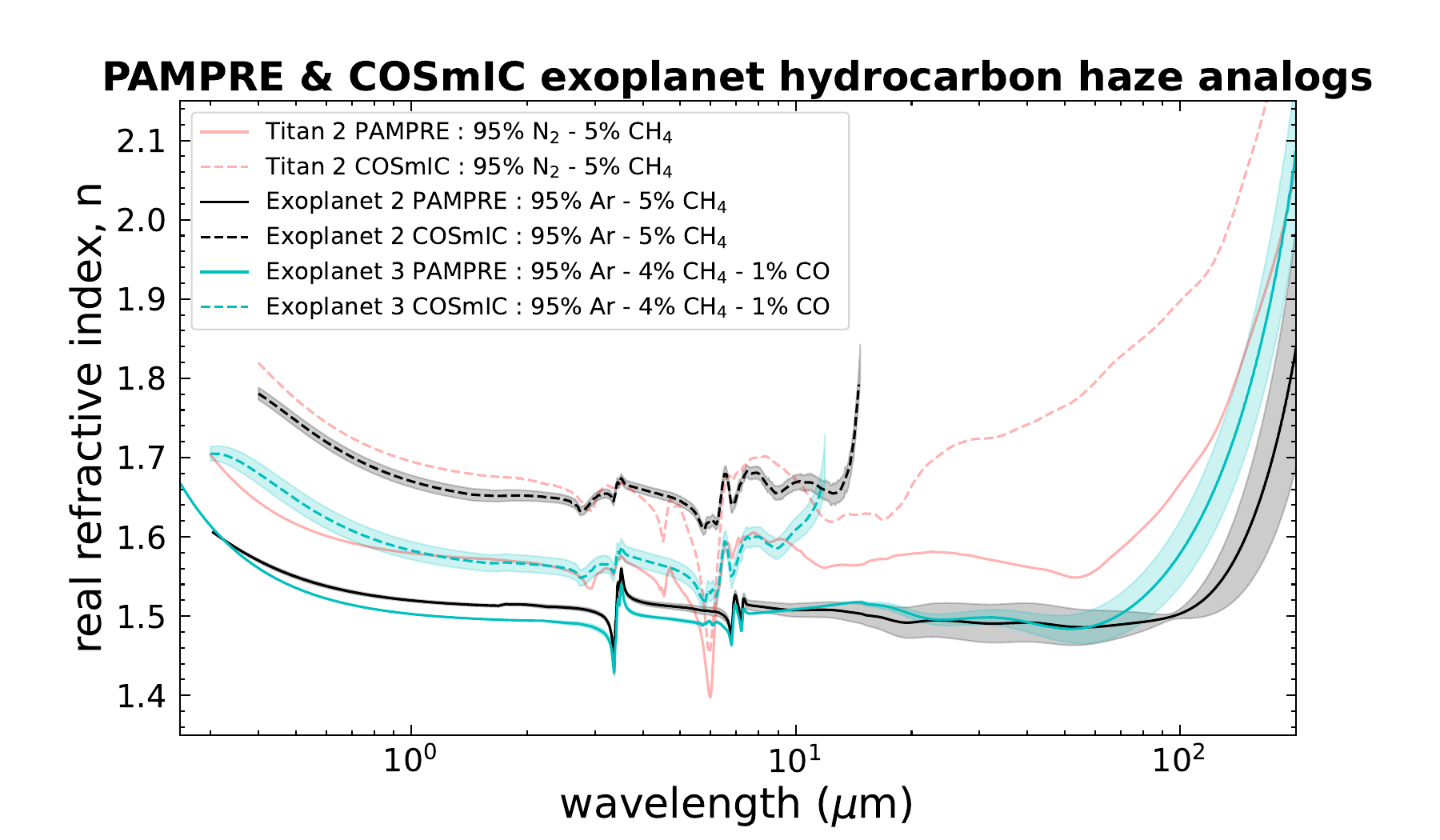}
\includegraphics[width=1\columnwidth]{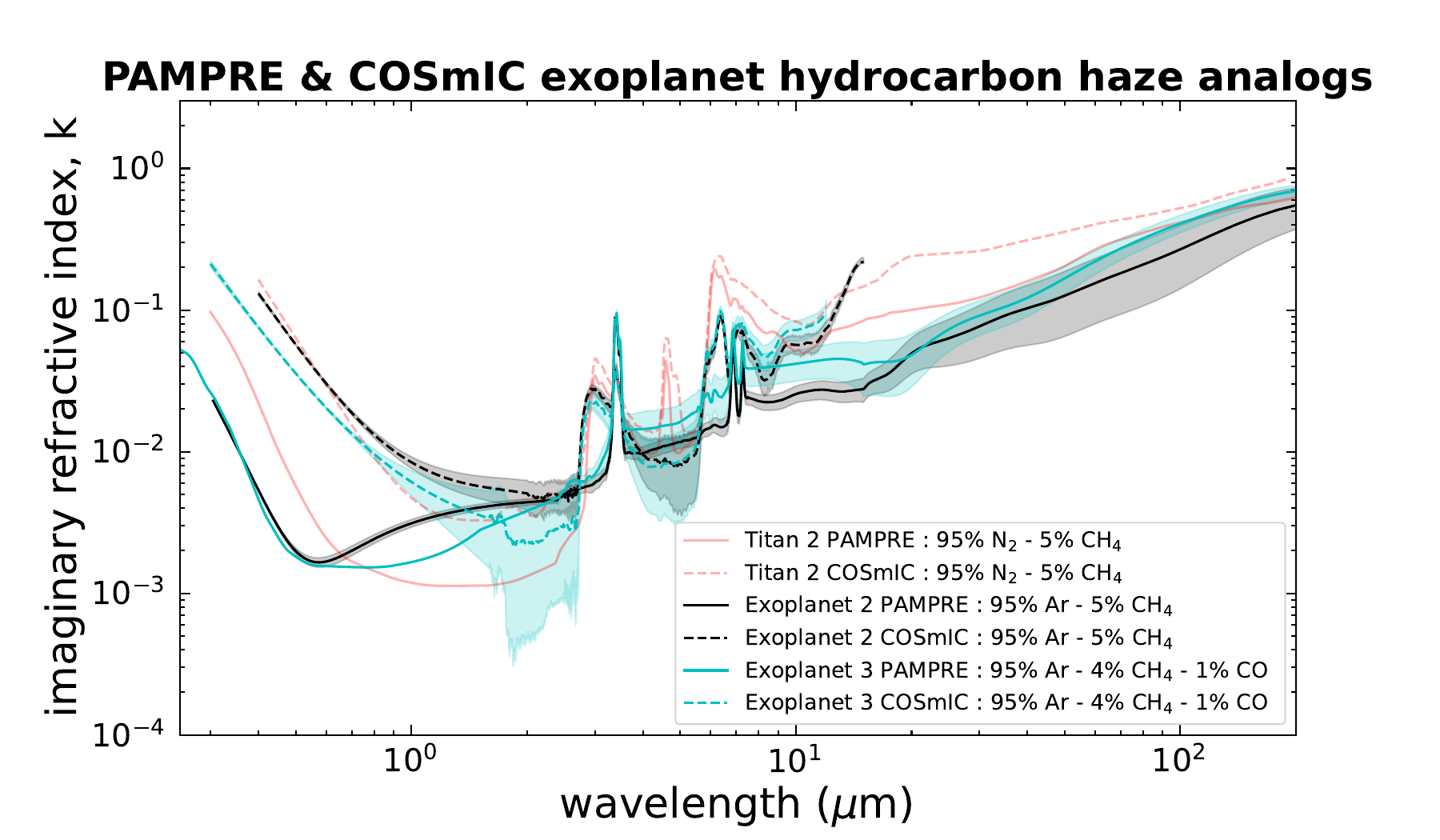}
\includegraphics[width=1\columnwidth]{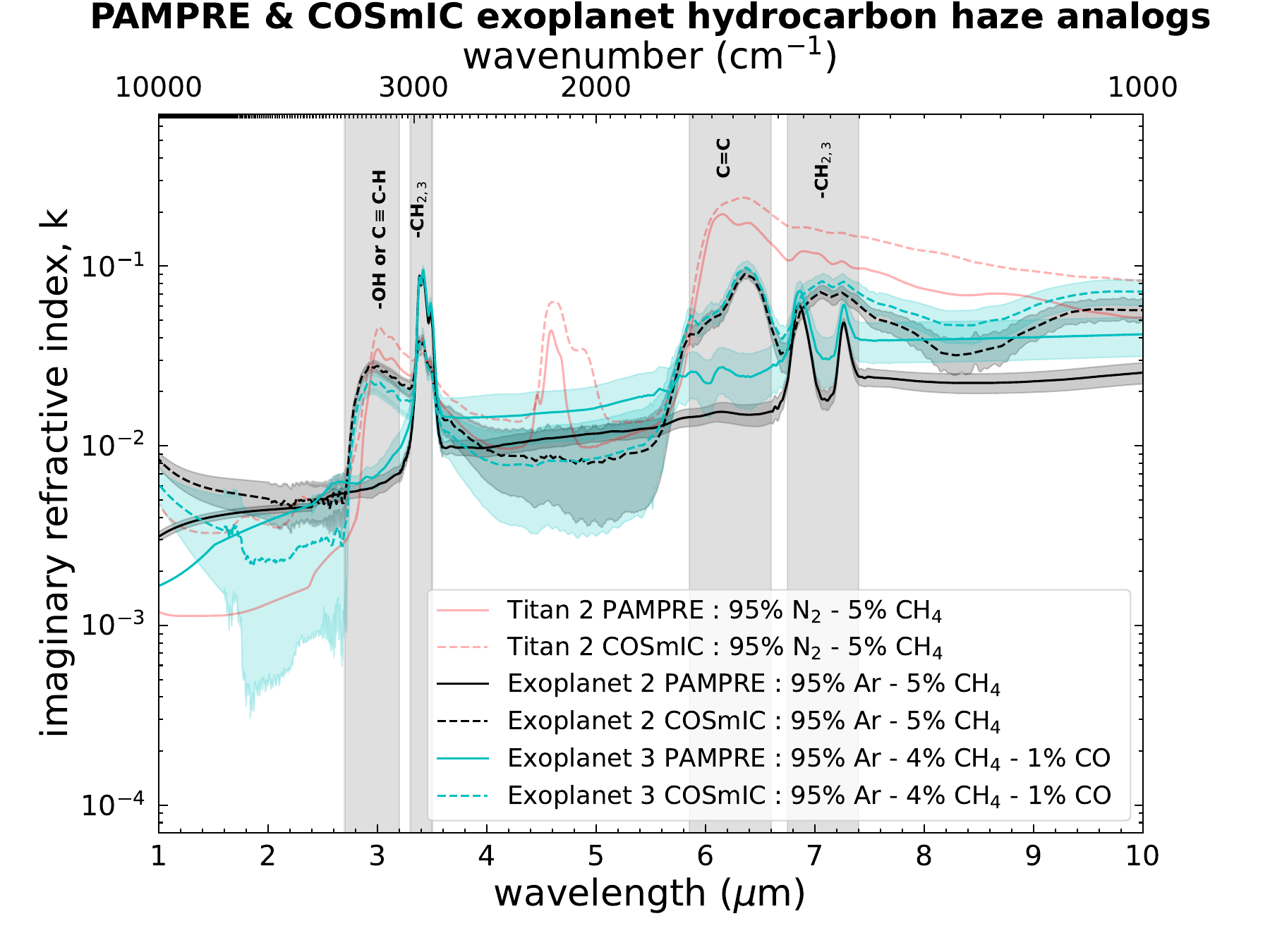}
\caption{Refractive indices n (top) and k (middle \& bottom) from UV to FIR of hydrocarbon exoplanet haze analogs produced without nitrogen with the PAMPRE and COSmIC set-ups. Data for the Exoplanet 2 and Exoplanet 3 analogs produced from different gas relative abundances CH$_4$/CO (see Table \ref{tab:samples}) are shown.       }
\label{fig:Figure12}
\end{figure}

Fig. \ref{fig:Figure12} shows the refractive indices from UV to FIR of the Exoplanet 2 and 3 haze analogs produced without nitrogen using different gas relative abundances CH$_4$/CO (see Table \ref{tab:samples}) with the PAMPRE and COSmIC set-ups. 

We find that the UV-Visible k values of the PAMPRE Exoplanet 2 and 3 analogs are significantly lower compared to the haze analogs produced in N$_2$-dominated gas mixtures, i.e., compared to the PAMPRE Titan, Pluto and Exoplanet 1 haze analogs (Fig. \ref{fig:Figure12}, middle panel). The lack of nitrogen in the solid analog can explain these lower UV-Visible k values and the shift of the bandgap energy towards higher energies (shorter wavelengths) since the electronic transitions of these PAMPRE Exoplanet 2 and 3 analogs are now solely attributed to methyl functional groups, without the contribution from amine functional groups. The difference between the UV-Visible k values of the PAMPRE Exoplanet 2 and Titan 2 analogs confirm our previous interpretation that the incorporation of nitrogen increases absorption properties in the UV-Visible range (Fig. \ref{fig:Figure9} and \ref{fig:Figure12}, middle panels). In addition, the n values of the PAMPRE Exoplanet 2 and 3 analogs are lower than the n values obtained on the PAMPRE Titan, Pluto and Exoplanet 1 analogs (Fig. \ref{fig:Figure12}, top panel) which suggest that the absence of nitrogen in the solid sample also leads to a decrease of n. \cite{Sciamma23} reported a similar behavior between their Titan analogs and their analog produced without nitrogen. 

The spectral variations of n in the IR are small for the PAMPRE and COSmIC Exoplanet 2 and 3 analogs (Fig. \ref{fig:Figure12}, top panel). This is explained by the weaker MIR features of these analogs compared to the analogs produced in N$_2$-dominated gas mixtures (see Fig. \ref{fig:Figure12}, middle \& bottom panels). The important variations of n in the IR are mainly caused by the strong C=C and C=N resonances around 6 $\mu$m in the Titan, Pluto and Exoplanet 1 analogs (Fig. \ref{fig:Figure9}, \ref{fig:Figure10} and \ref{fig:Figure11}). The alkene/aromatic C=C features are observed on the COSmIC Exoplanet 2 and 3 analogs but their k values are lower than those observed on the COSmIC Titan, Pluto and Exoplanet 1 analogs. These C=C features are however not observed on the PAMPRE Exoplanet 2 and 3 analogs. In addition, a large feature at 3 $\mu$m is observed on the COSmIC Exoplanet 2 and 3 analogs but not on the PAMPRE analogs. This feature is likely attributed to -C$\equiv$C-H, although the possibility of it being caused by O-H cannot be ruled out based on our current data. The MIR absorption properties of the PAMPRE Exoplanet 2 and 3 analogs are dominated by alkane/aliphatic CH$_{2,3}$ stretching (3.4 - 3.7 $\mu$m) and bending ($\approx$ 6.7 - 7.2 $\mu$m) modes. These C-H features are also observed on the COSmIC Exoplanet 2 and 3 analogs although they are stronger for the PAMPRE analogs (higher k values) between 3.4 and 3.7 $\mu$m (Fig. \ref{fig:Figure12}, middle \& bottom panels). The differences in MIR features between the PAMPRE and COSmIC Exoplanet 2 and 3 analogs strongly suggest a difference in the solid composition despite the similar initial gas composition. It seems that the COSmIC Exoplanet 2 and 3 analogs contain more aromatic/alkene compounds than the PAMPRE analogs produced with similar gas compositions.

The comparison of the Exoplanet 2 and Exoplanet 3 haze analogs further supports our previous interpretation that that the refractive indices are poorly sensitive to changes of the gas relative abundance CH$_4$/CO. The slope of k in the UV-Visible is indeed similar for both PAMPRE analogs (Fig. \ref{fig:Figure12}, middle panel). The comparison of haze refractive indices between the COSmIC and PAMPRE analogs once again reveals strong variations, with higher UV-Visible k values for the COSmIC analogs. These data obtained on haze analogs produced without nitrogen suggest that the difference in UV-Visible k values reported in Fig. \ref{fig:Figure9}, \ref{fig:Figure10} and \ref{fig:Figure11} is not only attributed to a change in nitrogen incorporation but also likely caused by stronger methyl electronic transitions in the COSmIC analogs compared to the PAMPRE analogs. \\

These new state-of-the-art data presented in Fig. \ref{fig:Figure12} reveal that photochemical hazes produced in N-poor environments exhibit very different optical properties in the entire spectral range from UV to FIR compared to hazes produced in N-rich environments. We emphasize that the refractive indices of these hydrocarbon hazes produced at low temperature via photochemistry are significantly different than the refractive indices of soot material (\cite{Lavvas21} and references therein) which are generally more absorbing in the entire spectral range. These new data should be used to consider the radiative impact of hazes in the N-poor stratosphere of Solar System gas giants (e.g., \cite{Guerlet20}) and the detectability of haze features in sub-Neptune exoplanet atmospheres (e.g., \cite{Gao23} and \cite{He23}). As suggested by the study of \cite{Khare87}, the strength of the C-H IR features and the UV-Visible absorption slope may be different in H$_2$-dominated gas mixtures which will be the focus of future work.

\begin{figure}
\centering
\includegraphics[width=1.\columnwidth]{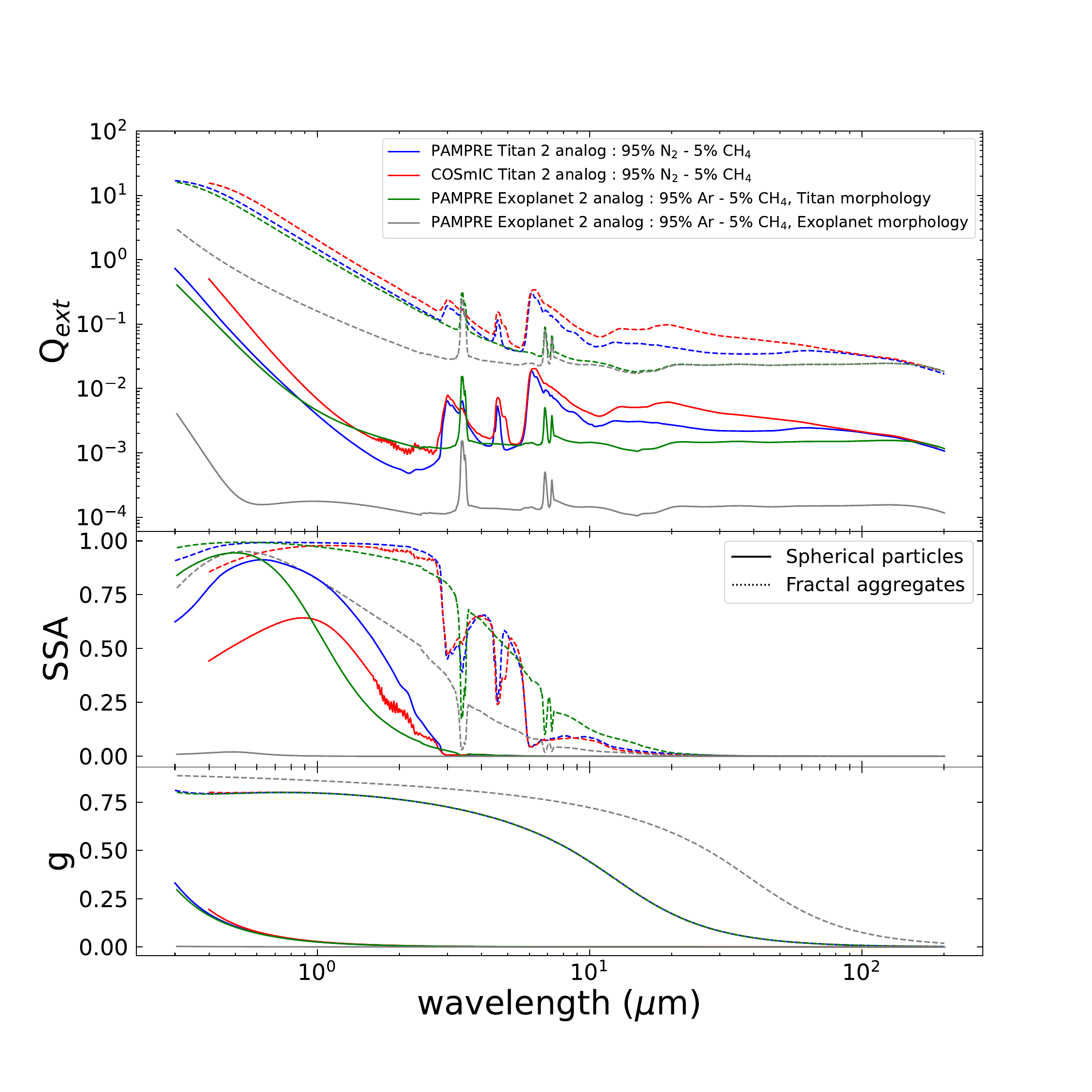}
\caption{Radiative properties including extinction efficiency, single-scattering albedo and asymmetry parameter of haze particles using the refractive indices derived on PAMPRE and COSmIC Titan and Exoplanet haze analogs. The calculations consider simple spherical particles (solid curves) and fractal aggregates of spherical monomers (dashed curves), both expected in hazy planetary atmospheres. Two different morphologies are shown. A first morphology similar to Titan hazes is considered with D$_f$ = 2, N$_m$ = 3000 and r$_m$ = 50 nm, for the three different haze analogs shown (blue, red, green curves). A second morphology to represent exoplanet hazes is shown (gray curves) only for the PAMPRE Exoplanet 2 analog using D$_f$ = 2, N$_m$ = 3.10$^6$ and r$_m$ = 5 nm. The extinction efficiency is here defined by the extinction cross section divided by the surface area of the sphere of equivalent volume.    }
\label{fig:Figure13}
\end{figure}

\subsection{Influence of the refractive indices on the radiative properties of haze particles}

The refractive indices of laboratory-generated haze analogs are used as input in light scattering models to consider the influence of the solid chemical composition and derive the optical properties of spherical and/or aggregate particles observed in planetary atmospheres and surfaces. We performed calculations using the refractive indices of the COSmIC Titan 2, PAMPRE Titan 2 and PAMPRE Exoplanet 2 analogs which exhibit very different n and k values from UV to FIR. The extinction efficiency (Q$_{ext}$), single-scattering albedo (SSA), and asymmetry parameter (g) of haze particles with a similar composition and thus similar refractive indices to our analogs are shown in Fig. \ref{fig:Figure13}. We consider simple spherical particles (solid curves, Fig. \ref{fig:Figure13}) and fractal aggregates of spherical monomers (dashed curves, Fig. \ref{fig:Figure13}) since both are expected to be present at different altitudes in planetary atmospheres and their optical properties are known to differ significantly (e.g., \cite{Adams19}). Our calculations for spherical haze particles follow the theory of \cite{Wiscombe79} (open-source miepython python package\footnote{https://miepython.readthedocs.io/en/latest/}). We further verify the results with the LX-MIE model\footnote{https://github.com/NewStrangeWorlds/LX-MIE/}, described in \cite{Kitzmann18}, which uses a slightly different numerical procedure. We determined the optical properties of spherical particles with a radius of 50 nm ($\pm$ 10\%) guided by observational constraints on the size of Titan aerosols (e.g., \cite{Tomasko08} and \cite{Tomasko09}). Since photochemical hazes are also known to evolve as fractal aggregates (e.g., \cite{West91}, \cite{Cabane92} and \cite{Lavvas08a}), we performed additional calculations following the Mean-Field approximation for scattering by fractal aggregates of identical Mie spheres, first described in \cite{Botet97}, to emphasize the difference in optical properties compared to simple spherical particles. We used an improved version of the model (MFT-M+, \cite{Rannou24}) that accounts for a correction in the limit of the geometrical optics, presented in \cite{Tazaki18} on the one hand, and for the effect of monomer adjacency in the correlation pair function, presented in \cite{Rannou24} on the other hand. The input parameters used, i.e., the particle fractal dimension (D$_f$), the number of monomers (N$_m$) and the monomer radius (r$_m$), are chosen following previous analyses of Titan and exoplanet observations. For a first Titan-like particle morphology, we used D$_f$ = 2, N$_m$ = 3000  and r$_m$ = 50 nm using constraints from Titan observations \citep{Tomasko08,Tomasko09}. For a second particle morphology more representative of exoplanet hazes, we considered three million monomers with a 5-nm radius, still with a D$_f$ of 2, following \cite{Adams19} and \cite{Lavvas17}. In Fig. \ref{fig:Figure13}, we used the Titan-like haze morphology for the three different haze analog refractive indices (red, green and blue curves), and we used the second exoplanet morphology only for the PAMPRE Exoplanet 2 analog (gray curve).

Fig. \ref{fig:Figure13} shows that the extinction efficiency is strongly influenced by the refractive indices and by the aerosol morphology in the entire spectral range from UV to FIR.  The single-scattering albedo, that controls the fraction of light scattered by aerosols, is also very sensitive to the refractive indices while the asymmetry parameter, that reflects the angular distribution of the scattered light, is rather insensitive to the refractive indices but very sensitive to the particle morphology. Fig. \ref{fig:Figure13} emphasizes that the optical properties of exoplanet hazes are significantly different because of the change of refractive indices but also because of the change in morphology. 

The variations of the UV-Visible extinction slope, combined with the variation in the single scattering albedo, suggest that the composition of the haze particles will significantly influence the planetary albedo and the atmospheric heating-cooling properties. This will directly control the absorption and transmission of diffuse light through the hazy atmosphere. These differences will likely lead to significant variations of the temperature in the upper region of the atmosphere which is probed during spectroscopic exoplanet observations (e.g., \cite{Steinrueck23}). 

Fig. \ref{fig:Figure13} also shows that the MIR absorption features of hazes significantly impact the extinction efficiency and may therefore be used to infer the presence of hazes and constrain other atmospheric properties from JWST observations, as proposed by \cite{He23}. Given the transparency of N$_2$ in the spectral range of JWST, the differences in the MIR haze features shown in Fig. \ref{fig:Figure13} may help constrain the presence or absence of nitrogen, in the atmospheres of exoplanets for instance, using the different N-H (3.1 - 3.3 $\mu$m), -C$\equiv$N/-N$\equiv$C (4.4 - 4.7 $\mu$m) and C=N (6.1 - 6.6 $\mu$m) solid features. 

Recent work revealed that the addition of sulfur in the gas mixture increases the haze production rate \citep{He20,Reed22} leading to solid analogs with a composition rich in sulfur \citep{Vuitton21} and higher n and k values at Visible wavelengths \citep{Reed23}. On the other hand, a gas mixture rich in sulfur and CO$_2$, as expected for the atmosphere of early-Earth, would produce hazes with lower absorption properties at visible wavelengths likely caused by a solid composition richer in inorganic salts \citep{Jansen25}. These recent studies emphasize the need for additional experimental measurements to understand the influence of different gases on the optical properties of hazes in a broad spectral range.

\section{Conclusions}

In the present work, we derive the refractive indices from UV to FIR (up to 200 $\mu$m) of Titan, Pluto and Exoplanet haze analogs produced with various gas compositions and two different experimental set-ups, namely PAMPRE at LATMOS (France) and COSmIC at the NASA Ames Research Center (USA). Using different approaches to determine the refractive indices, we found that the Cauchy analytical functions are more reliable than the Tauc-Lorentz physical description to retrieve accurate n and k values in the UV-Visible-NIR. We report strong agreement between the refractive indices determined by ellipsometry and spectroscopy. Discrepancies in the n and k values determined with both techniques arise when the sample are weakly absorbing, mainly in the NIR spectral range, as the retrieval of low k values becomes very sensitive to measurement biases such as beam focalization. 

We report strong differences between the refractive indices of PAMPRE and COSmIC haze analogs, even for those produced from similar gas compositions, which is caused by the variations of the experimental conditions (i.e., gas temperature, pressure, energy of plasma discharge and gas residence time) influencing the gas phase haze precursors. Using transmission measurements performed at different temperatures from 40 to 288K, we confirm that the measurement temperature does not influence the refractive indices of the haze analogs in the mid-IR. The COSmIC analogs generally present higher n and k values in the entire spectral range. Using different gas compositions where we varied the N$_2$/CH$_4$ and CH$_4$/CO abundance ratios in the initial gas mixture, we observed that changes in the gas N$_2$/CH$_4$ ratio strongly influence the haze analogs k values in the entire spectral range. On the other hand, a decrease of the gas relative abundance CH$_4$/CO from $\infty$ to 4 does not affect the refractive indices of the organic analogs significantly. Based on the data reported in previous work, it suggests that CO and CO$_2$ influence the gas chemistry and haze composition differently which leads to variations of the haze refractive indices. Using analogs produced without any nitrogen, we found that C-H-O haze analogs present refractive indices very different than nitrogenous haze analogs. They are generally more transparent with lower n values in the entire spectral range from the UV to the IR. The variations of IR absorption features between hazes produced with and without nitrogen in the gas composition could help constrain the detection of gaseous N$_2$ in exoplanet atmospheres from telescope observations. 

For each application that requires the use of haze analog refractive indices, we recommend using both PAMPRE and COSmIC data sets to fully understand and assess the influence of the refractive indices on the outcome, whether they are used for modeling or observation studies. The large differences in the k values of our different analogs in the UV-Visible and IR suggest that the radiative impact of photochemical hazes may change significantly following variations of the haze composition. These new data should be used in future work to re-analyze previous observations of Titan and Pluto's surface and atmosphere, as well as to re-assess the influence of hazes on the cooling and heating properties of the atmosphere. The state-of-the-art refractive indices data reported in this work for C-H hazes produced without nitrogen should be used to analyze observations of the Solar System gas giants (e.g., Jupiter) and exoplanet atmospheres (e.g., sub-Neptunes), as well as to assess the impact of the haze composition on the detectability of haze signatures and the resulting degeneracy on retrieved haze properties (e.g., production rate or particle number density) during the analysis of JWST data.

\begin{acknowledgements}
T.D. acknowledges support by the "ADI 2021" project funded by the IDEX Paris-Saclay (ANR-11-IDEX-0003-02). T.D. thanks the NOMIS Foundation ETH Fellowship Programme and the respective research made possible thanks to the support of the NOMIS Foundation. T.D. thanks the SOLEIL synchrotron for the AILES beamtime provided (proposal ID 20231358 and 20221651). T.D., E.S.-O. and L.M. thank the entire staff at AILES for their support and assistance during the optical experiments. T.D., L.M. and L.V. thank Prof. N. Carrasco and the European Research Council for funding via the ERC OxyPlanets project (grant agreement No. 101053033). E.S.-O., L.J., and C.L.R. acknowledge support from the Internal Scientist Funding Model (ISFM) NASA Ames Laboratory Astrophysics Directed Work Package (22-A22ISFM-0009). E.S.-O., C.L.R., T.L.R., D.H.W. acknowledge support from the Internal Scientist Funding Model (ISFM) NASA Center for Optical Constants.    
\end{acknowledgements}

\section*{Data Availability}
The refractive indices data provided in this article are available upon request to tdrant@ethz.ch.

\begin{appendix}

\section{Transmission spectra of the different PAMPRE haze analogs from UV to near-IR}

Transmission spectroscopy was used to characterize the optical properties of the PAMPRE haze analogs from UV to near-IR, up to 2.5 $\mu$m (see Sect. 3.1). Fig. \ref{fig:S1} presents the transmission spectra of the four other PAMPRE analogs (Titan 2, Pluto, Exoplanet 1, Exoplanet 3) not shown in Fig. \ref{fig:Figure3}. The analysis with the Swanepoel method is also shown with the constructed envelopes used for the calculations. Table \ref{tab:S1} presents the film thicknesses of each analog derived during the data analysis. We performed measurements at different positions on each analog sample to derive an uncertainty on the refractive indices. Table \ref{tab:S1} confirms that the thickness of the PAMPRE analog samples is generally homogeneous and does not vary significantly between the different positions. Fig. \ref{fig:S2} presents the determination of n and k with the Swanepoel method for the PAMPRE Titan 2 haze analog.

\renewcommand{\thefigure}{A.1} % change le label en "Figure S1"
\begin{figure*}
\centering
\includegraphics[width=0.9\columnwidth]{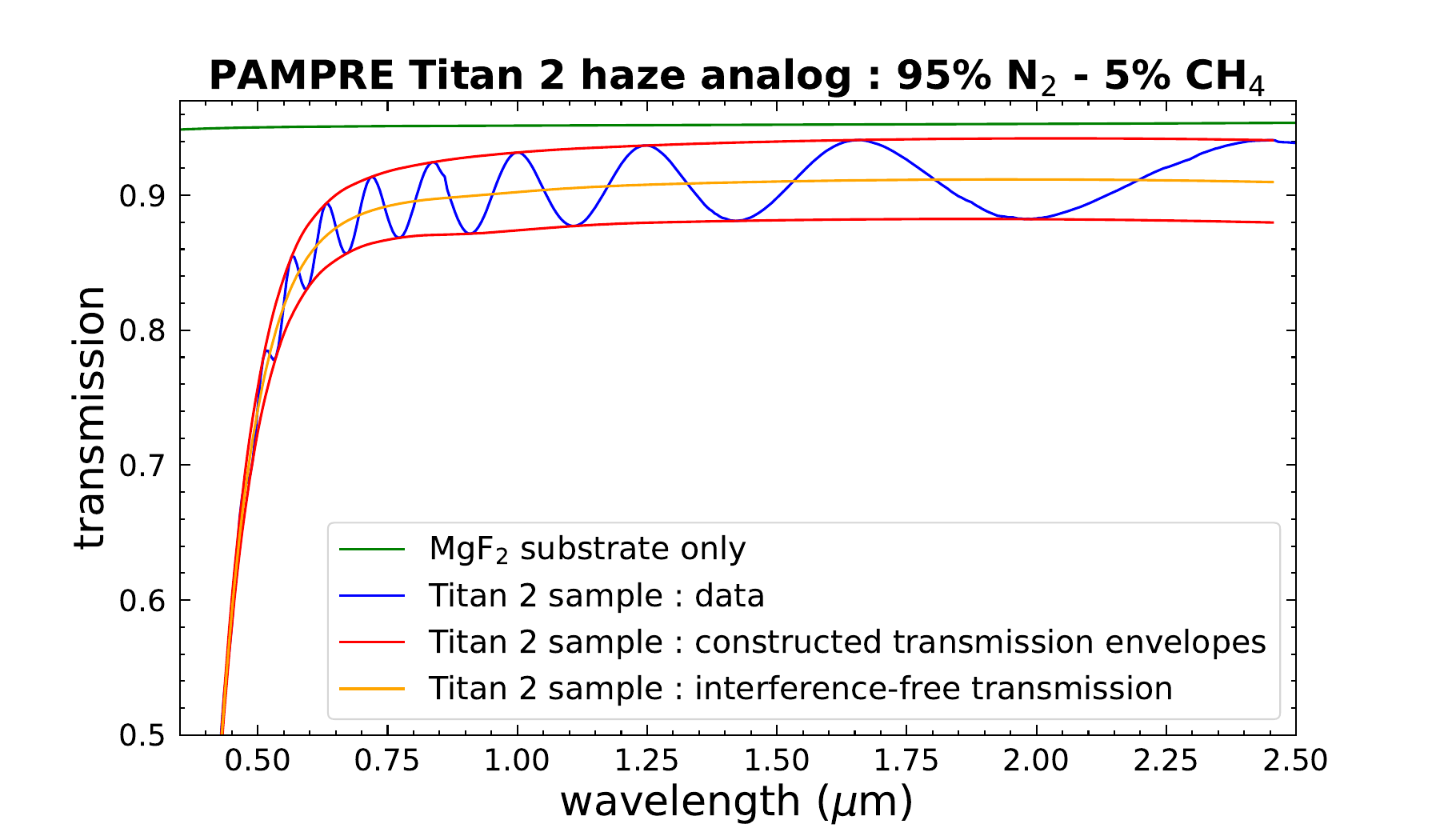}
\includegraphics[width=0.9\columnwidth]{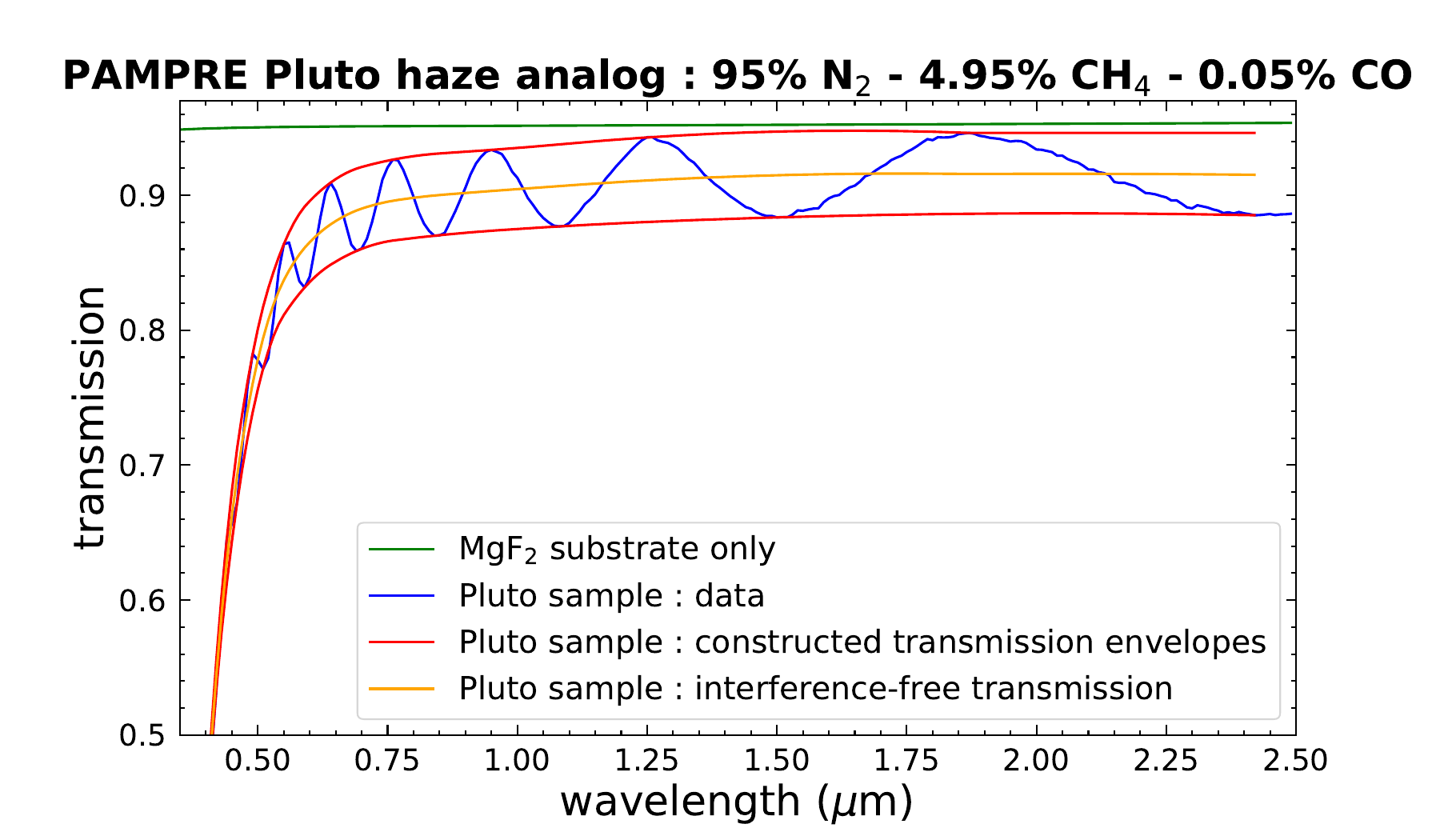}
\includegraphics[width=0.9\columnwidth]{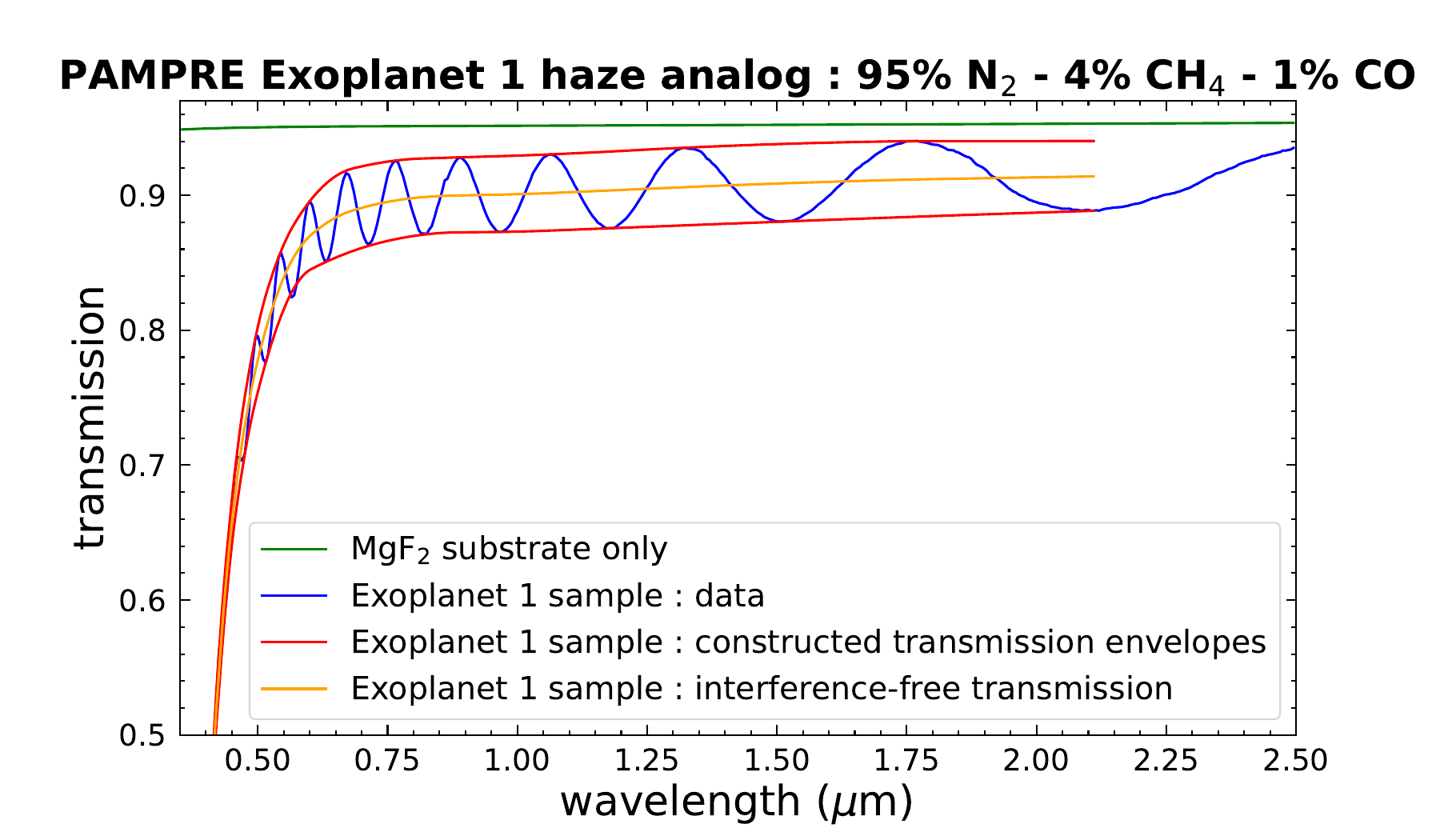}
\includegraphics[width=0.9\columnwidth]{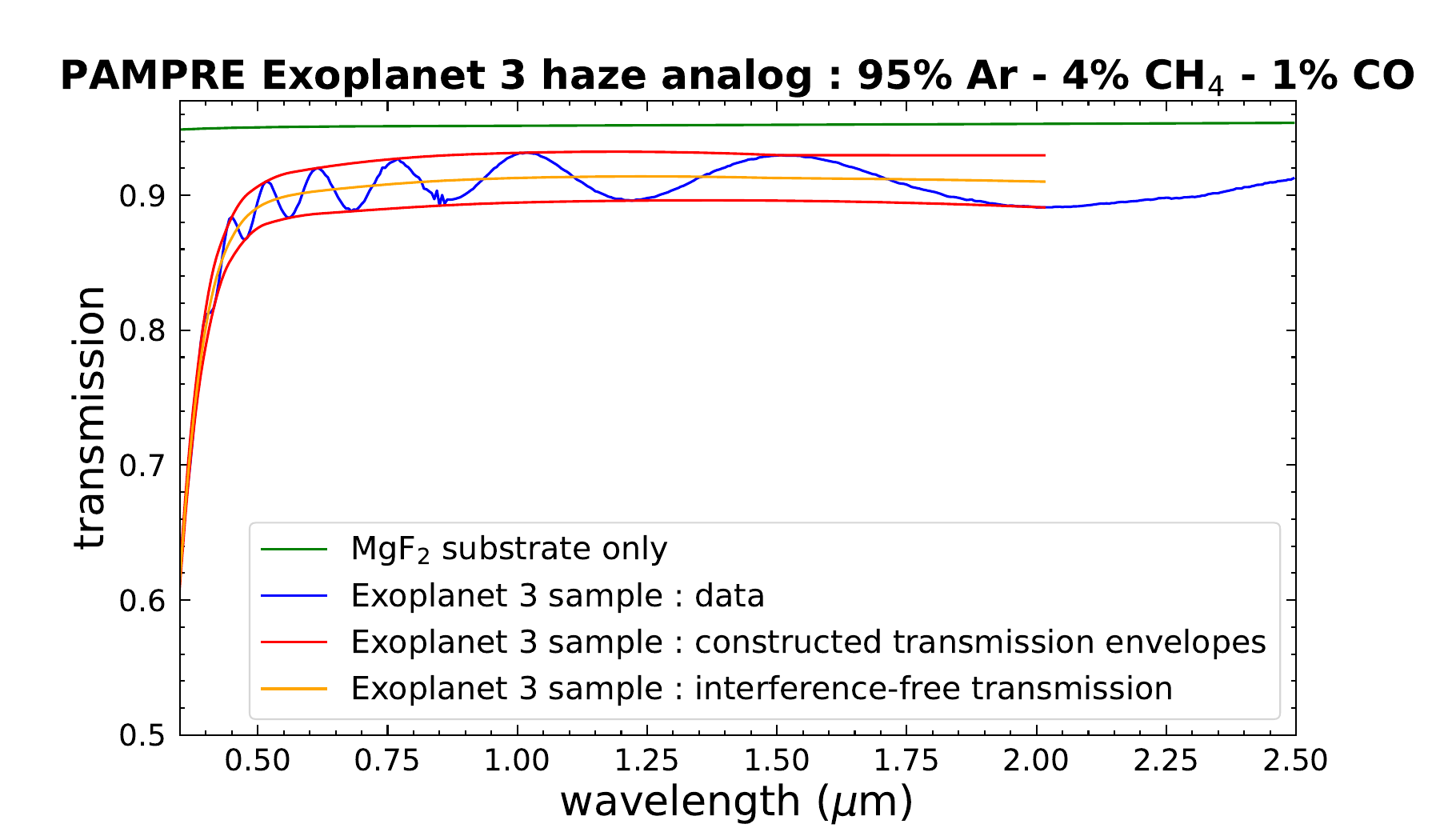}
\caption{Transmission spectra obtained for the different PAMPRE haze analogs from UV to near-IR. For the Exoplanet 3 analog, the absorption slope in the UV-Visible is shifted toward shorter wavelengths compared to the other analogs, which indicates that this sample without nitrogen is  more transparent in that spectral range.} 
\label{fig:S1}
\end{figure*}

\renewcommand{\thefigure}{A.2} % change le label en "Figure S1"
\begin{figure*}
\centering
\includegraphics[width=0.9\columnwidth]{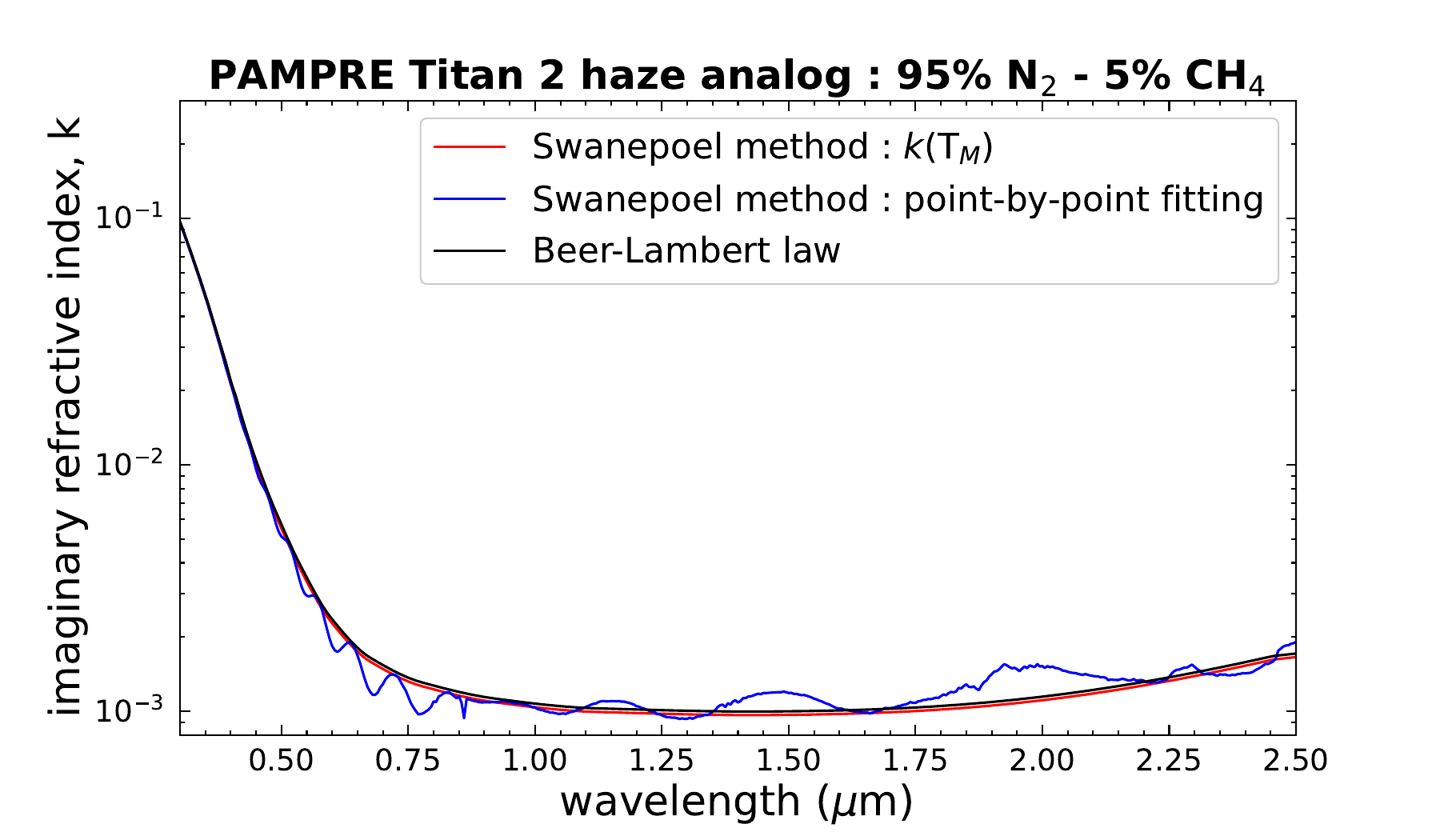   }
\includegraphics[width=0.9\columnwidth]{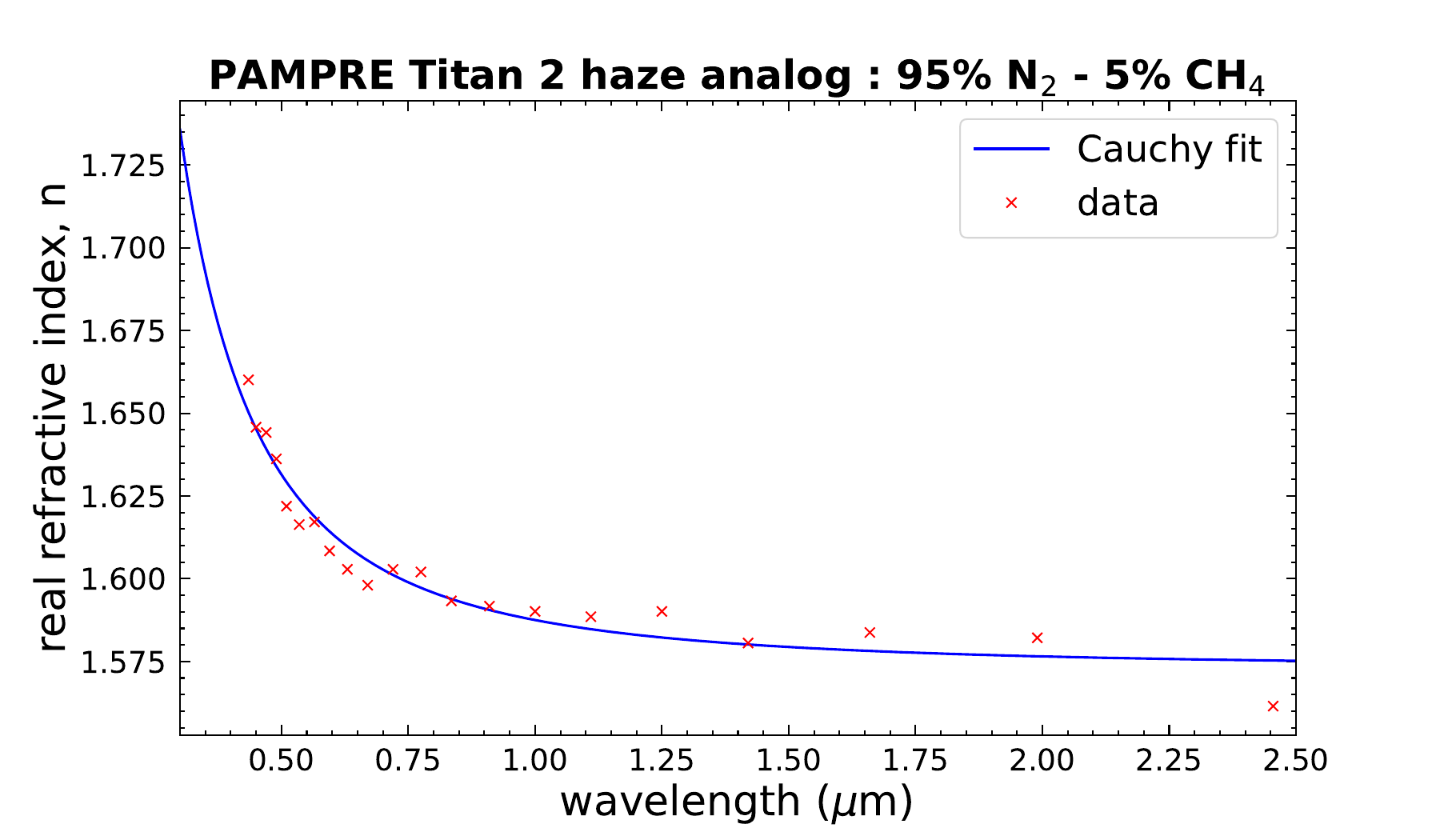   }
\label{fig:Figure5}
\caption{Determination of the refractive indices with the Swanepoel method for the PAMPRE Titan 2 haze analog. The point-by-point fitting approach (left) shows small variations of k in the Visible-NIR and reflects the propagation of uncertainties on n and the film thickness into the calculation of k. For n, a value can be derived for each fringe extremum (red points) and the different data points are then fitted with a Cauchy function to obtain a continuous spectrum in the entire range from UV to near-IR. }
\label{fig:S2}
\end{figure*}

\renewcommand{\thetable}{A.1} % change le label en "Table S1"
\begin{table*}
\centering
\caption{Film thicknesses derived for each PAMPRE analog sample from UV to near-IR using the transmission data.    } 
\begin{tabular}{lccccccc}
 \hline
 \hline
  \rule{0pt}{2.5ex}PAMPRE analog   & Gas composition & Spectral range & Position &  Film thickness$^a$ (nm) \\
 \hline 
\rule{0pt}{2.5ex}Titan 1  &  90\% N$_2$ - 10\% CH$_4$ & 0.3 - 2.5 $\mu$m & 1 &  1737.0 $\pm$ 12.0  \\
\rule{0pt}{2.5ex}  &  &  & 2 & 1745.2 $\pm$ 11.7   \\
\rule{0pt}{2.5ex}  &  &  & 3 & 1740.2 $\pm$ 13.0   \\
\hline
\rule{0pt}{2.5ex}Titan 2  &  95\% N$_2$ - 5\% CH$_4$ & 0.3 - 2.5 $\mu$m & 1 &  1572.2 $\pm$ 11.3   \\
\rule{0pt}{2.5ex} &   &  & 2 &  1672.6 $\pm$ 25.1    \\
\hline
\rule{0pt}{2.5ex}Pluto  &  95\% N$_2$ - 4.95\% CH$_4$ - 0.05\% CO & 0.27 - 2.5 $\mu$m & 1 &  1183.9 $\pm$ 15.4   \\
\rule{0pt}{2.5ex}  &  &  & 2 &  1184.9 $\pm$ 20.1    \\
\hline
\rule{0pt}{2.5ex}Exoplanet 1  &  95\% N$_2$ - 4\% CH$_4$ - 1\% CO & 0.285 - 2.5 $\mu$m & 1 &  1670.2 $\pm$ 13.4   \\
\rule{0pt}{2.5ex}  &  &  & 2 &  1672.6 $\pm$ 12.5    \\
\rule{0pt}{2.5ex}  &  &  & 3 &  1676.7 $\pm$ 13.6    \\
\hline
\rule{0pt}{2.5ex}Exoplanet 2  &  95\% Ar - 5\% CH$_4$ & 0.3 - 2.5 $\mu$m & 1 & 606.0  \\
\rule{0pt}{2.5ex} &   &  & 2 & 606.6    \\
\rule{0pt}{2.5ex} &   &  & 3 & 607.9    \\
\hline
\rule{0pt}{2.5ex}Exoplanet 3  &  95\% Ar - 4\% CH$_4$ - 1\% CO & 0.25 - 2.5 $\mu$m & 1 &  1008.7 $\pm$ 16.1   \\
\rule{0pt}{2.5ex}  &   &  & 2 &  1009.3 $\pm$ 13.8    \\
\hline
\end{tabular}
\begin{tablenotes}
      \small
      \item \textbf{Notes.} 
      \item $^a$ With the Swanepoel method, a thickness value can be derived for each adjacent interference fringes and an uncertainty can thus be obtained using the different pairs of fringes observed in the spectrum. The error given here corresponds to the maximum deviation from the mean value of the thickness.  
      
\end{tablenotes}
\vspace{0.1cm}

\label{tab:S1}
\end{table*}

\renewcommand{\thefigure}{A.3} % change le label en "Figure S1"
\begin{figure*}
\centering
\includegraphics[width=0.9\columnwidth]{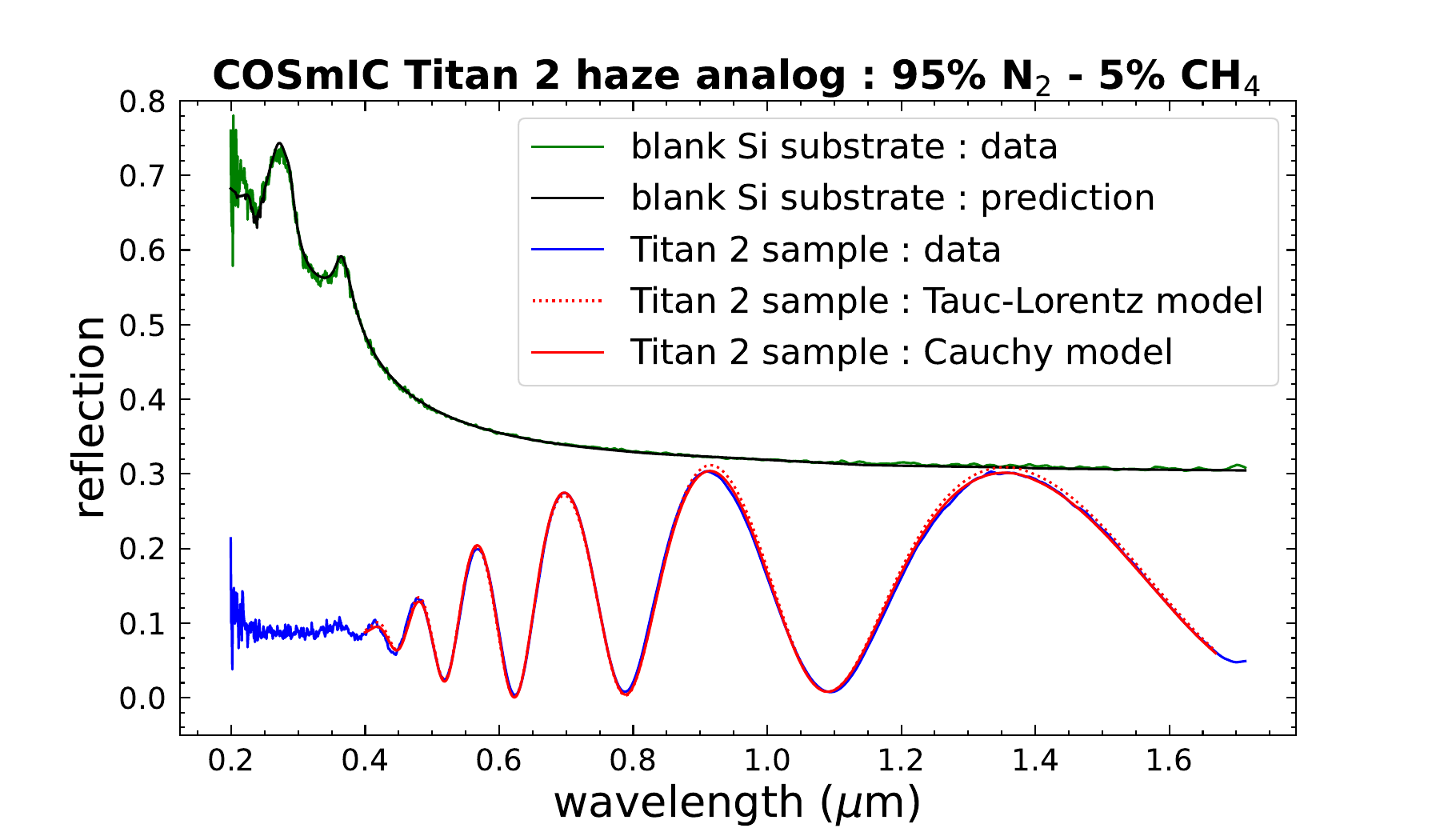   }
\includegraphics[width=0.9\columnwidth]{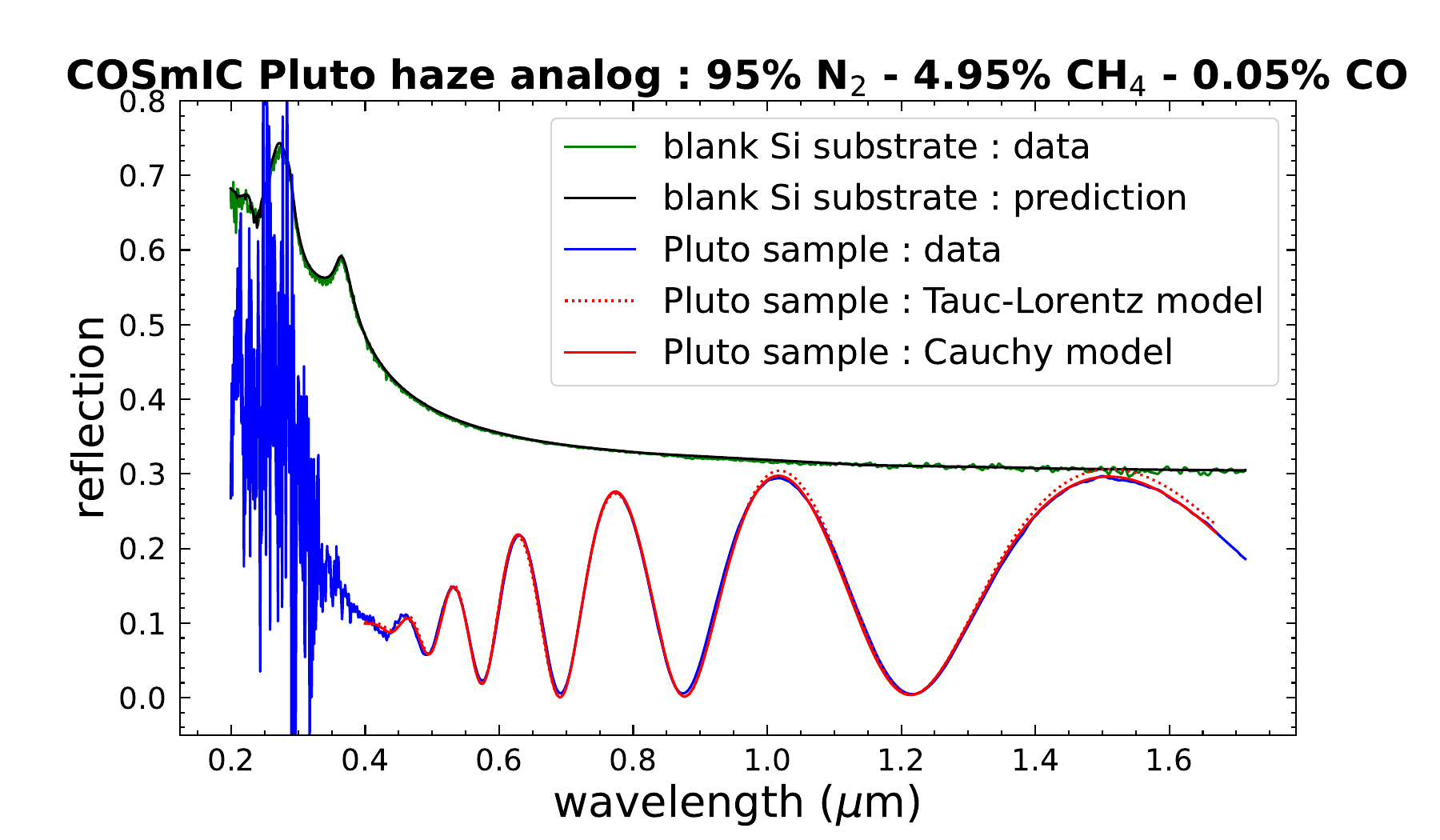}
\includegraphics[width=0.9\columnwidth]{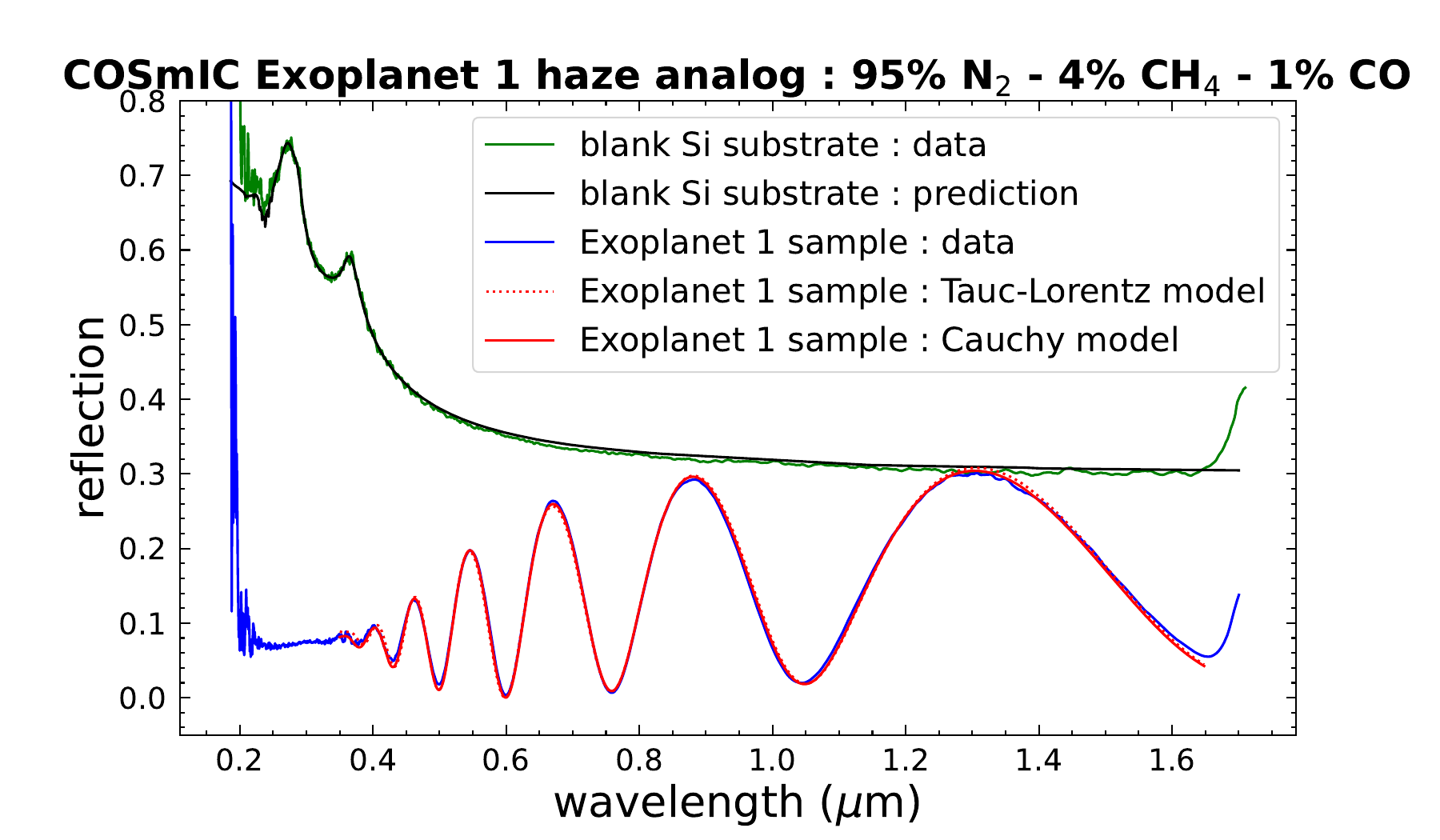}
\includegraphics[width=0.9\columnwidth]{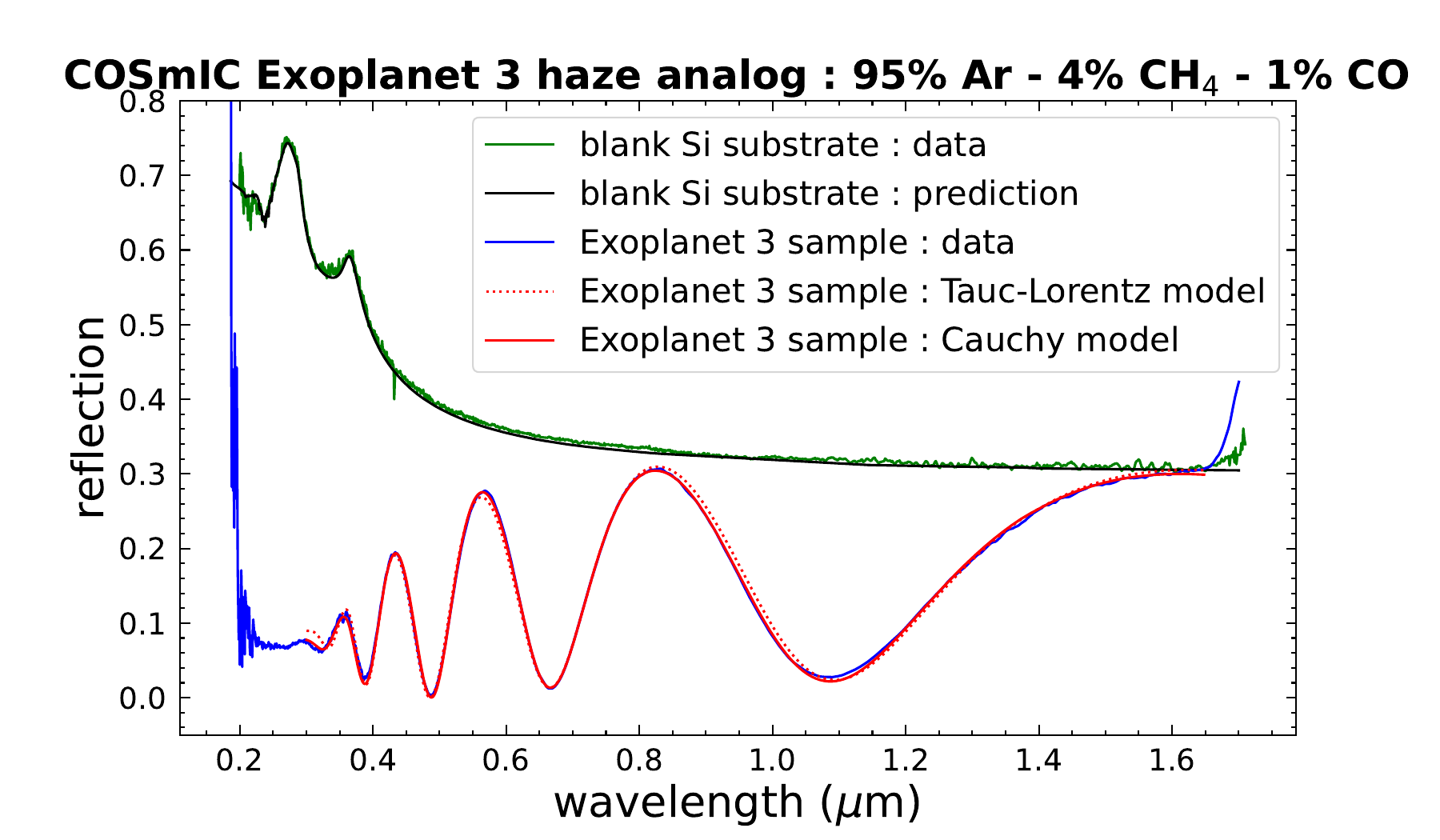}
\caption{Reflection spectra obtained for the different COSmIC haze analogs from UV to near-IR. The measured spectra are directly fitted to a theoretical model with Cauchy and Tauc-Lorentz equations for the refractive indices.  }
\label{fig:S3}
\end{figure*}

\renewcommand{\thetable}{A.2} % change le label en "Table S1"
\begin{table*}
\caption{Film thicknesses determined for each COSmIC haze analog from the reflection spectra obtained in the UV-Vis-NIR spectral range.     } 
\centering
\begin{tabular}{lccccccc}
 \hline
 \hline
  \rule{0pt}{2.5ex}COSmIC analog  & Gas composition & Spectral range & Position$^a$ &  Film thickness (nm) \\
 \hline 
\rule{0pt}{2.5ex}Titan 1  &  90\% N$_2$ - 10\% CH$_4$ & 0.4 - 1.67 $\mu$m & 1 &  1134.0   \\
\rule{0pt}{2.5ex}  &   &  & 2 &  1031.1   \\
\rule{0pt}{2.5ex}  &   &  & 3 &  990.1   \\
\rule{0pt}{2.5ex}  &   &  & 4 &  1108.9   \\
\hline
\rule{0pt}{2.5ex}Titan 2  &  95\% N$_2$ - 5\% CH$_4$ & 0.4 - 1.67 $\mu$m & 1 &  798.2  \\
\rule{0pt}{2.5ex}  &  &  & 2 &  1023.0    \\
\rule{0pt}{2.5ex}  &  &  & 3 &  802.6    \\
\hline
\rule{0pt}{2.5ex}Pluto  &  95\% N$_2$ - 4.95\% CH$_4$ - 0.05\% CO & 0.4 - 1.67 $\mu$m & 1 &  762.4   \\
\rule{0pt}{2.5ex}  &  &  & 2 &  624.5    \\
\rule{0pt}{2.5ex}  &  &  & 3 &  574.8    \\
\rule{0pt}{2.5ex}  &  &  & 4 &  875.0    \\
\hline
\rule{0pt}{2.5ex}Exoplanet 1  &  95\% N$_2$ - 4\% CH$_4$ - 1\% CO & 0.35 - 1.65 $\mu$m & 1 &  813.8   \\
\rule{0pt}{2.5ex}  &  &  & 2 &  688.5    \\
\rule{0pt}{2.5ex}  &  &  & 3 &  722.7    \\
\rule{0pt}{2.5ex}  &  &  & 4 &  809.2    \\
\hline
\rule{0pt}{2.5ex}Exoplanet 2  &  95\% Ar - 5\% CH$_4$ & 0.35 - 1.67 $\mu$m & 1 &  447.8   \\
\rule{0pt}{2.5ex} &  &  & 2 &  464.6   \\
\hline
\rule{0pt}{2.5ex}Exoplanet 3  &  95\% Ar - 4\% CH$_4$ - 1\% CO & 0.3 - 1.67 $\mu$m & 1 &  611.2  \\
\rule{0pt}{2.5ex} &  &  & 2 &  647.0    \\
\rule{0pt}{2.5ex} &  &  & 3 &  559.0    \\
\rule{0pt}{2.5ex} &  &  & 4 &  611.4    \\
\rule{0pt}{2.5ex} &  &  & 5 &  509.8    \\
\hline
\end{tabular}
\begin{tablenotes}
      \small
      \item \textbf{Notes.} 
      \item $^a$ Several spectra were acquired at different positions on the sample to assess thickness variability and obtain an uncertainty on the refractive indices.
      
\end{tablenotes}
\vspace{0.1cm}
\label{tab:S2}
\end{table*}

\renewcommand{\thefigure}{A.4} % change le label en "Figure S1"
\begin{figure*}
\centering
\includegraphics[width=0.9\columnwidth]{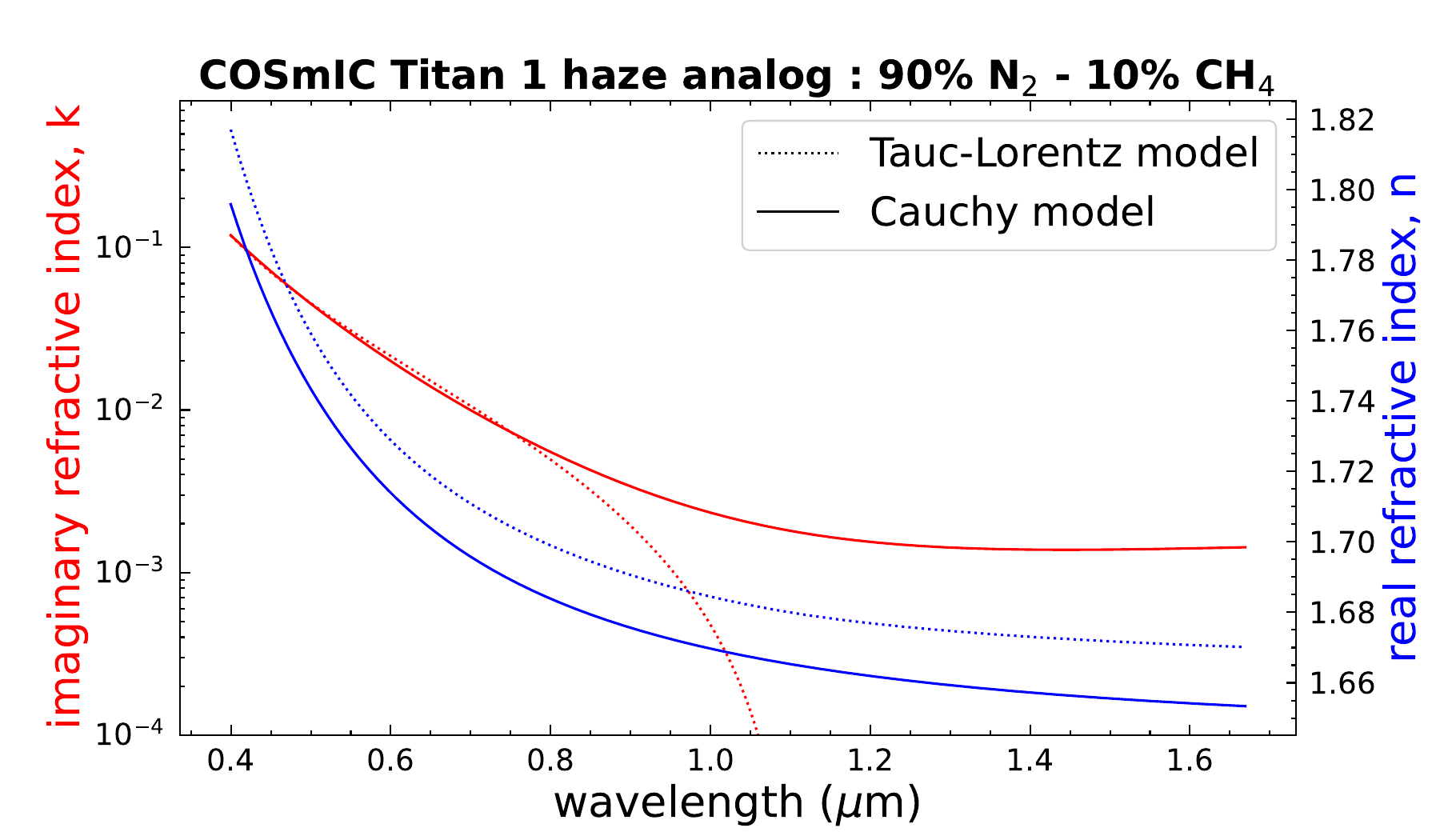   }
\includegraphics[width=0.9\columnwidth]{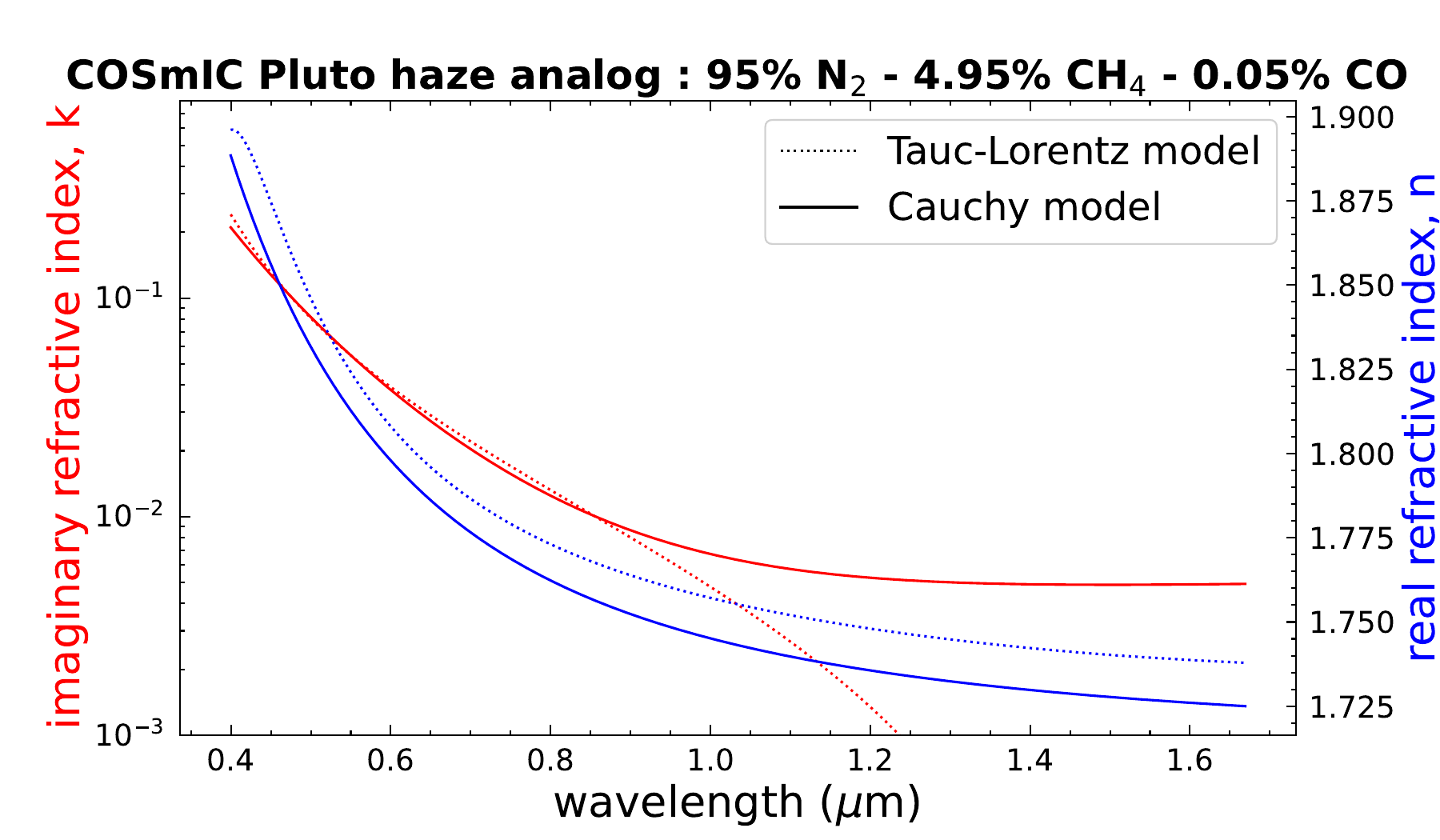   }
\caption{Refractive indices n and k derived from the UV-Vis-NIR reflection spectra for the COSmIC Titan 1 haze analog (left) and COSmIC Pluto haze analog (right). The results obtained with both Cauchy and Tauc-Lorentz functions are shown.  }
\label{fig:S4}
\end{figure*}

\renewcommand{\thefigure}{A.5} % change le label en "Figure S1"
\begin{figure*}[p]
\centering
\includegraphics[width=0.9\columnwidth]{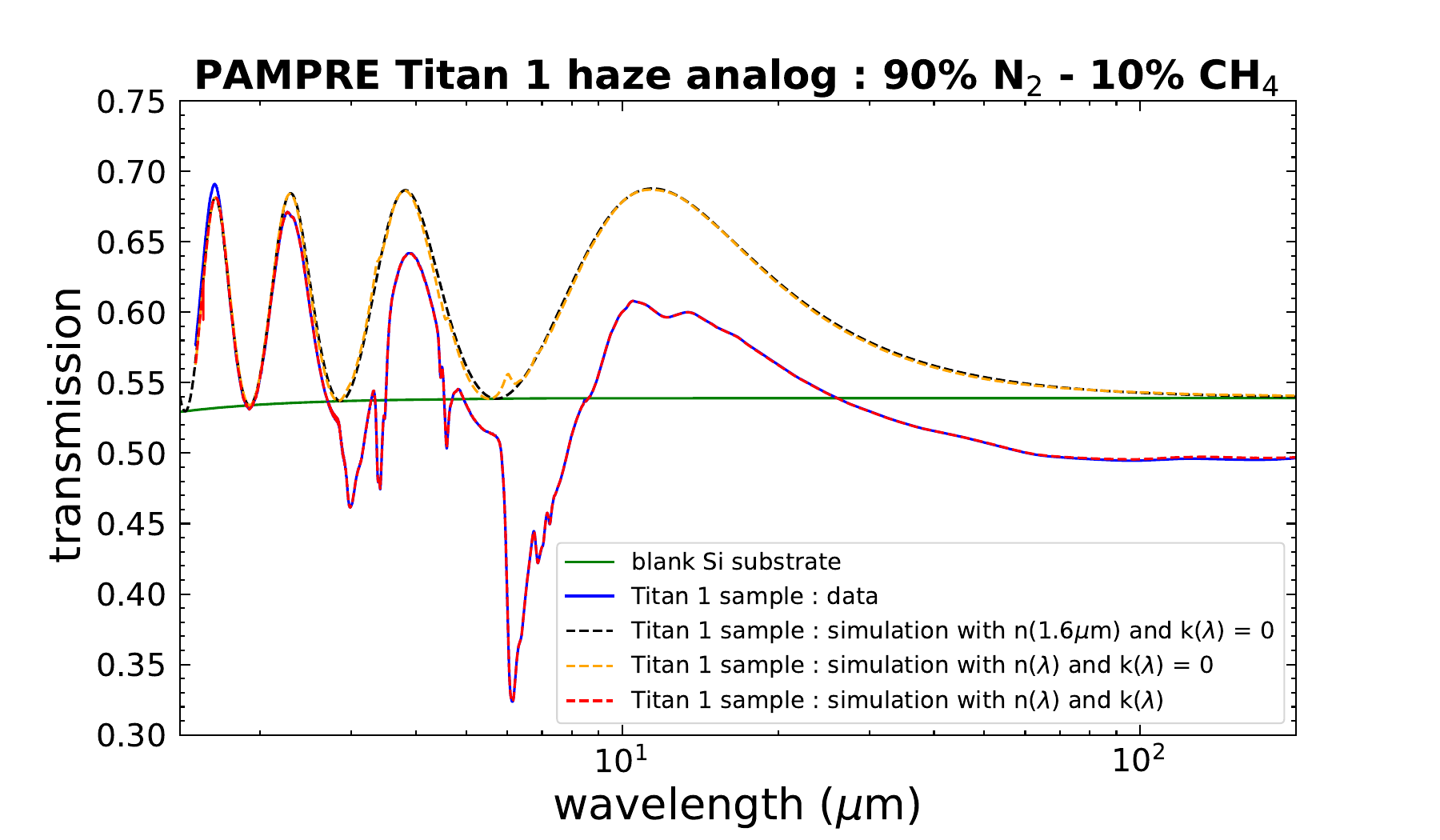   }
\includegraphics[width=0.9\columnwidth]{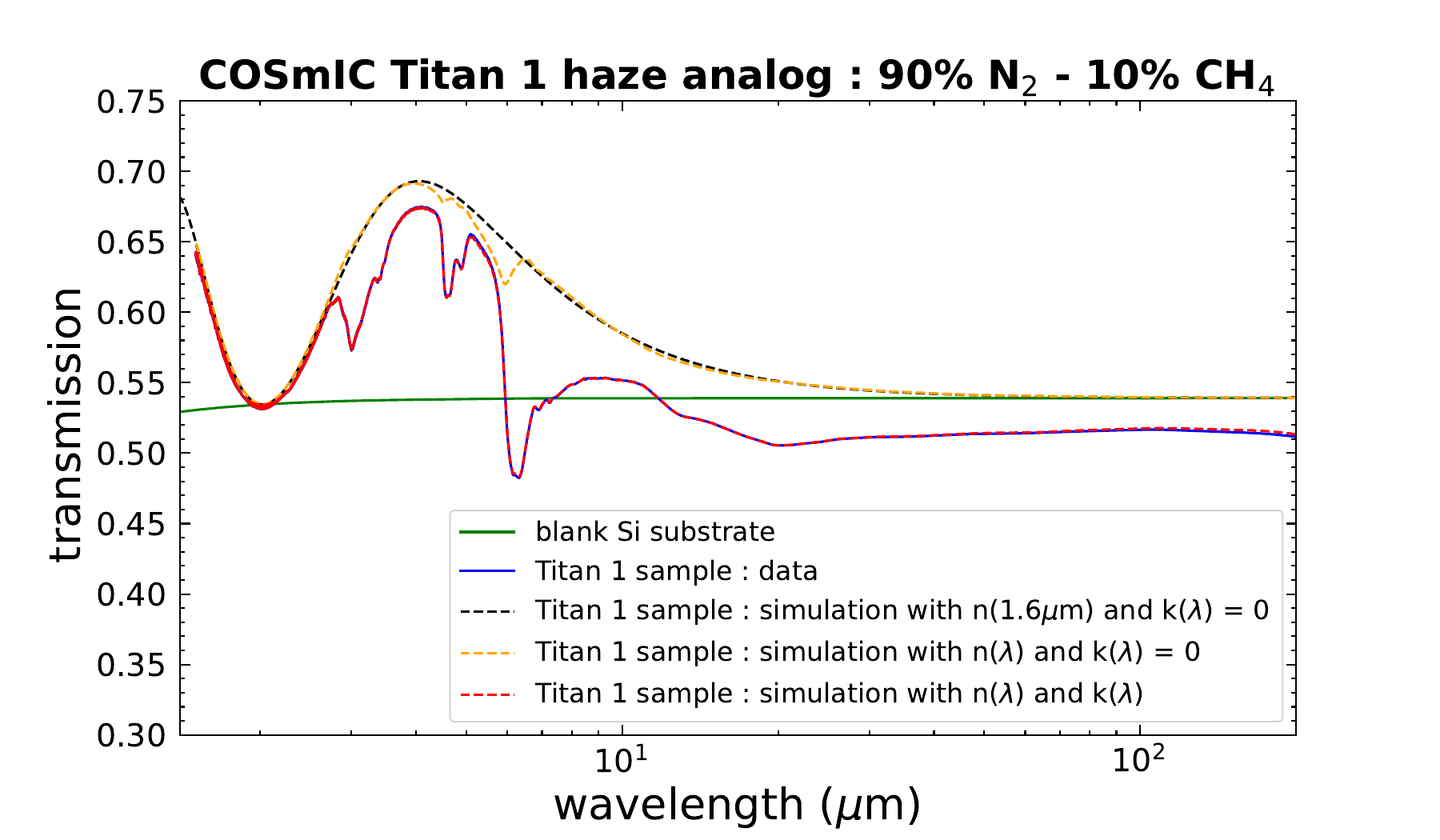   }
\includegraphics[width=0.9\columnwidth]{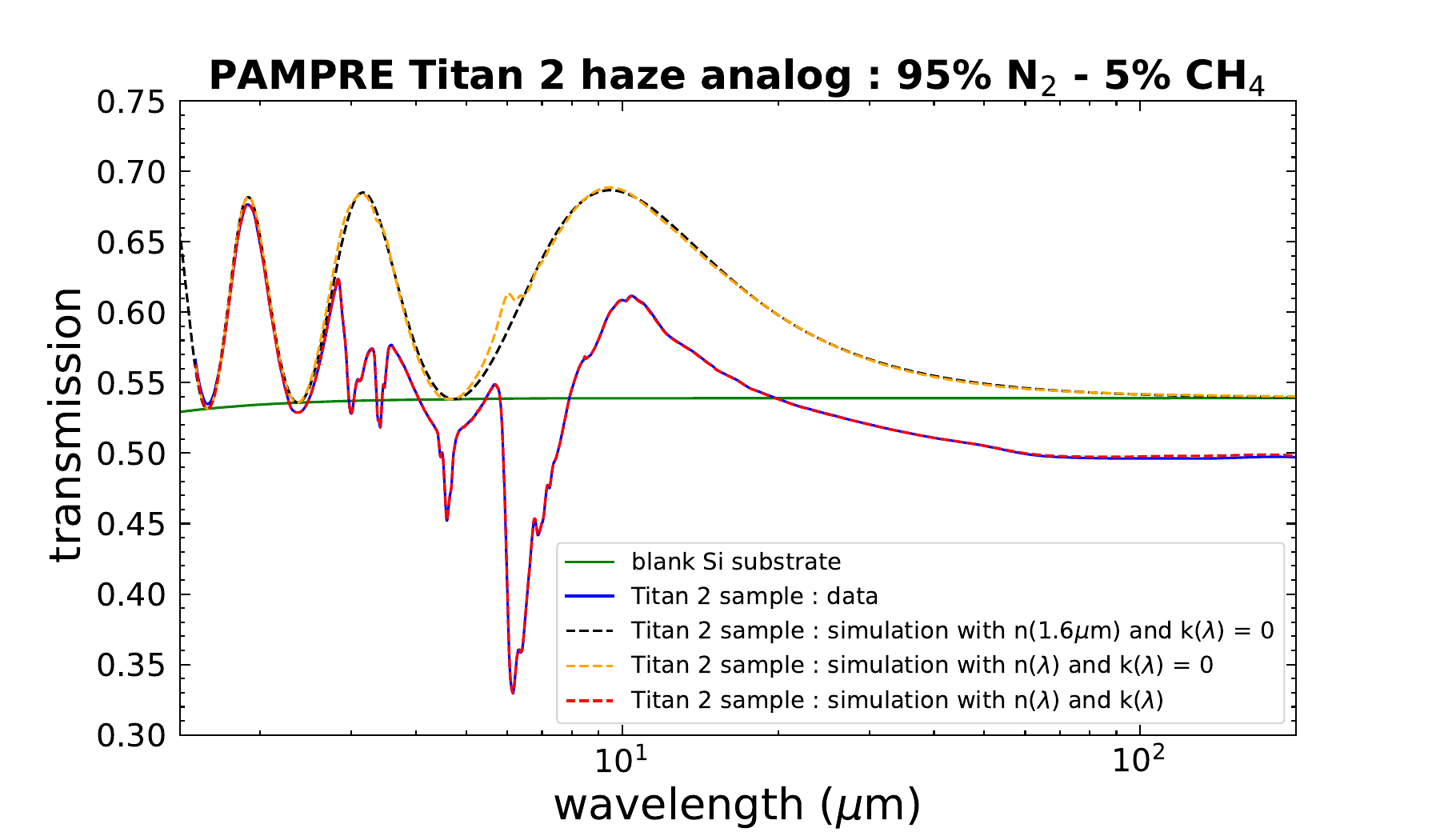   }
\includegraphics[width=0.9\columnwidth]{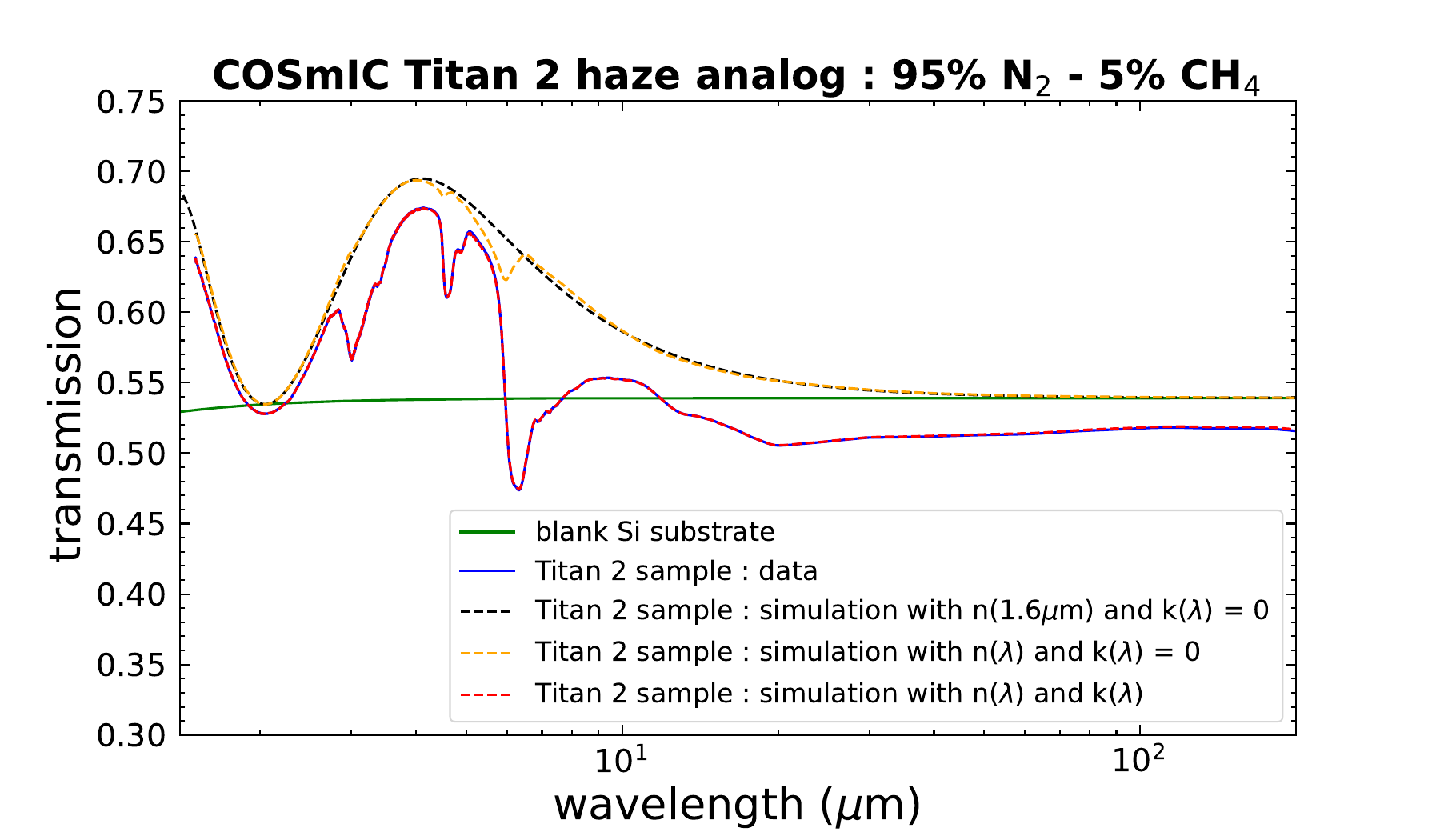   }
\includegraphics[width=0.9\columnwidth]{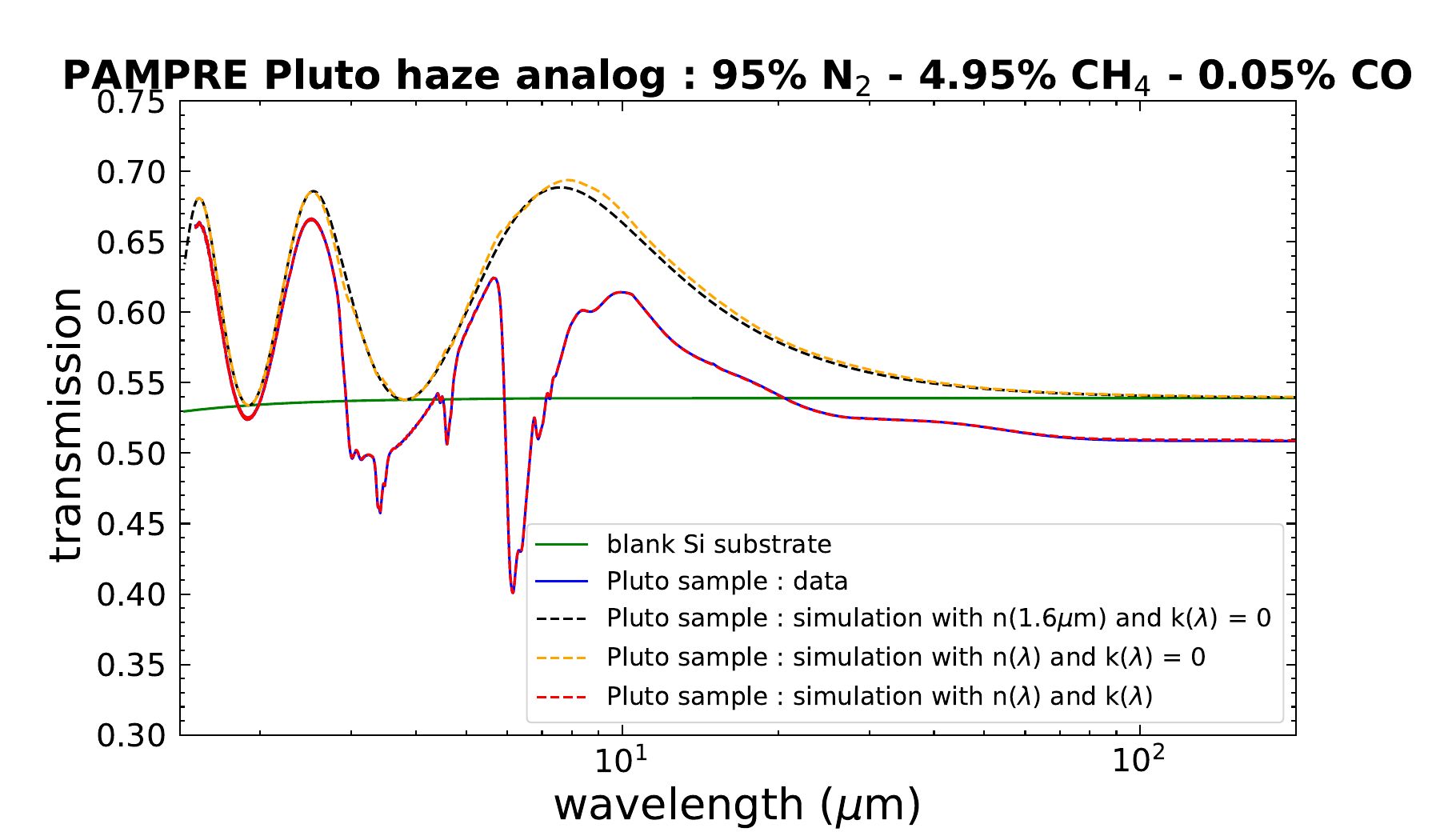   }
\includegraphics[width=0.9\columnwidth]{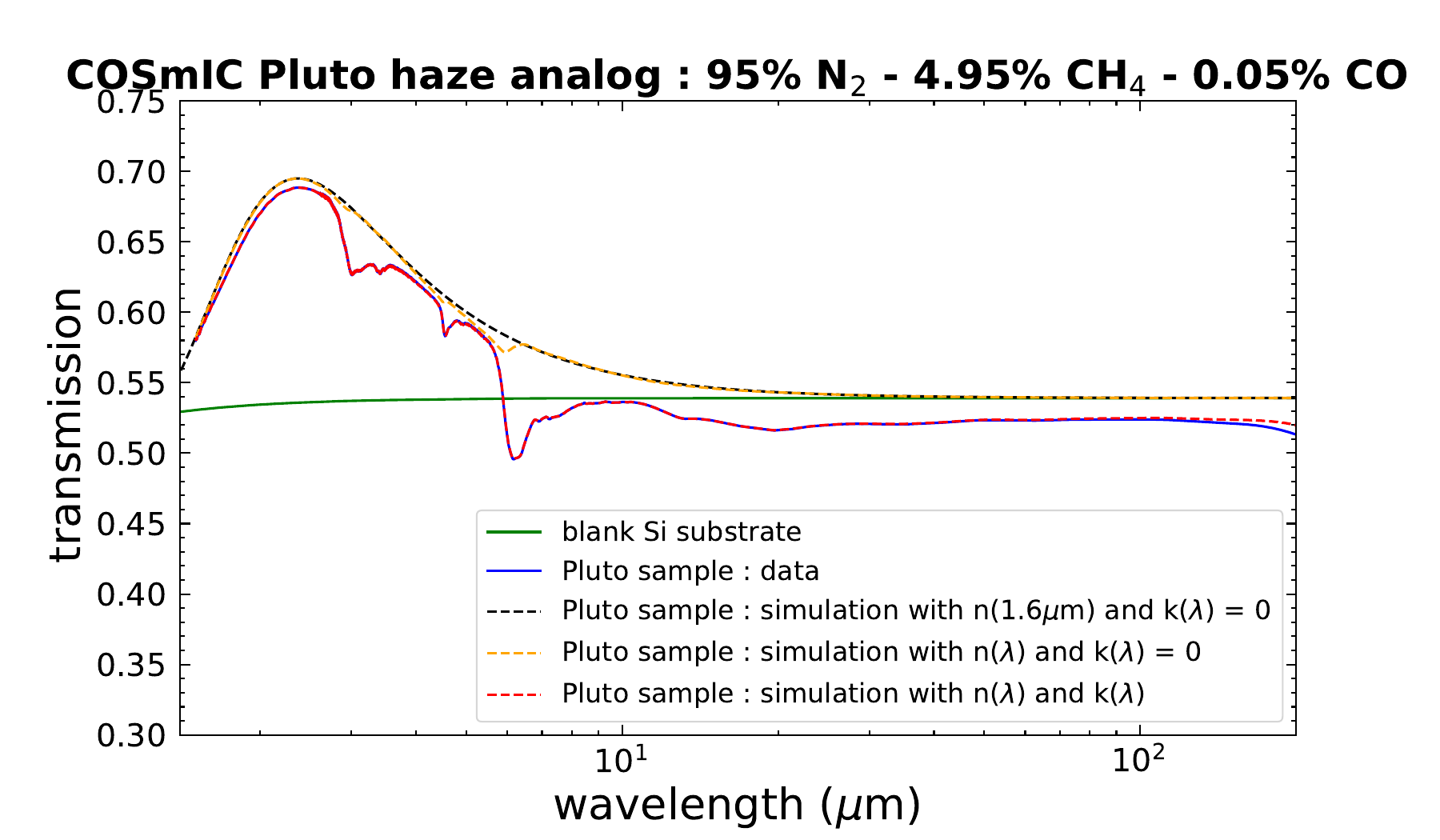   }
\caption{Transmission spectra obtained for the different PAMPRE and COSmIC haze analogs from near-IR to far-IR. The data is analyzed using the iterative model with singly-subtractive Kramers-Kronig (SSKK) loop to ensure a physical correlation between n and k. The first step presents a simulation considering the film thickness and a real refractive index constant in the entire IR range (black curve). The second step now considers variations of n in the IR after calculation with the SSKK model (orange curve). The final step presents the complete fit to the data with n and k correlated by the SSKK model (red curve).  }
\label{fig:s5}
\end{figure*}
\begin{figure*}[p]
\centering
\includegraphics[width=0.9\columnwidth]{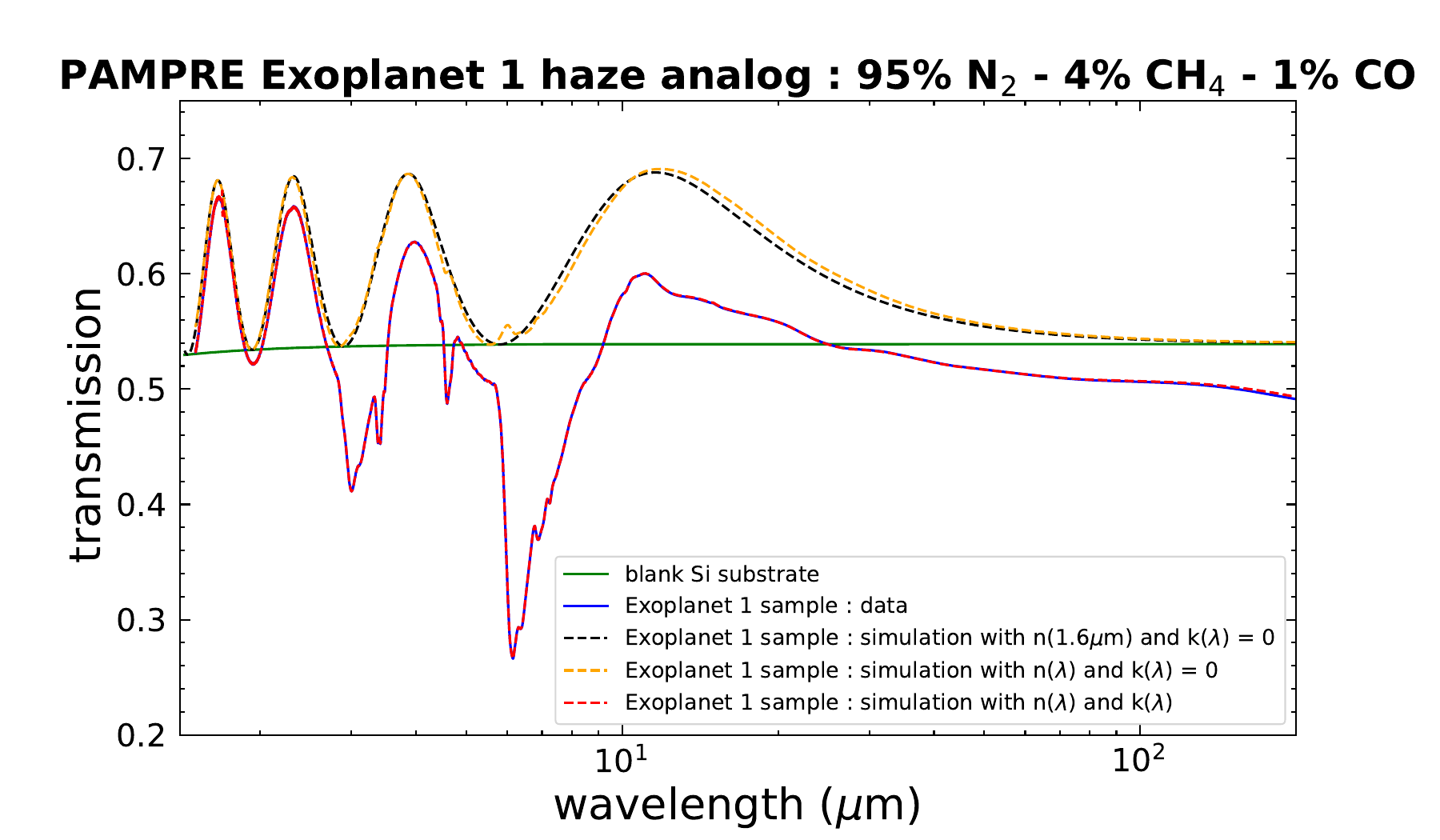   }
\includegraphics[width=0.9\columnwidth]{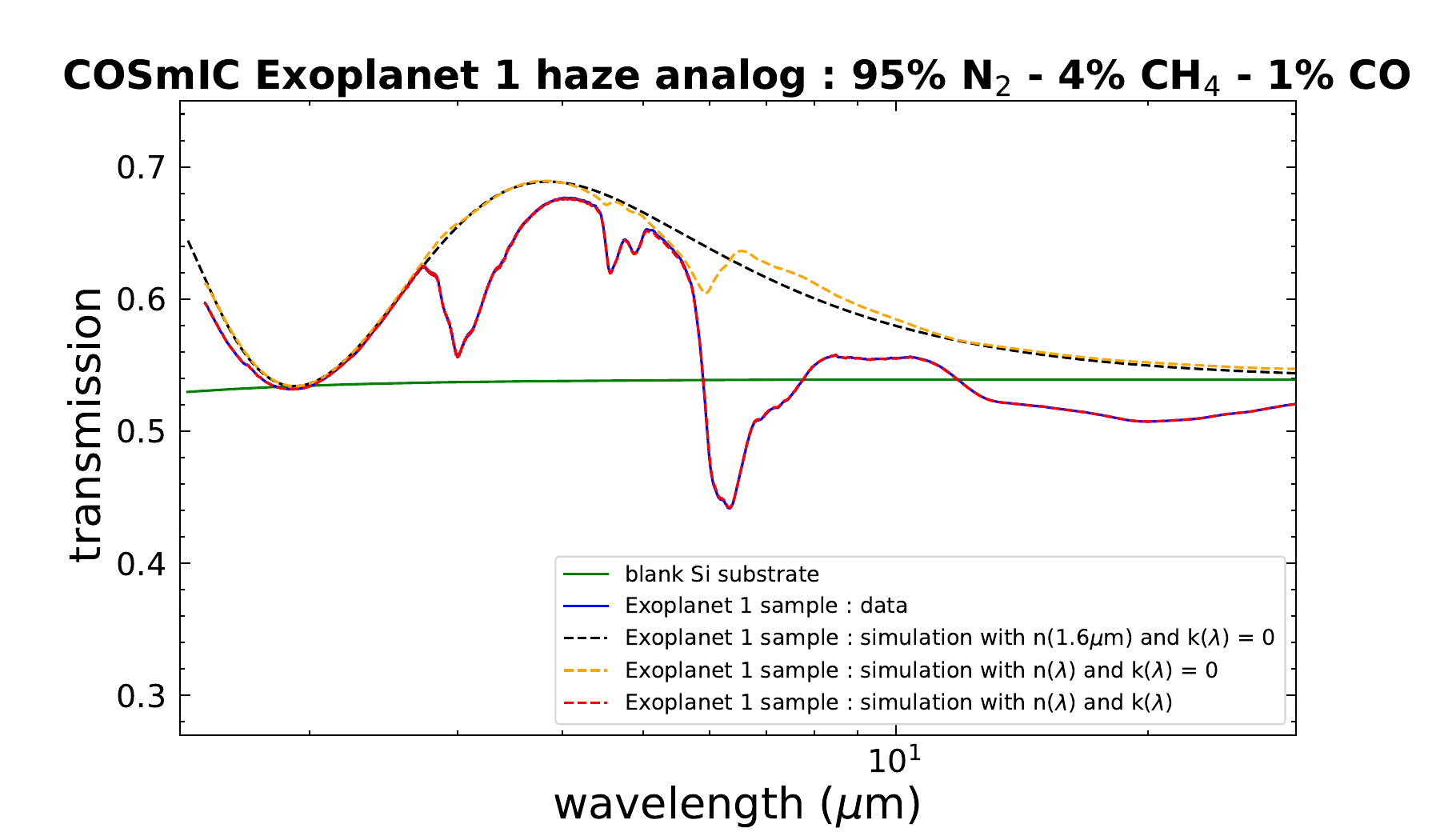   }
\includegraphics[width=0.9\columnwidth]{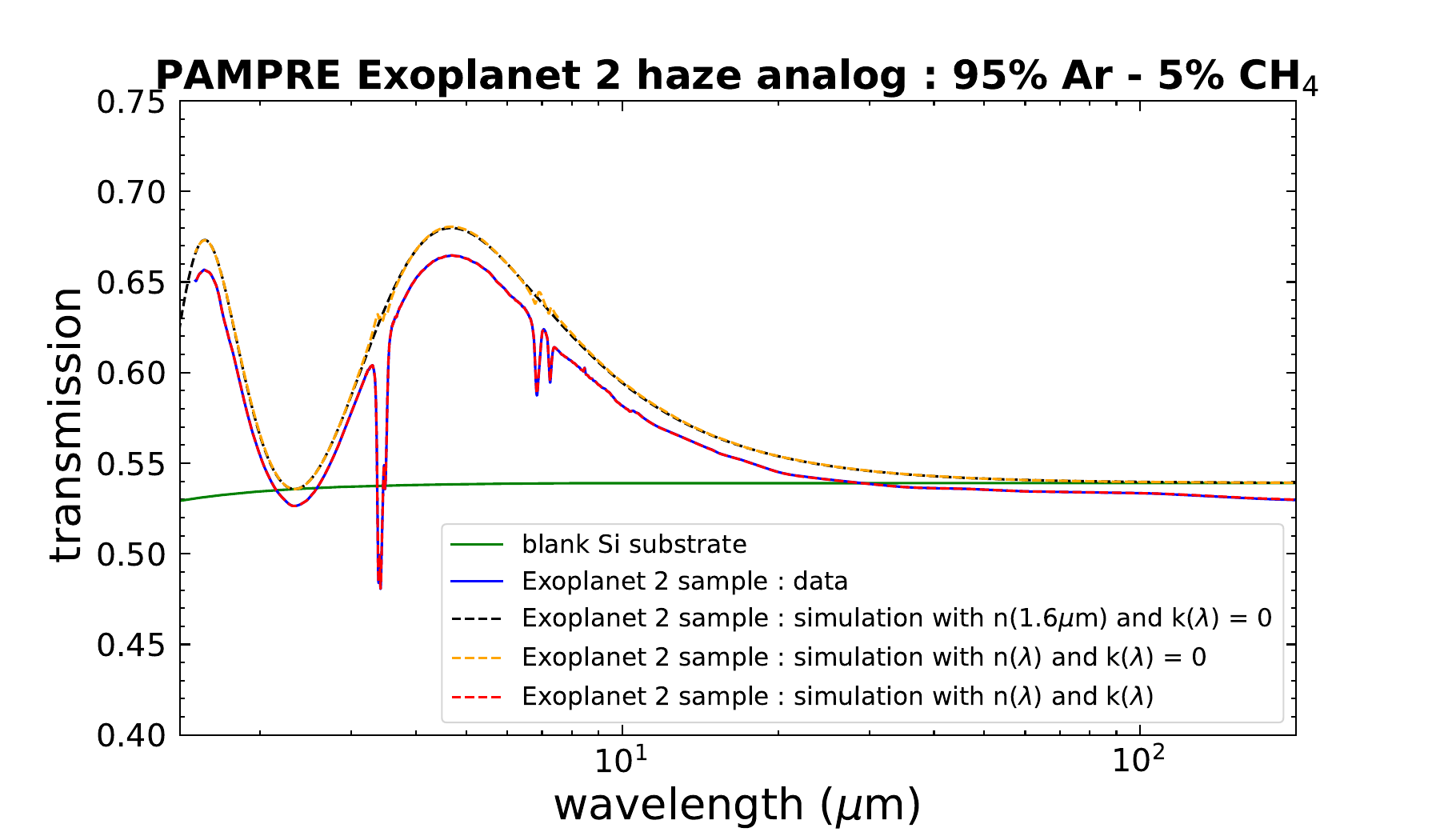   }
\includegraphics[width=0.9\columnwidth]{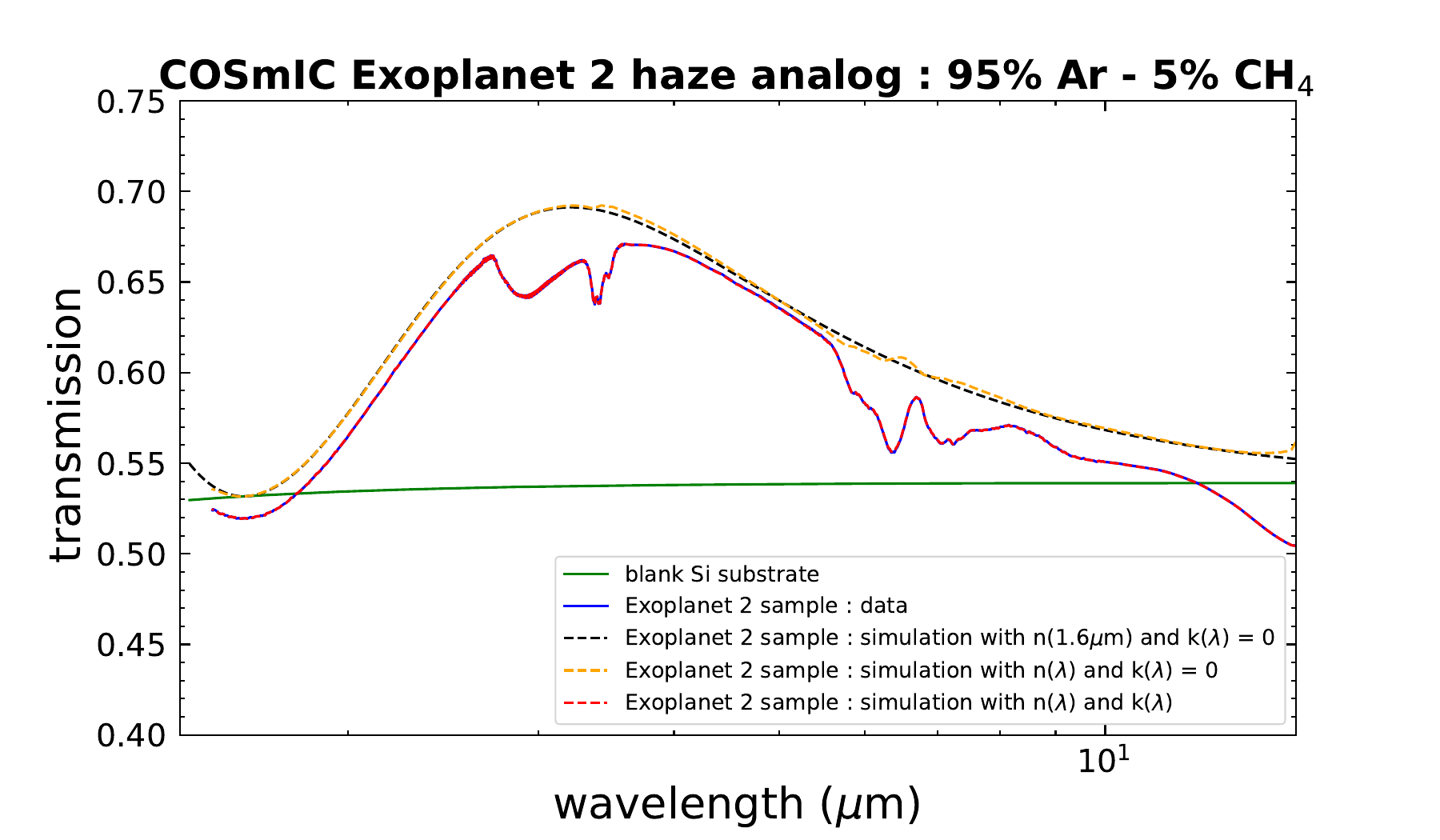   }
\includegraphics[width=0.9\columnwidth]{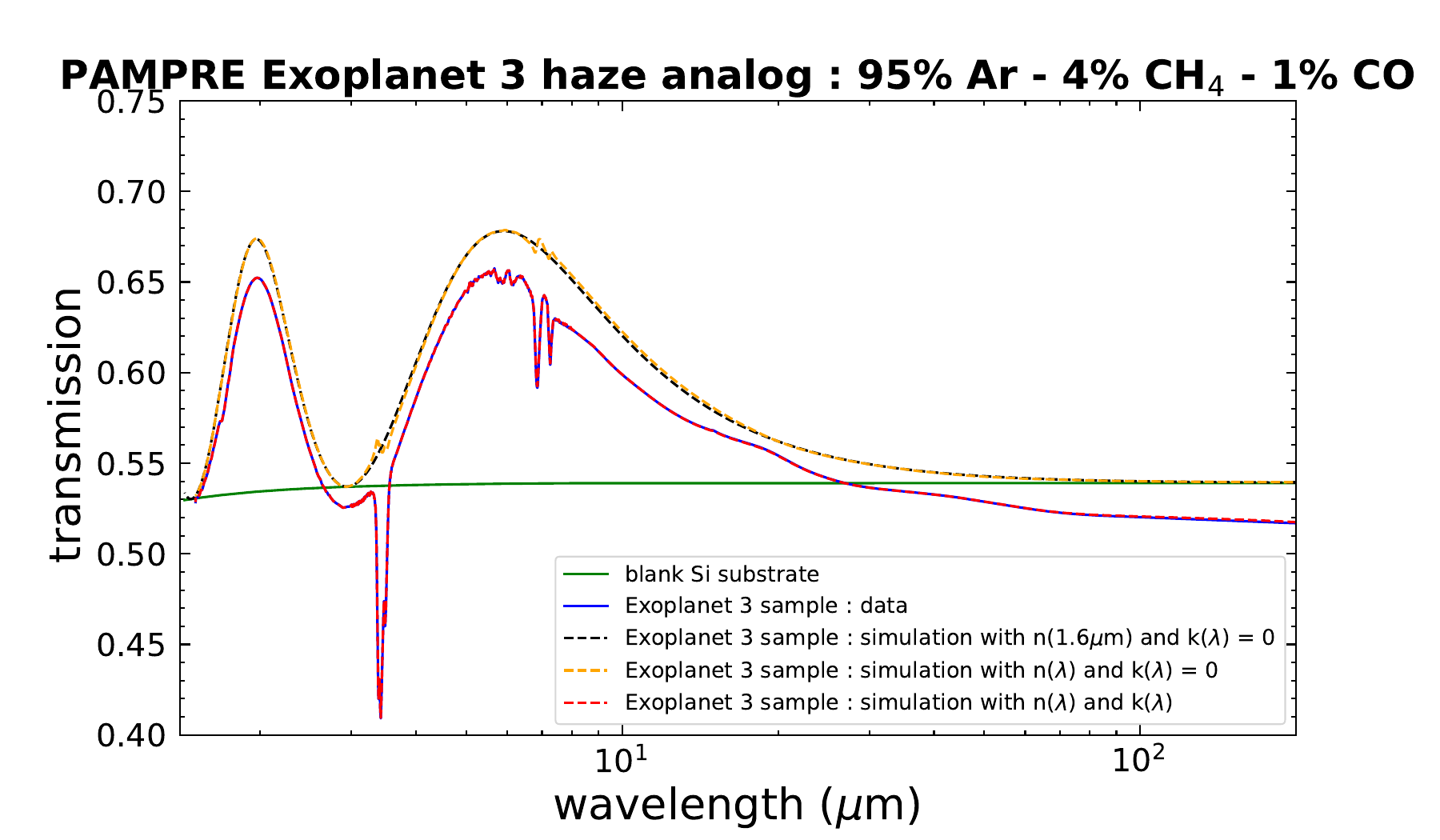   }
\includegraphics[width=0.9\columnwidth]{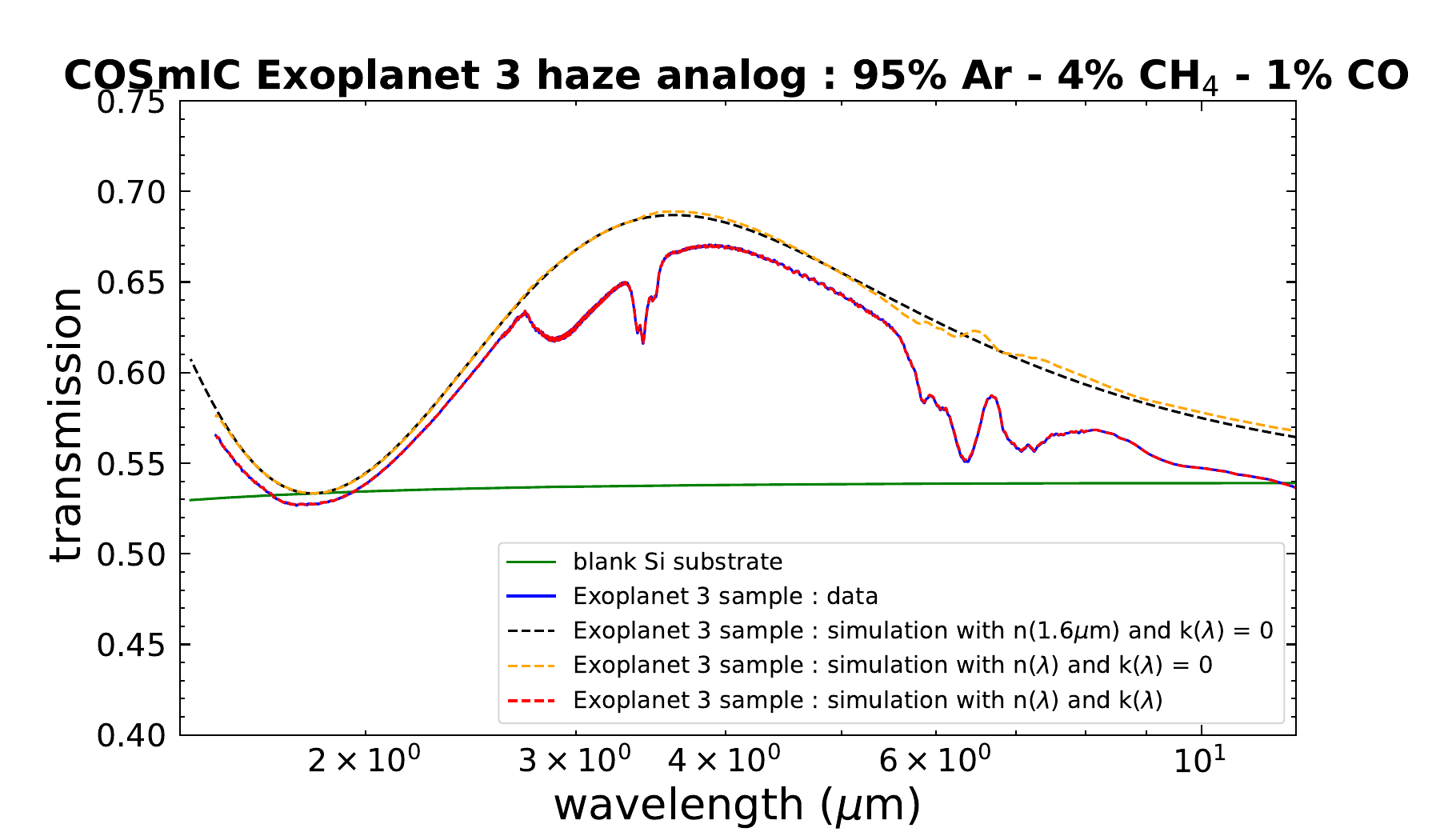   }
\label{fig:s5}
\caption{Continued   }
\label{fig:s5}
\end{figure*}

\clearpage

\renewcommand{\thefigure}{A.6} % change le label en "Figure S1"
\begin{figure}[!h]
\includegraphics[width=0.9\columnwidth]{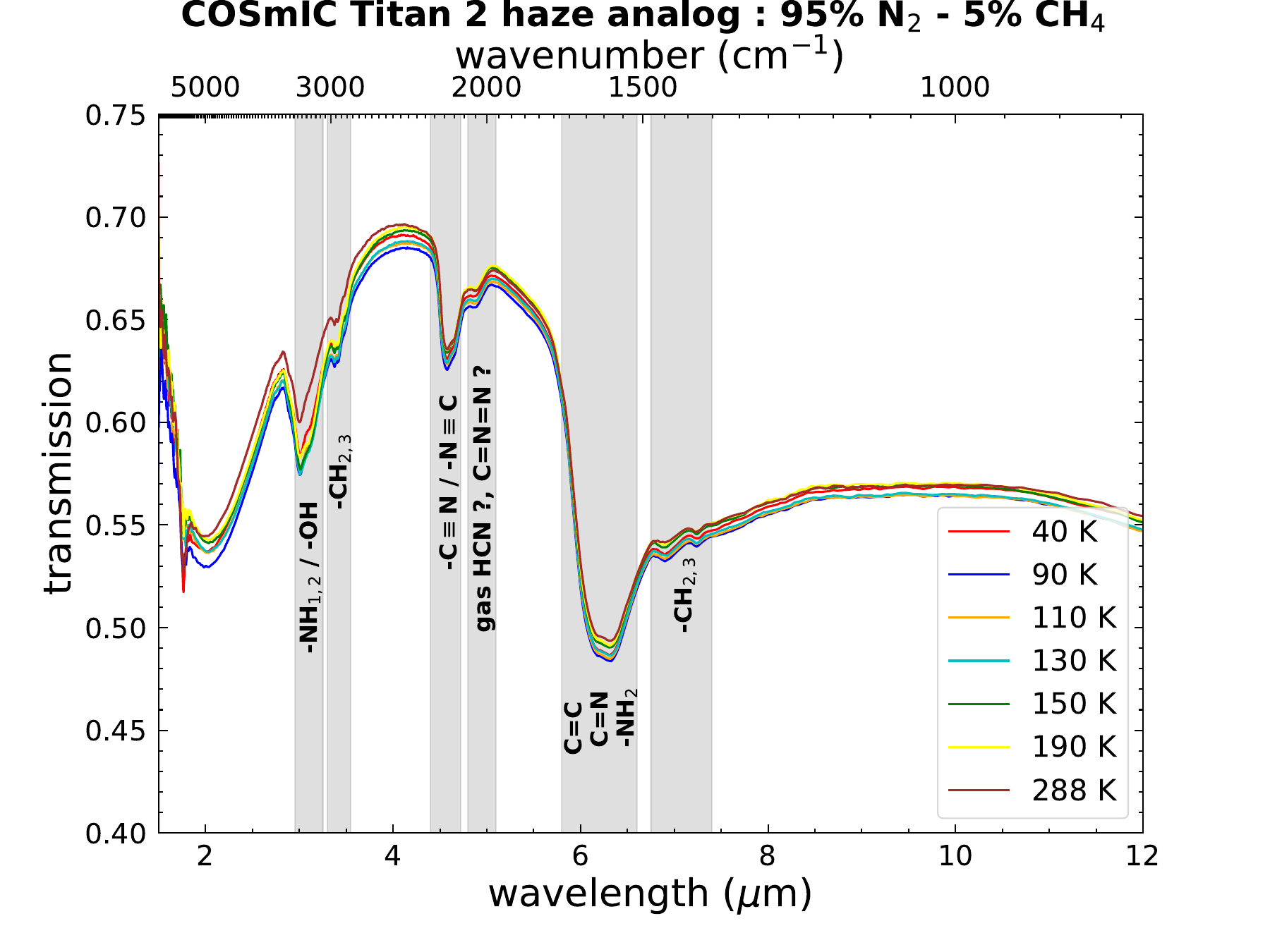   }
\includegraphics[width=0.9\columnwidth]{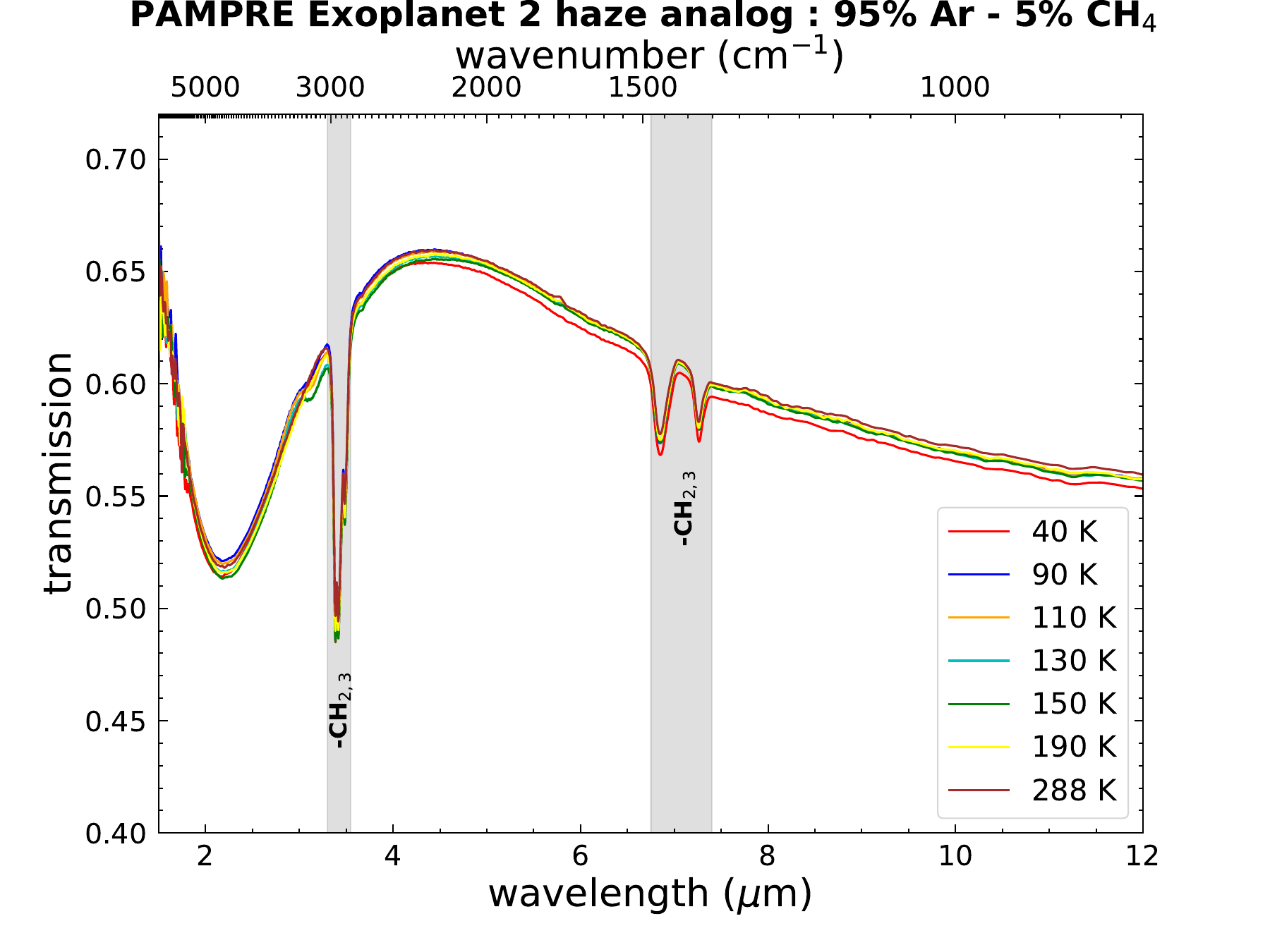   }
\includegraphics[width=0.9\columnwidth]{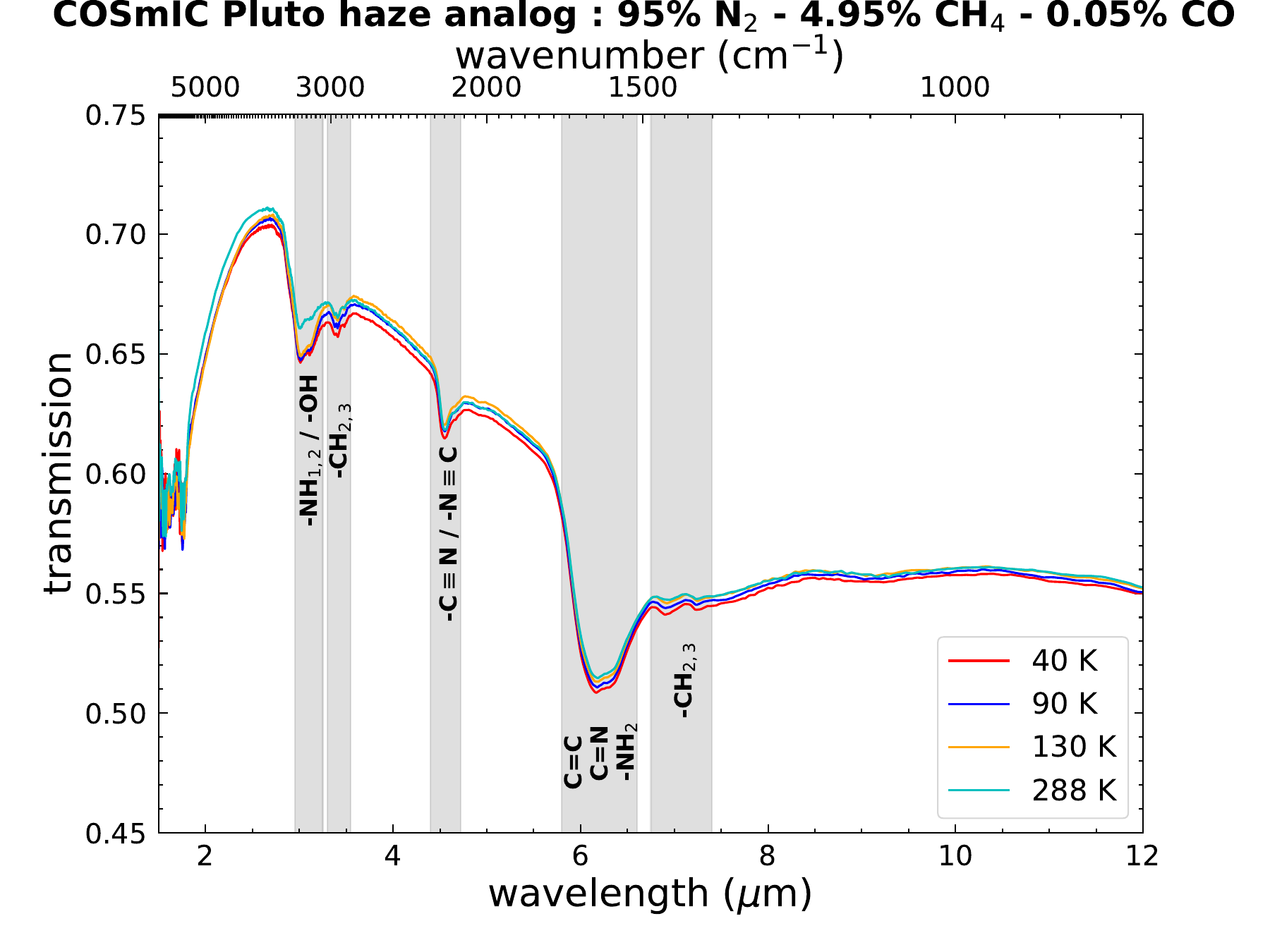   }
\captionof{figure}{Transmission spectra obtained, in the mid-IR, at low temperature with the cryostat set-up for a few PAMPRE and COSmIC haze analogs.  }
\label{fig:S6}
\FloatBarrier
\end{figure}

\section{Reflection spectra of the different COSmIC haze analogs from UV to near-IR}

Reflection measurements, at normal incidence and with high spatial resolution, were performed on the COSmIC haze analogs from Visible to near-IR, up to 1.67 $\mu$m (see Sect. 3.2). Fig. \ref{fig:S3} presents the reflection spectra for the COSmIC haze analogs (Titan 2, Pluto, Exoplanet 1, Exoplanet 3) not shown in Fig. \ref{fig:Figure4}. The spectra are fitted with a theoretical model (see Sect. 4.3) using the Cauchy and Tauc-Lorentz equations to express the refractive indices (see Fig. \ref{fig:S3}). The film thicknesses determined on each analog during the fit are shown in Table \ref{tab:S2}. Since several spectra were obtained at different positions on each sample, Table \ref{tab:S2} confirms a more significant variability of thickness across the surface of the COSmIC analogs compared to the PAMPRE analogs. Fig. \ref{fig:S4} presents the refractive indices determined for the COSmIC Titan 1 and COSmIC Pluto haze analogs with both Cauchy and Tauc-Lorentz equations. The results show that the Tauc-Lorentz model is often unable to constrain positive values of k in the NIR. Both descriptions however provide reliable k values in the region of strong absorption (i.e., in the Visible). The assumption of k = 0 below the bandgap energy in the Tauc-Lorentz description also leads to over-estimations of n (see Fig. \ref{fig:S4}) as discussed in Sect. 5.1. 

\section{Transmission spectra of the different PAMPRE and COSmIC haze analogs from near-IR to far-IR}

We performed transmission measurement from near-IR to far-IR (1.5 - 200 $\mu$m) using the Bruker IFS125HR Fourier-Transform IR (FTIR) spectrometer at the Ailes beamline of synchrotron SOLEIL (see Sect. 3.4). Fig. \ref{fig:s5} presents the transmission spectra acquired on all the PAMPRE and COSmIC haze analogs. It also shows the analysis with the iterative model, using singly-subtractive Kramers-Kronig calculations, detailed in the Sect. 4.5. Additional measurements were performed at low temperature using a cryostat set-up (details in Sect. 3.4). The measurements were performed on the COSmIC Titan 2 analog, PAMPRE Titan 2 analog, COSmIC Pluto analog and PAMPRE Exoplanet 2 analog. The spectra acquired from 40 K to 288 K are shown in Fig. \ref{fig:S6} for the analogs not shown in Fig. \ref{fig:Figure6}. For the COSmIC Pluto haze analog, less spectra were acquired since these experiments are expensive in time.

\end{appendix}


\begin{thebibliography}{3}

\bibitem[Adams et al. (2019)]{Adams19} Adams, D., Gao, P., de Pater, I., Morley, C.V. 2019, ApJ, 874, 61

\bibitem[Alcouffe et al. (2010)]{Alcouffe10} Alcouffe, G., Cavarroc, M., Cernogora, G., et al. 2010, Plasma Sources Sci. Technol., 19

\bibitem[Alves et al. (2012)]{Alves12} Alves, L.L., Marques, L., Pintassilgo, C.D., et al. 2012, Plasma Sources Sci. Technol., 21

\bibitem[Arney et al. (2016)]{Arney16} Arney, G., Domagal-Goldman, S.D., Meadows, V.S., et al. 2016, Astrobiology, 16, 11

\bibitem[Aspnes \& Studna (1983)]{Aspnes83} Aspnes, D.E. \& Studna, A.A. 1983, Phys. Rev. B, 27, 985

\bibitem[Atreya et al. (2005)]{Atreya05} Atreya, S.K., Wong, A.S., Baines, K.H., Wong, M.H., Owen, T.C. 2005, Planetary and Space Science, 53, 498

\bibitem[Baines et al. (2005)]{Baines05} Baines, K.H., Drossart, P., Momary, T.W., et al. 2005, Earth Moon Planet, 96, 119

\bibitem[Bellucci et al. (2009)]{Bellucci09} Bellucci, A., Sicardy, B., Drossart, P., et al. 2009, Icarus, 201, 198

\bibitem[Botet et al. (1997)]{Botet97} Botet, R., Rannou, P., Cabane, M. 1997, Applied Optics, 36, 8791

\bibitem[Brassé et al. (2015)]{Brasse15} Brassé, C., Muñoz, O., Coll, P., Raulin, F. 2015, Planetary and Space Science, 109, 159

\bibitem[Cabane et al. (1992)]{Cabane92} Cabane, M., Chassefière, E., Israel, G. 1992, Icarus, 96, 176

\bibitem[Campi \& Coriasso (1988)]{Campi88} Campi, D. \& Coriasso, C. 1988, Materials Letters, 7, 134

\bibitem[Carlson et al. (2016)]{Carlson16} Carlson, R.W., Baines, K.H., Anderson, M.S., Filacchione, G., Simon, A.A. 2016, Icarus, 274, 106

\bibitem[Carrasco et al. (2016)]{Carrasco16} Carrasco, N., Jomard, F., Vigneron, J., Etcheberry, A., Cernogora, G. 2016, Planetary and Space Science, 128, 52

\bibitem[Corrales et al. (2023)]{Corrales23} Corrales, L., Gavilan, L., Teal, D.J., Kempton, E.M.-R. 2023, ApJL, 943, L26

\bibitem[Courtin et al. (2015)]{Courtin15} Courtin, R., Kim, S.J., Bar-Nun, A. 2015, A\&A, 573, A21

\bibitem[Dodge (1984)]{Dodge84} Dodge, M.J. 1984, Appl. Opt., 23, 1980

\bibitem[Drant et al. (2024)]{Drant24} Drant, T., Garcia-Caurel, E., Perrin, Z., et al. 2024, A\&A, 682, A6

\bibitem[Drant et al. (2025)]{Drant25} Drant, T., Tian, M., Carrasco, N., Heng, K. 2025, A\&A, 698, A76

\bibitem[Dubois et al. (2020)]{Dubois20} Dubois, D., Carrasco, N., Jovanovic, L., et al. 2020, Icarus, 338, 113437

\bibitem[Edwards \& Ochoa (1980)]{Edwards80} Edwards, D.F. \& Ochoa, E. 1980, Appl. Opt., 19, 4130

\bibitem[Fayolle et al. (2021)]{Fayolle21} Fayolle, M., Quirico, E., Schmitt, B., et al. 2021, Icarus, 367, 114574

\bibitem[Fujiwara (2007)]{Fujiwara07} Fujiwara, H. 2007, Spectroscopic Ellipsometry : Principles and Applications, John Wiley \& Sons Ltd, The Atrium, Southern Gate, Chichester, West Sussex PO19 8SQ, England 

\bibitem[Gao et al. (2023)]{Gao23} Gao, P., Piette, A.A.A., Steinrueck, M.E., et al. 2023, ApJ, 951, 96

\bibitem[Gao et al. (2020)]{Gao20} Gao, P., Thorngren, D.P., Lee, E.K.H., et al. 2020, Nat Astron, 4, 951

\bibitem[Gao et al. (2021)]{Gao21} Gao, P., Wakeford, H.R., Moran, S.E., Parmentier, V. 2021, JGR planets, 126, 4

\bibitem[Gautier et al. (2011)]{Gautier11} Gautier, T., Carrasco, N., Buch, A., et al. 2011, Icarus, 213, 625

\bibitem[Gautier et al. (2012)]{Gautier12} Gautier, T., Carrasco, N., Mahjoub, A., et al. 2012, Icarus, 221, 320

\bibitem[Gautier et al. (2024)]{Gautier24} Gautier, T., Serigano, J., Das, K., et al. 2024, A\&A, 690, A165

\bibitem[Gavilan Marin et al. (2020)]{Gavilan20} Gavilan Marin, L., Bejaoui, S., Haggmark, M., et al. 2020, ApJ, 889, 101

\bibitem[Gavilan et al. (2017)]{Gavilan17} Gavilan, L., Broch, L., Carrasco, N., Fleury, B., Vettier, L. 2017, ApJL, 848, L5

\bibitem[Gavilan et al. (2018)]{Gavilan18} Gavilan, L., Carrasco, N., Hoffmann, S.V., Jones, N.C., Mason, N.J. 2018, ApJ, 861, 110

\bibitem[Gladstone et al. (2016)]{Gladstone16} Gladstone, G.R., Stern, S.A., Ennico, K., et al. 2016, Science, 351, 6279

\bibitem[Gladstone \& Young (2019)]{Gladstone19} Gladstone, G.R. \& Young, L.A. 2019, Annu. Rev. Earth Planet. Sci., 47, 119

\bibitem[Grundy et al. (2018)]{Grundy18} Grundy, W.M., Bertrand, T., Binzel, R.P., et al. 2018, Icarus, 314, 232

\bibitem[Guerlet et al. (2020)]{Guerlet20} Guerlet, S., Spiga, A., Delattre, H., Fouchet, T. 2020, Icarus, 351, 113935

\bibitem[Harbecke (1986)]{Harbecke86} Harbecke, B. 1986, Appl. Phys. B, 39, 165

\bibitem[Hasenkopf et al. (2010)]{Hasenkopf10} Hasenkopf, C.A., Beaver, M.R., Trainer, M.G., et al. 2010, Icarus, 207, 903

\bibitem[Hawranek et al. (1976)]{Hawranek76} Hawranek, J.P., Neelakantan, P., Young, R.P., Jones, R.N. 1976, Spectrochimica Acta, 32, 85

\bibitem[He et al. (2018)]{He18} He, C., Hörst, S.M., Lewis, N.K., et al. 2018, ApJL, 856, L3

\bibitem[He et al. (2020)]{He20} He, C., Hörst, S.M., Lewis, N.K., et al. 2020, Nat Astron, 4, 986

\bibitem[He et al. (2022a)]{He22b} He, C., Hörst, S.M., Radke, M., Yant, M. 2022a, Planet. Sci. J., 3, 25

\bibitem[He et al. (2024)]{He23} He, C., Radke, M., Moran, S.E., et al. 2024, Nat astron, 8, 182

\bibitem[He et al. (2022b)]{He22} He, C., Serigano, J., Hörst, S.M., Radke, M., Sebree, J.A. 2022b, ACS Earth Space Chem., 6, 2295

\bibitem[Hörst (2017)]{Horst17} Hörst, S.M. 2017, JGR Planets, 122, 432

\bibitem[Hörst \& Tolbert (2014)]{Horst14} Hörst, S.M. \& Tolbert, M.A. 2014, ApJ, 781, 53

\bibitem[Hörst et al. (2018)]{Horst18} Hörst, S.M., Yoon, Y.H., Ugelow, M.S., et al. 2018, Icarus, 301, 136

\bibitem[Hu et al. (2024)]{Hu24} Hu, R., Bello-Arufe, A., Zhang, M., et al. 2024, Nature, 630, 609

\bibitem[Imanaka et al. (2012)]{Imanaka12} Imanaka, H., Cruikshank, D.P., Khare, B.N., McKay, C.P. 2012, Icarus, 218, 247

\bibitem[Imanaka et al. (2004)]{Imanaka04} Imanaka, H., Khare, B.N., Elsila, J.E., et al. 2004, Icarus, 168, 344

\bibitem[Jansen et al. (2025)]{Jansen25} Jansen, K.T., Reed, N.W., Browne, E.C., Tolbert, M.A. 2025, Astrobiology, 25, 395

\bibitem[Jellison \& Modine (1996)]{Jellison96} Jellison, G.E. \& Modine, F.A. 1996, Appl. Phys. Lett., 69, 371

\bibitem[Jenkins \& White (1981)]{Jenkins81} Jenkins, F.A. \& White, H.E. 1981, Fundamentals of Optics (Auckland:McGraw-Hill), 482

\bibitem[Jovanovic et al. (2021)]{Jovanovic21} Jovanovic, L., Gautier, T., Broch, L., et al. 2021, Icarus, 362, 114398

\bibitem[Jovanovic et al. (2020)]{Jovanovic20} Jovanovic, L., Gautier, T., Vuitton, V., et al. 2020, Icarus, 346, 113774

\bibitem[Khare et al. (1984)]{Khare84} Khare, B.N., Sagan, C., Arakawa, E.T., et al. 1984, Icarus, 60, 127

\bibitem[Khare et al. (1994)]{Khare94} Khare, B.N., Sagan, C., Thompson, W.R., et al. 1994, Canadian journal of chemistry, 72, 678

\bibitem[Khare et al. (1987)]{Khare87} Khare, B.N., Sagan, C., Thompson, W.R., Arakawa, E.T., Votaw, P. 1987, J Geophys Res., 92, 15067

\bibitem[Kim \& Courtin (2013)]{Kim13} Kim, S.J. \& Courtin, R. 2013, A\&A, 557, L6

\bibitem[Kim et al. (1991)]{Kim91} Kim, S.J., Drossart, P., Caldwell, J., et al. 1991, Icarus, 91, 145

\bibitem[Kim et al. (2011)]{Kim11} Kim, S.J., Jung, A., Sim, C.K., et al. 2011, Planetary and Space Science, 59, 699

\bibitem[Kim et al. (2018)]{Kim18} Kim, S.J., Lee, D.W., Sim, C.K., et al. 2018, Journal of Quantitative Spectroscopy \& Radiative Transfer, 210, 197

\bibitem[Kitzmann \& Heng (2018)]{Kitzmann18} Kitzmann, D. \& Heng, K. 2018, MNRAS, 475, 94

\bibitem[Kramers (1927)]{Kramers27} Kramers, M.H.A. 1927, Atti. Cong. Intern. Fisica (Transactions of Volta Centenary Congress) Como, 2, p. 545.

\bibitem[Krasnopolsky (2020)]{Krasnopolsky20} Krasnopolsky, V.A. 2020, Icarus, 335, 113374

\bibitem[Kronig (1926)]{Kronig26} Kronig, R. de L. 1926, Journal of the optical society of America, 12, 547

\bibitem[Lauer et al. (2021)]{Lauer21} Lauer, T.R., Spencer, J.R., Bertrand, T., et al. 2021, Planet. Sci. J., 2, 214

\bibitem[Lavvas \& Arfaux (2021)]{Lavvas21} Lavvas, P. \& Arfaux, A. 2021, MNRAS, 502, 5643

\bibitem[Lavvas et al. (2008)]{Lavvas08a} Lavvas, P.P., Coustenis, A., Vardavas, I.M. 2008, Planetary and Space Science, 56, 27

\bibitem[Lavvas \& Koskinen (2017)]{Lavvas17} Lavvas, P. \& Koskinen, T. 2017, ApJ, 847, 32

\bibitem[Lellouch et al. (2011)]{Lellouch11} Lellouch, E., de Bergh, C., Sicardy, B., Käufl, H.U., Smette, A. 2011, A\&A, 530, L4

\bibitem[Lellouch et al. (2017)]{Lellouch16} Lellouch, E., Gurwell, M., Butler, B., et al. 2017, Icarus, 286, 289

\bibitem[Liggins et al. (2023)]{Liggins23} Liggins, P., Jordan, S., Rimmer, P.B., Shorttle, O. 2023, JGR Planets, 128, 3

\bibitem[Liggins et al. (2020)]{Liggins20} Liggins, P., Shorttle, O., Rimmer, P.B. 2020, EPSL, 550, 116546

\bibitem[Lora et al. (2018)]{Lora18} Lora, J.M., Kataria, T., Gao, P. 2018, ApJ, 853, 58

\bibitem[Madhusudhan et al. (2023)]{Madhusudhan23} Madhusudhan, N., Sarkar, S., Constantinou, S., Holmberg, M., Piette, A.A.A., Moses, J.I. 2023, ApJL, 956, L13

\bibitem[Mahjoub et al. (2012)]{Mahjoub12} Mahjoub, A., Carrasco, N., Dahoo, P.-R., et al. 2012, Icarus, 221, 670

\bibitem[Mikal-Evans (2022)]{Mikal22} Mikal-Evans, T. 2022, MNRAS, 510, 980

\bibitem[Mills et al. (2021)]{Mills21} Mills, F.P., Moses, J.I., Gao, P., Tsai, S.-M. 2021, Space Sci Rev, 217, 43

\bibitem[Moran et al. (2022)]{Moran22} Moran, S.E., Hörst, S.M., He, C., et al. 2022, JGR Planets, 127, 1

\bibitem[Niemann et al. (2005)]{Niemann05} Niemann, H.B., Atreya, S.K., Bauer, S.J., et al. 2005, Nature, 438, 779

\bibitem[Niemann et al. (2010)]{Niemann10} Niemann, H.B., Atreya, S.K., Demick, J.E., et al. 2010, JGR Planets, 115, E12

\bibitem[Nuevo et al. (2022)]{Nuevo22} Nuevo, M., Sciamma-O'Brien, E., Sandford, S.A., et al. 2022, Icarus, 376, 114841

\bibitem[Ohno \& Kawashima (2020)]{Ohno20} Ohno, K. \& Kawashima, Y. 2020, ApJL, 895, L47

\bibitem[Ohta \& Ishida (1988)]{Ohta88} Ohta, K. \& Ishida, H. 1988, Applied Spectroscopy, 42, 952

\bibitem[Olkin et al. (2017)]{Olkin17} Olkin, C.B., Ennico, K., Spencer, J. 2017, Nat Astron, 1, 663

\bibitem[Ortiz et al. (1996)]{Ortiz96} Ortiz, J.L., Moreno, F., Molina, A. 1996, Icarus, 119, 53

\bibitem[Perrin et al. (2021)]{Perrin21} Perrin, Z., Carrasco, N., Chatain, A., et al. 2021, Processes, 9, 6

\bibitem[Perrin et al. (2025)]{Perrin25} Perrin, Z., Carrasco, N., Gautier, T., Ruscassier, N., Maillard, J., Afonso, C., Vettier, L. 2025, Icarus, 429, 116418

\bibitem[Protopapa et al. (2020)]{Protopapa20} Protopapa, S., Olkin, C.B., Grundy, W.M., et al. 2020, AJ, 159, 74

\bibitem[Ramirez et al. (2002)]{Ramirez02} Ramirez, S.I., Coll, P., da Silva, A., Navarro-González, R., Lafait, J., Raulin, F. 2002, Icarus, 156, 515

\bibitem[Rannou et al. (2024)]{Rannou24} Rannou, P., Botet, R., Tazaki, R. 2024, Icarus, 424, 116247

\bibitem[Rannou et al. (2010)]{Rannou10} Rannou, P., Cours, T., Le Mouélic, S., et al. 2010, Icarus, 208, 850

\bibitem[Rannou et al. (2022)]{Rannou22} Rannou, P., Coutelier, M., Rey, M., Vinatier, S. 2022, A\&A, 666, A140

\bibitem[Reed et al. (2023)]{Reed23} Reed, N.W., Jansen, K.T., Schiffman, Z.R., Tolbert, M.A., Browne, E.C. 2023, ApJL, 954, L44

\bibitem[Reed et al. (2022)]{Reed22} Reed, N.W., Wing, B.A., Tolbert, M.A., Browne, E.C. 2022, Geophysical Research Letters, 49, 9

\bibitem[Salama et al. (2018)]{Salama18} Salama, F., Sciamma-O’Brien, E., Contreras, C.S., Bejaoui, S. 2018, in IAU Proc., 13, 364

\bibitem[Salzberg \& Villa (1957)]{Salzberg57} Salzberg, C.D. \& Villa, J.J. 1957, Journal of the Optical Society of America, 47, 244

\bibitem[Sánchez-Lavega et al. (2023)]{Sanchez-Lavega23} Sánchez-Lavega, A., Irwin, P., García Muñoz, A. 2023, Astron Astrophys Rev, 31, 5

\bibitem[Sciamma-O'Brien et al. (2012)]{Sciamma12} Sciamma-O'Brien, E., Dahoo, P.-R., Hadamcik, E., et al. 2012, Icarus, 218, 356

\bibitem[Sciamma-O'Brien et al. (2024)]{Sciamma24} Sciamma-O'Brien, E., Drant, T., Wogan, N. 2024, Nat Rev Chem, 8, 157

\bibitem[Sciamma-O'Brien et al. (2014)]{Sciamma14} Sciamma-O'Brien, E., Ricketts, C.L., Salama, F. 2014, Icarus, 243, 325

\bibitem[Sciamma-O'Brien et al. (2023)]{Sciamma23} Sciamma-O'Brien, E., Roush, T.L., Rannou, P., Dubois, D., Salama, F. 2023, Planet. Sci. J., 4, 121

\bibitem[Sciamma-O'Brien \& Salama (2020)]{Sciamma20} Sciamma-O'Brien, E. \& Salama, F. 2020, ApJ, 905, 45

\bibitem[Sciamma-O'Brien et al. (2017)]{Sciamma17} Sciamma-O'Brien, E., Upton, K.T., Salama, F. 2017, Icarus, 289, 214

\bibitem[Scipioni et al. (2021)]{Scipioni21} Scipioni, F., White, O., Cook, J.C., et al. 2021, Icarus, 359, 114303

\bibitem[Shorttle et al. (2024)]{Shorttle24} Shorttle, O., Jordan, S., Nicholls, H., Lichtenberg, T., Bower, D.J. 2024, ApJL, 962, L8

\bibitem[Socrates (2001)]{Socrates01} Socrates, G. 2001, 'Infrared and Raman characteristic group frequencies', John Wiley \& Sons LTD

\bibitem[Steinrueck et al. (2023)]{Steinrueck23} Steinrueck, M.E., Koskinen, T., Lavvas, P., et al. 2023, ApJ, 951, 117

\bibitem[Stenzel et al. (1991)]{Stenzel91} Stenzel, O., Hopfe, V., Klobes, P. 1991, J. Phys. D: Appl. Phys., 24, 2088

\bibitem[Stern et al. (2015)]{Stern15} Stern, S.A., Bagenal, F., Ennico, K., et al. 2015, Science, 350, 6258

\bibitem[Swanepoel (1983)]{Swanepoel83} Swanepoel, R. 1983, J. Phys. E: Sci. Instrum., 16, 1214

\bibitem[Szopa et al. (2006)]{Szopa06} Szopa, C., Cernogora, G., Boufendi, L., Correia, J.J., Coll, P. 2006, Planetary and Space Science, 54, 394

\bibitem[Tauc et al. (1966)]{Tauc66} Tauc, J., Grigorovici, R., Vancu, A. 1966, phys. stat. sol., 15, 627

\bibitem[Tazaki \& Tanaka (2018)]{Tazaki18} Tazaki, R. \& Tanaka, H. 2018, ApJ, 860, 79

\bibitem[Tian \& Heng (2024)]{Tian24} Tian, M. \& Heng, K. 2024, ApJ, 963, 157

\bibitem[Tomasko et al. (2009)]{Tomasko09} Tomasko, M.G., Doose, L.R., Dafoe, L.E., See, C. 2009, Icarus, 204, 271

\bibitem[Tomasko et al. (2008)]{Tomasko08} Tomasko, M.G., Doose, L., Engel, S., et al. 2008, Planetary and Space Science, 56, 669

\bibitem[Trainer et al. (2004a)]{Trainer04b} Trainer, M.G., Pavlov, A.A., Curtis, D.B., et al. 2004a, Astrobiology, 4, 409

\bibitem[Trainer et al. (2004b)]{Trainer04} Trainer, M.G., Pavlov, A.A., Jimenez, J.L., et al. 2004b, Geophysical Research Letters, 31, 17

\bibitem[Tran et al. (2003)]{Tran03} Tran, B.N., Joseph, J.C., Ferris, J.P., Persans, P.D., Chera, J.J. 2003, Icarus, 165, 379

\bibitem[Ugelow et al. (2018)]{Ugelow18} Ugelow, M.S., De Haan, D.O., Hörst, S.M., Tolbert, M.A. 2018, ApJL, 859, L2

\bibitem[Ugelow et al. (2017)]{Ugelow17} Ugelow, M.S., Zarzana, K.J., Day, D.A., Jimenez, J.L., Tolbert, M.A. 2017, Icarus, 294, 1

\bibitem[Vinatier et al. (2012)]{Vinatier12} Vinatier, S., Rannou, P., Anderson, C.M., et al. 2012, Icarus, 219, 5

\bibitem[Vuitton et al. (2009)]{Vuitton09} Vuitton, V., Lavvas, P., Yelle, R.V., et al. 2009, Planetary and Space Science, 57, 1558

\bibitem[Vuitton et al. (2021)]{Vuitton21} Vuitton, V., Moran, S.E., He, C., et al. 2021, Planet. Sci. J., 2, 2

\bibitem[Wan et al. (2021)]{Wan21} Wan, L., Zhang, X., Bertrand, T. 2021, ApJ, 922, 244

\bibitem[Wan et al. (2023)]{Wan23} Wan, L., Zhang, X., Hofgartner, J.D. 2023, ApJ, 955, 108

\bibitem[West \& Smith (1991)]{West91} West, R.A. \& Smith, P.H. 1991, Icarus, 90, 330

\bibitem[Wilson \& Atreya (2004)]{Wilson04} Wilson, E.H \& Atreya, S.K. 2004, JGR Planets, 109, E6

\bibitem[Wiscombe (1979)]{Wiscombe79} Wiscombe, W.J. 1979, University Corporation for Atmospheric Research

\bibitem[Wogan et al. (2024)]{Wogan24} Wogan, N.F., Batalha, N.E., Zahnle, K.J., Krissansen-Totton, J., Tsai, S.-M., Hu, R. 2024, ApJL, 963, L7

\bibitem[Wong et al. (2003)]{Wong03} Wong, A., Yung, Y.L., Friedson, A.J. 2003, Geophy. Res. Let., 30, 8

\bibitem[Zhang et al. (2017)]{Zhang17} Zhang, X., Strobel, D.F., Imanaka, H. 2017, Nature, 551, 352


\end{thebibliography}
\end{document}